\documentclass[10pt,journal,compsoc]{IEEEtran}
\usepackage{lmodern}
\usepackage[T1]{fontenc}
\usepackage{amsmath}
\usepackage{amsthm}
\usepackage{amssymb}
\usepackage{upgreek}
\usepackage{graphicx}
\usepackage[caption=false]{subfig}
\usepackage{color}
\usepackage{resizegather}
\usepackage{enumerate}
\usepackage{hhline}
\usepackage[shortlabels]{enumitem}
\usepackage{bm}
\usepackage{array}
\usepackage[ruled,vlined]{algorithm2e}
\usepackage{soul}
\usepackage{algorithmic}
\usepackage{multirow}

\makeatletter

\theoremstyle{definition}

\theoremstyle{plain}
\newtheorem{thm}{\protect\theoremname}

\theoremstyle{plain}
\newtheorem{lemma}{\protect\lemmaname}

\theoremstyle{plain}
\newtheorem{coro}{\protect\coroname}

\theoremstyle{plain}

\theoremstyle{definition}

\theoremstyle{definition}
\newtheorem*{defn*}{\protect\definitionname}

\theoremstyle{plain}
\newtheorem{assm}{\protect\assumptionname}

\theoremstyle{plain}
\newtheorem{remark}{\protect\remarkname}

\theoremstyle{definition}
\newtheorem*{hypo}{\protect\hyponame}

\makeatother

\usepackage[english]{babel}

\providecommand{\definitionname}{Definition}
\providecommand{\examplename}{Example}
\providecommand{\propositionname}{Proposition}
\providecommand{\theoremname}{Theorem}
\providecommand{\lemmaname}{Lemma}
\providecommand{\coroname}{Corollary}
\providecommand{\assumptionname}{Assumption}
\providecommand{\remarkname}{Remark}
\providecommand{\hyponame}{Working hypothesis}

\newcommand{\abs}[1]{\left|#1\right|} 

\newcommand{\probmod}{P(\Delta x_{i,j}\neq0)}
\newcommand{\probeq}{P(\Delta x_{i,j}=0)}
\newcommand{\probmodhat}{P(\Delta\hat x_{i,j}\mbox{$\neq$}0)}
\newcommand{\probeqhat}{P(\Delta\hat x_{i,j}\mbox{$=$}0)}
\newcommand{\prob}[1]{P\left(#1\right)}
\newcommand{\FN}{\mathrm{FN}}
\newcommand{\FP}{\mathrm{FP}}

\newcommand{\stego}{\boldsymbol S} 
\newcommand{\pred}{\boldsymbol  C}

\begin{document}

% ----> Arxiv page
\thispagestyle{empty}
\begin{minipage}[t][5cm][t]{1.\textwidth}

{\Large

{\huge \textbf{Citation for published version:} }

~

\copyright IEEE Transactions on Dependable and Secure Computing (2022).
The definitive, peer reviewed and edited version of this article is publised in:

~

David Meg\'{\i}as, Daniel Lerch-Hostalot, ``Subsequent embedding in targeted image steganalysis: Theoretical framework and practical applications''. doi: 10.1109/TDSC.2022.3154967.

~

~

{\huge \textbf{DOI} }

~

10.1109/TDSC.2022.3154967

~

~

{\huge \textbf{Document Version} }

~

This is the Accepted Manuscript version.

~

~

{\huge \textbf{Copyright and Reuse} }

~

This manuscript version is made under the terms of the Creative Commons Attribution Non Commercial No Derivatives license (CC-BY-NC-ND) http://creativecommons.org/licenses/by-nc-nd/3.0/es, which permits others to download it and share it with others as long as they credit you, but they can't change it in any way or use them commercially.

}

\end{minipage}
\newpage
\setcounter{page}{0}
% <---- Arxiv page

\title{Subsequent embedding in \textcolor{black}{targeted} image steganalysis: Theoretical framework and practical applications}

\author{David Meg\'{\i}as,~\IEEEmembership{Member,~IEEE,} and Daniel Lerch-Hostalot
\IEEEcompsocitemizethanks{\IEEEcompsocthanksitem D. Meg\'{\i}as and D. Lerch-Hostatlot are with the Internet Interdisciplinary Institute (IN3), Universitat Oberta de Catalunya (UOC), CYBERCAT-Center for Cybersecurity Research of Catalonia, Castelldefels (Barcelona), Catalonia, Spain.\protect\\
E-mail: \{dmegias,dlerch\}@uoc.edu}}

	%\markboth{IEEE TRANSACTIONS ON DEPENDABLE AND SECURE COMPUTING, VOL. X, NO. X, JANUARY 2022}%
	%{Authors: title}
	
	\maketitle

\begin{abstract}
Steganalysis is a collection of techniques used to detect whether secret information is embedded in a carrier using steganography. Most of the existing steganalytic methods are based on machine learning, which requires training a classifier with ``laboratory'' data. However, applying machine-learning classification to a new data source is challenging, since there is typically a mismatch between the training and the testing sets. In addition, other sources of uncertainty affect the steganlytic process, including the mismatch between the targeted and the actual steganographic algorithms, unknown parameters --such as the message length-- and having a mixture of several algorithms and parameters, which would constitute a realistic scenario. This paper presents subsequent embedding as a valuable strategy that can be incorporated into modern steganalysis. Although this solution has been applied in previous works, a theoretical basis for this strategy was missing. Here, we cover this research gap by introducing the ``directionality'' property of features concerning data embedding. Once a consistent theoretical framework sustains this strategy, new practical applications are also described and tested against standard steganography, moving steganalysis closer to real-world conditions. 
\end{abstract}


\IEEEpeerreviewmaketitle{}

\section{Introduction}
\label{sec:intro}

\bstctlcite{IEEEexample:BSTcontrol}

The main objective of steganalysis is to detect the presence of secret data embedded into apparently innocent objects using steganography. These objects are mainly digital media and the most commonly used carriers for steganography are digital images.
{\color{black} Currently, most state-of-the-art methods are adaptive \cite{Li:2014:hill, Holub:2014:uniward, Liao2020a, Liao2020b}}. The most successful detectors are based on machine learning. The usual approach is to prepare a database of images and use them for training a classifier. This classifier will be used to classify future images in order to predict whether they contain a hidden message or not.

The first few machine learning-based methods used in steganalysis \cite{Avcibas:2003,Farid:2003} work in two steps. The first step is to extract sensitive features to changes made when hiding a message in an image. The second step is to train a classifier for predicting if future images are \emph{cover} (which do not contain an embedded message) or \emph{stego} (which contain an embedded message). These features were tailored to be effective against the specific steganographic methods to be detected. There are various feature extraction methods and their development has led up to what we know as Rich Models \cite{Fridrich:2012:RM,
kodovsky:2012:JPG,
%kodovsky:2012:JPG,
Song:2015:GFR}. Different types of classifiers can be used for steganalysis, although Support Vector Machines \cite{Farid:2003} and Ensemble Classifiers \cite{Kodovsky:2012:EC} were common a few years ago.

During the last few years, we have seen how those technologies are being replaced by classification in a single step using Convolutional Neural Networks (CNN). Most of CNN's proposals for steganalysis consist of custom-designed networks, often extending the ideas previously used in Rich Models \cite{qian:2015,Xu:2016,Ye:2017,
Yedroudj:2018,
%Yedroudj:2018,
Boroumand:2019:SRNet, Yousfi:2020:onehot}. Lately, better results have been obtained using networks designed for computer vision, pre-trained with Imagenet, such as EfficientNet, MixNet or ResNet \cite{Yousfi:2019:breaking_alaska1, Yousfi:2020:alaska2}.

With the use of machine learning in steganalysis, the problem of Cover Source Mismatch (CSM) arises. This problem refers to classifying images that come from a different source than those used to train the classifier. The first references about CSM in the literature can be found in \cite{Goljan:2006,Cancelli:2008:csm}. Among the main reasons that cause CSM are the ISO sensitivity and the processing pipeline. A recent analysis shows that the processing pipeline could have the greatest impact \cite{Giboulot:2020}.

Different approaches to CSM have been tried in recent years. In \cite{Fridrich:2014:csm_mitig}, different strategies are proposed to mitigate the problem, such as training with a mix of sources, using classifiers trained with different sources each, and using the classifier with the closest source. In \cite{Pasquet:2014}, the \emph{islet approach} method is proposed, which organizes the images into clusters and assigns a classifier to each one. Nowadays, the usual approach to face CSM is trying to build a sufficiently large and diverse training database with images from different cameras and with different processing pipelines \cite{Xu:2015:csm, cogranne:2019:alaska:into_the_wild,
% cogranne:2019:alaska:into_the_wild
cogranne:2020:alaska2,ruiz:2021}. With this approach, the main aim is that the machine learning method focuses on the features that are universal to all images and allow classifying images that come from different sources correctly. However, to date, there is no sufficiently complete database to allow sufficient mitigation of the CSM. In fact, given a training database, it is usually straightforward to find a set of images with CSM such that the classifier error increases significantly \cite{Lerch-Hostalot:2019}.  

CSM is not the only source of uncertainty in the steganalytic process. In targeted steganalysis, for example, the steganalyst makes an assumption about the steganographic scheme used by the steganographer. If this assumption is wrong, and the steganographic scheme assumed by the steganalyst differs from the true one, stego source mismatch (SSM)\footnote{\color{black} In this paper, the term SSM is used only if the steganographic scheme applied  by the steganalyst to create stego samples differs from the one used by the steganographer (for the testing set).} occurs \cite{Butora:2020:compat}. If the steganographic algorithm is guessed correctly, but the length of the message (or the embedding bit rate) is unknown, there is another source of uncertainty, referred to as unknown message length \cite{westfeld,Fridrich2003, Pevny:2011}.
%,Pevny:2011}). 
Finally, we can have a mixture of those situations: CSM, SSM, and unknown message lengths, with different image sources, algorithms and message lengths in the same testing set. This situation is typically referred to as steganalysis ``into the wild'' or real-world steganalysis \cite{Ker:2013:real_world, cogranne:2019:alaska:into_the_wild}. These situations are also considered in this paper. 

A deeper analysis of related work, the comparison of the proposed approach with prior art, and the main contributions of this work are detailed in Appendix A (supplemental material).

\emph{Subsequent embedding} has been suggested as a tool to increase the accuracy of steganalysis. This technique is effective in dealing with some cases of CSM. In \cite{Lerch-Hostalot:2015}, the use of subsequent embedding to create an unsupervised detection method was shown, making it possible to bypass the CSM problem by skipping the training step. In \cite{Lerch-Hostalot:2019}, this technique is used to predict the classification error, thus detecting when the classifier is not appropriate to classify a given testing set. \textcolor{black}{With the methods described in this paper, some images can be labeled as ``inconsistent'' during steganalysis. If they are not classified as either cover or stego, the classification accuracy increases compared to the standard steganalysis approach, thus increasing the accuracy obtained with standard steganalysis.}

The basis of those methods is to carry out additional (random) data embeddings --using steganography-- to the images of the  training and the testing sets. These additional data embeddings are assumed to distort some relevant features in the same direction as the first application of steganography. In other words, a set of relevant features of the images that have two subsequent messages embedded into them are expected to be significantly different from the corresponding features of the images with a single embedded message. This property is called ``directionality'' of the features concerning steganography and has not been explored from a theoretical standpoint in the state-of-the-art. This paper takes a theoretical perspective on the directionality property and establishes sufficient conditions for it to be satisfied for a simple feature model. Then, the analysis is extended empirically to illustrate the property in state-of-the-art feature models. After that, some applications of subsequent embedding, beyond the published ones, are presented and analyzed. 

The rest of this paper is organized as follows. Section \ref{sec:defs} provides basic definitions and notations that are used throughout the paper. Section \ref{sec:simple} presents the simple feature model used for the theoretical analysis and provides several theorems and proofs related to directional features.
\textcolor{black}{Section \ref{sec:theoretical} provides a theoretical analysis of the simple model}. Section \ref{sec:validation} validates the theoretical model and analyzes the directionality of features for true feature models. Section \ref{sec:practical} presents four different practical applications of subsequent embedding in image steganalysis. Finally, Section \ref{sec:conclusion} concludes the paper and suggests future research directions.

\section{Basic definitions}
\label{sec:defs}
This section introduces the definitions that are used throughout the paper. %and describes a simple model for feature-based steganalysis that is analyzed in other sections.

\subsection{Cover image}
Let $X$ be an image of $w\times h$ pixels, i.e. $$X=\left(x_{i,j}\right) : {(i,j)=(1,1),\dots,(w,h)},$$
where $w$ and $h$ are, respectively, the width and the height of the image. The values of the pixels are assumed to be $n$-bit unsigned integers, i.e. $x_{i,j}\in[0,\dots,2^n-1]$.

\subsection{Stego image}
Let $\stego$ be a steganographic embedding scheme that takes the cover image $X=\left(x_{i,j}\right)$ as input and outputs a stego image $X'=\left(x'_{i,j}\right)$ for $(i,j)=(1,1),\dots,(w,h)$. Typically, the steganographic embedding function requires some parameters $p$ (e.g. the embedding bit rate or payload), a (secret) stego key $K\in\mathbf K$ and a secret message $m\in\mathbf M$, where $\mathbf K$ is the key space and $\mathbf M$ is the message space\footnote{Note that the set of all possible embedded messages ($\mathbf M$) is finite, since digital images are limited in size and, hence, the embeddable message sizes are also limited.}. Thus, we can write $X'={\stego}_p^K(X,m)$. In the sequel, we use ${\stego}_p$ always to remark that the steganographic embedding function is considered for a particular selection of parameters.

It is worth pointing out that the parameters $p$ and the message $m$ are not completely independent. For an image of $h\times w$ pixels and a message $m$ of length $\abs{m}$, the embedding bit rate (included in $p$) should match $\abs{m}/(h\times w)$. For this reason, $\mathbf M$ represents the set of all possible messages of \textbf{appropriate size} to be compatible with the embedding bit rate and the image size.

%{\color{black}Estem indicant bitrate y missatge com a parametres diferents, i son el mateix. De fet, hi ha tres coses que son el mateix: el payload, el bitrate y el missatge. El que mes es fa servir es el payload, aixi que l'hauriem d'esmentar, que correspon al percentatge de la capacitat total de la imatge. Com que es fa servir algun tipus de matrix embedding, no coincideix amb el bitrate. En les eines reals es fa servir directament el missatge, i el bitrate es calcula, tot i que a vegades no es pot i part del missatge no entra. En el nostre cas, en base a la teoria, ens interessa el bitrate, pero penso que no estaria de mes anomenar els altres, per evitar confusions.}

\subsection{Small modification and low probability  of modification}
\label{sec:lowprob}
We assume that the steganographer tries to maximize statistical undetectability and, hence, ${\stego}_p$ limits the changes to pixels to $\pm 1$ operations (at most), i.e. $\left|x'_{i,j}-x_{i,j}\right|\leq 1$. In fact, if pixel changes were greater than $\pm 1$, we would expect that the statistical properties of the stego image would be distorted to a larger extent and the steganographic scheme would possibly be more detectable. Hence, from the steganalysis point of view, $\pm 1$ changes can be considered a worst-case scenario.

In addition, the probability that a given pixel is selected by the embedding algorithm is $\alpha\in(0,1]$ (where 0 means no pixel is selected and 1 means all pixels are selected). Typically, $\alpha$ is close to zero for  highly undetectable methods. For the pixels that are selected, on average, half of them will have the correct LSB, whereas the other half will require a $\pm 1$ change (i.e. $\left|x'_{i,j}-x_{i,j}\right|= 1$). Hence, the probability that a pixel is changed is $\beta=\alpha/2$ (and the probability that a pixel is unchanged is $1-\beta$).

Note that $\alpha$ does not necessarily match the embedding bit rate or payload. Some methods use matrix embedding or Syndrome-Trellis Codes (STC) \cite{Fil:2011} and can embed more bits than pixels are selected for modification. For example, using the HIgh-pass, Low-pass, Low-pass (HILL) cost function for steganography \cite{Li:2014:hill}, a payload of $0.4$ bits per pixel (bpp) would be obtained with $\alpha\approx 0.17$
\cite{Lerch-Hostalot:2019}.

%{\color{black}Com deia abans, a la literatura normalment aixo s'anomena payload. Per altra banda HILL es una funcio de cost que es limita a dir quin cost te amagar a cada pixel, no te res a veure amb la incrustacio. La incrustacio es fa, en teoria, amb els STC, pero tampoc es aixi en aquest cas ($\alpha=0.17$), doncs els experiments estan fets amb el simulador, i el simulador fa servir el "bound" teoric. Segurament amb STC el $\alpha$ sera inferior.  }

\subsection{Domains of cover, stego and ``steganalyzable'' images}
Formally, we define the domain of ``steganalyzable'' images $\mathcal I$ as the union of the domains of all cover ($\mathcal C$) and all stego ($\mathcal S$) images, or $\mathcal I=\mathcal C \cup \mathcal S$, where $\mathcal C \cap \mathcal S=\emptyset$. Any image either contains or does not contain an embedded message using the steganographic function ${\stego}_p$. Thus, the domains $\mathcal C$ and $\mathcal S$ are assumed to be disjoint.

\begin{remark}
\label{rem:effective}
The condition that $\mathcal C \cap \mathcal S=\emptyset$ is required for effective steganalysis. Actually, all cover images are also stego images, since all images contain some ``default'' (random-like) message for a given key $K\in \mathbf K$ and parameter vector $p$. However, such a ``default'' message is typically useless since it will not match the true message $m$ to be transmitted in a secret communication. The probability of finding an image that already contains the intended message $m$, known as steganography by image selection, decreases exponentially with $\abs{m}$, and is considered infeasible for typical (useful) message lengths. If the message is concise, it is possible to find a cover image that already contains it. However, in such a case, steganalysis cannot be successful. There would no way of classifying an image as cover or stego if the same image can be both cover and stego. Hence, we assume this ``effective'' disjointness for ``useful'' messages throughout the paper.
%{\color{black} Per a qualsevol imatge y qualsevol clau, existeix un missatge de llargada arbitraria que es pot amagar sense modificar cap bit (el missatge que coincideix amb l'LSB), no?. Es a dir, que totes les imatges cover, son tambe stego. }
\end{remark}

Now, we introduce the notation $\stego^*_p(\cdot)$ to denote the hypothesized application of the embedding algorithm to an image (or a set of images) for all possible secret keys and all possible embedded messages, i.e. given an image $X=(x_{i,j})$:
\[
\stego^*_p(X)= \left\{\stego^K_p(X,m) \mid \forall K\in\mathbf{K},\forall m \in\mathbf{M}\right\}.
\]
With this notation, all stego images for the embedding algorithm ${\stego}_p$ can be described as follows:
\[
\mathcal S=\stego^*_p(\mathcal C)=\left\{\stego^*_p(X) \mid \forall X\in\mathcal C
\right\}.
\]

\subsection{Feature-based steganalysis}
Steganalysis is defined as a classification problem that, given an image $Z=\left(z_{i,j}\right)$, tries to determine whether this image is cover or stego. 
This problem is typically overcome by using some classifier $\pred$. More formally, an ideal steganalytic classifier can be defined as a classification function
\[
\pred:\mathcal{I}\rightarrow\{0,1\},
\]
where $\pred(Z)=0$ if $Z\in\mathcal C$ and $\pred(Z)=1$ if $Z\in\mathcal S$. A real classifier will not perform as well as the ideal one, and some probability of misclassification exists (in the form of false positives or false negatives).

In this paper, we assume \textbf{targeted steganalysis}, since the classification is carried out for a specific steganographic algorithm $\stego$ subject to a set of parameters $p$. On the other hand, universal steganalysis refers to a classification function $\pred_*$ that would work for any steganography. 

Frequently, such a classification depends on a set of features, either explicit or implicit. In that case, the images are not classified directly, but a feature vector is first extracted from the images, using some function $\mathcal{F}$, and the classifier uses such features. $\mathcal F$ is a function that takes an image as input and outputs a feature vector. For an image $Z=\left(z_{i,j}\right)\in\mathcal{I}$, we have $\mathcal F(Z)=V\in\mathbb R^D$, where $D$ is the dimension of the feature vector:
\[
\mathcal F:\mathcal I \rightarrow \mathbb R^D.
\]
The classifier can now be denoted as 
$\pred_{{\stego}_p,\mathcal{F}}:\mathcal{I}\rightarrow\{0,1\}$, and the whole classification problem can be viewed as a two-step procedure:
\[
\left(z_{i,j}\right) \xrightarrow{\mathcal{F}(\cdot)}{V} \xrightarrow{\pred_{{\stego}_p,\mathcal{F}}(\cdot)}\{0,1\}.
\]

Finally, $\mathcal F$ can be detailed as a collection of $D$ real-valued functions $\mathcal F(\cdot)=\left[f_1(\cdot),\dots,f_D(\cdot)\right]^{\mathrm T}$. %The notation $f_i$ is somewhat abused in the sequel to refer to the values of the function $f_i(\cdot)$.

\subsection{Subsequent embedding}
Since a stego image is an image itself, it is thus possible to carry out a second embedding process to the resulting image. Most possibly, this will imply the loss of the message that was embedded the first time, but this is not relevant in this paper, since we are not using subsequent embedding to hide another message, but as a tool that the steganalyst may use during the steganalytic process.

In general, given an image $X\in \mathcal C$, the corresponding stego ($X'$) and ``double'' stego ($X''$) images are defined as follows:
\[
\begin{split}
X'&={\stego}_p^{K_1}(X,m_1),\\
X''&={\stego}_p^{K_2}(X',m_2)={\stego}_p^{K_2}\left({\stego}_p^{K_1}(X,m_1),m_2\right).
\end{split}
\]
We assume that the second embedding is carried out using the same parameters $p$, but with a different stego key and message since, in general, $K_1\neq K_2$ and $m_1\neq m_2$. Whereas the first key and message ($K_1$ and $m_1$) are selected by the steganographer, the second pair ($K_2$ and $m_2$) is selected by the steganalyst, who does not have access to either $K_1$ or $m_1$. Hence, throughout this paper, a second embedding and a ``double stego'' image are assumed to be this particular process. This is illustrated graphically below:
\[
X \xrightarrow{{\stego}_p^{K_1};m_1} X' \xrightarrow{{\stego}_p^{K_2};m_2} X''.
\]

%{\color{black} Pot ser seria millor fer servir $X$, $X'$ i $X''$. Normalment en ML es fa servir la $X$ per a les mostres y la $Y$ per a les etiquetes.}

If the parameters $p$ used in the first embedding differ from those of the second embedding, this is considered here either as SSM condition or unknown message length, which are sources of uncertainty in steganalysis.

\subsection{Domain of ``double stego'' images}
Given the domain $\mathcal I=\mathcal C \cup \mathcal S$ of cover and stego images, note that $\stego^*_p(\mathcal I)$ would contain stego and/or ``double stego'' images, but no cover images since 
%i.e $\stego^{\#}_p(\mathsf{Z})\subset \stego^*_p(\mathcal I)$, where
\[
\stego^*_p(\mathcal I)=\stego^*_p(\mathcal C) \cup \stego^*_p(\mathcal S) = \mathcal S \cup \mathcal D,
\]
with 
\[
\mathcal D=\stego^*_p(\mathcal S)=\left\{\stego^*_p\left(\stego^*_p(X_k)\right) \mid \forall X_k\in\mathcal C
\right\}.
\]
Here, $\mathcal D$ denotes the domain of ``double stego'' images.  

The following notation is introduced for this ``double embedding'' process. For an image $Z\in \mathcal I$, we define ${\stego^*_p}^{(2)} (Z)= \stego^*_p\left(\stego^*_p(Z)\right)$ and, hence:
\[
\mathcal D=\stego^*_p(\mathcal S)=
\stego^*_p\left(\stego^*_p (\mathcal C) \right)=
{\stego^*_p}^{(2)}(\mathcal C). 
\]

Similar to the assumption that $\mathcal C\cap \mathcal S=\emptyset$ for effective steganography (see Remark \ref{rem:effective}), here we need two additional assumptions, namely, $\mathcal S\cap \mathcal D=\emptyset$ and, of course, $\mathcal C\cap \mathcal D=\emptyset$. The second assumption can be readily accepted, since it means that the cover and ``double stego'' domains are disjoint. On the other hand, it may not be that obvious to assume that $\mathcal S$ and $\mathcal D$ are two distinct (and disjoint) domains since, in fact, both contain images with stego messages embedded into them. However, there are relevant differences between both domains. 

First, as described in Section \ref{sec:lowprob}, ${\stego}_p$ produces $\pm 1$ changes in some pixels (which are modified with probability $\beta$). This means that embedding another message into an already stego image will produce a few $\pm 2$ variations wrt the corresponding cover image. Such $\pm 2$ variations cannot be found in any stego image in $\mathcal S$ (let alone the cover images in $\mathcal C$). Second, even if the pixel selection in the second embedding occurred in such a way that none of the already modified pixels would be selected again for modification, the overall number of modified pixels wrt the corresponding cover image would be significantly larger for a ``double stego'' than for a stego image. For example, with LSB matching steganography and an embedding bit rate of $0.1$ bpp, a stego image would have 5\% of modified pixels compared to the cover image, whereas a ``double stego'' image for which a completely different set of pixels were modified in the second embedding would have approximately $5\%+5\%=10\%$ modified pixels\footnote{In this work, we focus on targeted steganalysis with fixed parameters $p$ and, hence, LSB matching at $0.1$ and $0.2$ bpp are considered as two distinct steganographic schemes ${\stego}_p$ and $\stego_{p'}$, with $p\neq p'$.} wrt the corresponding cover image. In general, the number of modified pixels in the second case will be somewhat lower than 10\%, since some pixels will be selected twice for modification.

%{\color{black} Enlloc de =10\% pot ser hauriem de dir "aproximadament", donat que alguns pixels es repetiran, donant lloc a els +-2 que comentes, i per tant sera <10\%.}

\begin{remark}
The above discussion does not apply to steganography for which the $\pm 1$ variations are not symmetric, like in LSB replacement, where even pixels may only be increased ($+1$ change) and odd pixels may only be decreased ($-1$ change). This would produce no $\pm 2$ variations after subsequent embedding.  In fact, a stego image and a ``double stego'' image with LSB replacement and 1 bpp embedding bit rate are completely indistinguishable. However, such non-symmetric changes are highly detectable using structural LSB detectors \cite{Dumitrescu:2002:structural, ker:2008:structural,Ker:2005:structural,Fillatre:2012:structural} and, hence, they are no longer used in modern steganographic schemes.
\end{remark}

\subsection{Individual embedding}
We denote as $\stego^{\#}_p(\cdot)$ \textbf{one application} of the embedding function for an image or a set of images, with a different (or random) key and a different (or random) message for each image. For an image $Z=(z_{i,j})\in \mathcal I$, we define:
\[
\exists K\in\mathbf K, \exists m\in\mathbf M: \stego^{\#}_p(Z)=\stego^K_p(Z,m).
\]
Similarly, for the set $\mathsf Z=\{Z_k\}\subset\mathcal{I}$, $\exists \mathsf K_{\mathsf Z_k} =\{K_{\mathsf Z_k}\} \subset \mathbf{K}$ and $\exists \mathsf M_{\mathsf Z_k} =\{m_{\mathsf Z_k}\} \subset \mathbf{M}$ such that
\begin{gather*}
  \stego^{\#}_p(\mathsf Z)=\left\{
  \stego^{K_{\mathsf Z_k}}_p(Z_k,m_{\mathsf Z_k}) \mid
  K_{\mathsf Z_k}\in \mathsf K_{\mathsf Z_k}, 
  m_{\mathsf Z_k} \in \mathsf M_{\mathsf Z_k},
  \forall Z_k\in\mathsf{Z}
  \right\}.
\end{gather*}

This operation transforms cover into stego images, and stego into ``double stego'' images, i.e. $\stego^{\#}_p(\mathsf{Z})\subset \left(\mathcal S \cup \mathcal D\right)$.%, where $\mathcal S \cap \mathcal D=\emptyset$. 

Note that each image in $\stego^{\#}_p(\mathsf Z)$ is obtained using, in principle, a different secret key $K_{Z_k}$ and a different embedded message $m_{Z_k}$. Thus, $\stego^{\#}_p(\mathsf{Z})$ is not unique, since a specific key and a specific message is chosen for each image $Z_k\in \mathsf Z$. Each selection for $K_{Z_k}$ and $m_{Z_k}$ leads to a different instance of $\stego^{\#}_p(\mathsf{Z})$.

The function $\stego^{\#}_p$ may be applied in two different scenarios. On the one hand, a steganographer would use it with specific keys and messages on a cover image (or a set of cover images) for
secret communication. In this case, a key may be reused for different images (although this is not advisable for security reasons), but the messages would be different in general. On the other hand,  $\stego^{\#}_p$ may be used within a steganalytic framework. In this case, the steganalyzer would choose a  random key and a  random message for each image. 

\begin{remark}
It is now convenient to distinguish $\stego^*_p$ from $\stego^{\#}_p$. Whereas $\stego^*_p$ is obtained by applying ${\stego}_p$ \textbf{for all possible} keys and messages, $\stego^{\#}_p$ is obtained using \textbf{one (different or random)  key and one (different or random) message} for each image. Hence, if $\abs{\cdot}$ denotes the cardinality of a set, we have $\abs{\stego^*_p({\mathsf Z})} \gg\abs{\mathsf Z}=\abs{\stego^{\#}_p({\mathsf Z})}$.
\end{remark}

\subsection{Primary and secondary steganalytic classifiers}
\label{sec:primary}
Given a collection of images ${\mathsf A}=\{A_k\}\subset \mathcal{I}$ to be classified (or predicted) as either cover or stego, we define the primary classifier as the function:
$\pred^{\mathsf A}_{{\stego}_p,\mathcal{F}}:\mathcal{I}\rightarrow\{0,1\}$,
where the fact that the classifier is built for the set $\mathsf A$ is explicitly noted. $\pred^{\mathsf A}_{{\stego}_p,\mathcal{F}}(A_k)=0$ means that $A_k$ is classified or predicted as cover (presumably $A_k\in \mathcal C$) and  $\pred^{\mathsf A}_{{\stego}_p,\mathcal{F}}(A_k)=1$ if $A_k$ is classified or predicted as stego  (presumably $A_k\in \mathcal S$).

Now, we can define the secondary set as $\mathsf B=\{B_k\}=\stego^{\#}_p(\mathsf{A})\subset{\stego^*_p(\mathcal I)}$, i.e. the set obtained after a new embedding process in all the images of $\mathsf A$.  If the sets of stego and ``double stego'' images are separable, we can thus define a secondary classification problem as follows:
$\pred^{\mathsf B}_{{\stego}_p,\mathcal{F}}:\stego^*_p\left(\mathcal{I}\right)\rightarrow\{0,1\}$,
where $\pred^{\mathsf B}_{{\stego}_p,\mathcal{F}}(B_k)=0$ means that  $B_k$ is classified or predicted as stego (presumably $B_k\in \mathcal S$) and $\pred^{\mathsf B}_{{\stego}_p,\mathcal{F}}(B_k)=1$ means that $B_k$ is classified or predicted as ``double stego'' (presumably $B_k\in \mathcal D$).

Hence, we make use of two different classifiers, on the one hand $\pred^{\mathsf A}_{{\stego}_p,\mathcal{F}}$, which is used to separate the images of $\mathsf A\subset \mathcal I$ as cover (predicted to be in $\mathcal C$) or stego (predicted to be in $\mathcal S$) and, on the other hand, $\pred^{\mathsf B}_{{\stego}_p,\mathcal{F}}$, which is used to separate the images of $\mathsf B=\stego^{\#}_p (\mathsf A)\subset \stego^*_p(\mathcal I)$ as stego (predicted to be in $\mathcal S$) or ``double stego'' (predicted to be in $\mathcal D$). The usefulness of this secondary classifier is clarified in Section \ref{sec:practical}. 

The superscripts ``$(\cdot)^{\mathsf A}$'' and ``$(\cdot)^{\mathsf B}$'' are used to denote the primary and the secondary classification problems, respectively, in the sequel.

%whereas the primary steganalytic problem consists in the classification of a collection of images as either cover or stego, using a some classifier $\pred_{{\stego}_p,\mathcal{F}}$, we can define a secondary classification problem as follows. If  are the images to be classified with $\pred_{{\stego}_p,\mathcal{F}}$, we have 

%If $\mathcal I=\mathcal C \cup \mathcal S$, then $\stego^*_p(\mathcal I)=\stego^*_p(\mathcal C) \cup \stego^*_p(\mathcal S)$, but $\stego^*_p(\mathcal C) \subset S$

%Now, the primary steganalyzer is a function that works on the set $\mathsf A$ and classifies images as cover (0, if they belong to $\mathcal C$) or stego (1, if they belong to $\mathcal S$). This classifier is denoted as $\pred^{\mathsf A}_{{\stego}_p,\mathcal{F}}$. 

\begin{figure*}[ht]
  \centering
  \subfloat[\textcolor{black}{Classification with directional features only.}] {\label{fig:hypo1} \includegraphics[width=.3\textwidth]{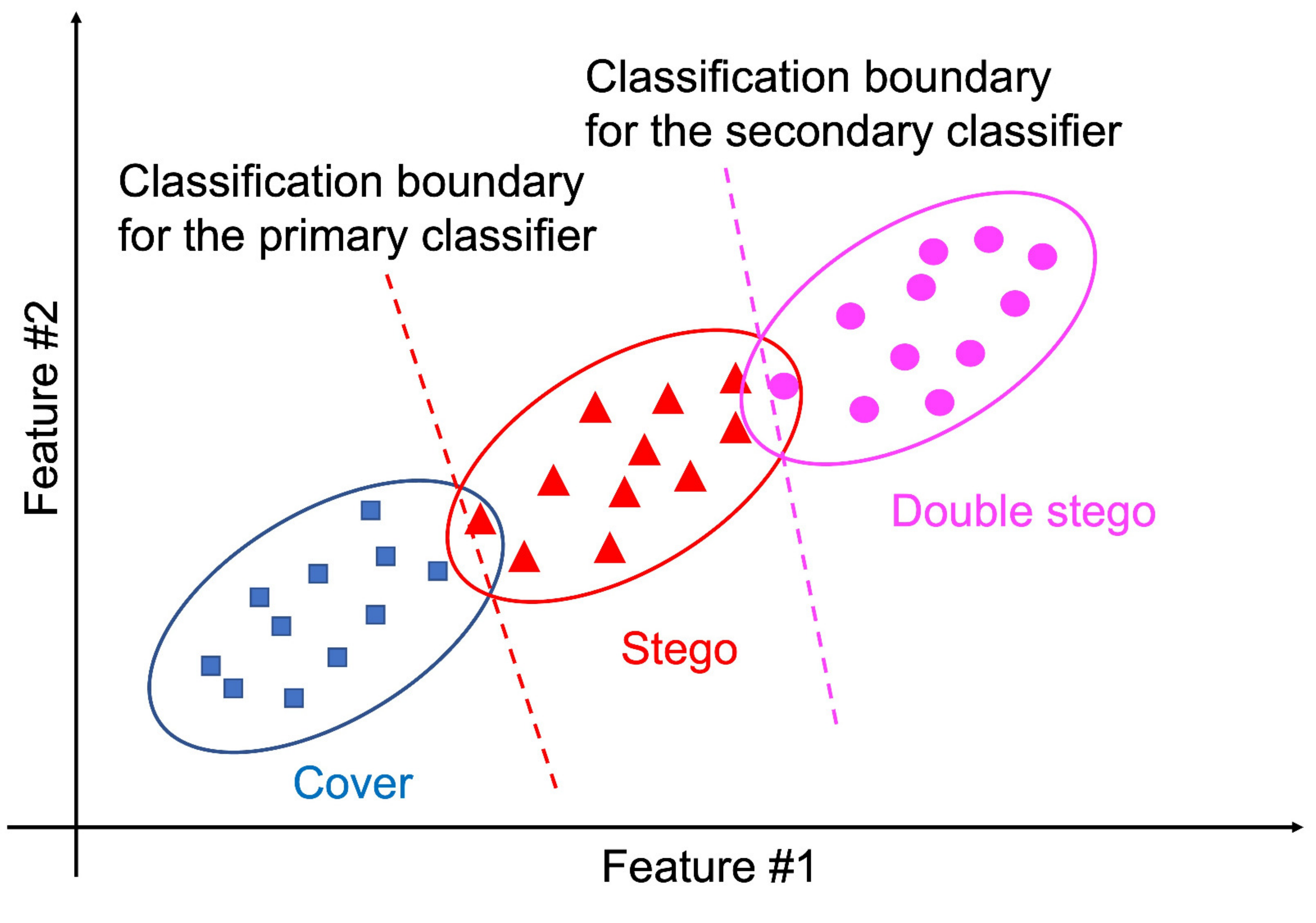}} \qquad
  \subfloat[\textcolor{black}{Classification with some directional features.}]{\label{fig:hypo2} \includegraphics[width=.3\textwidth]{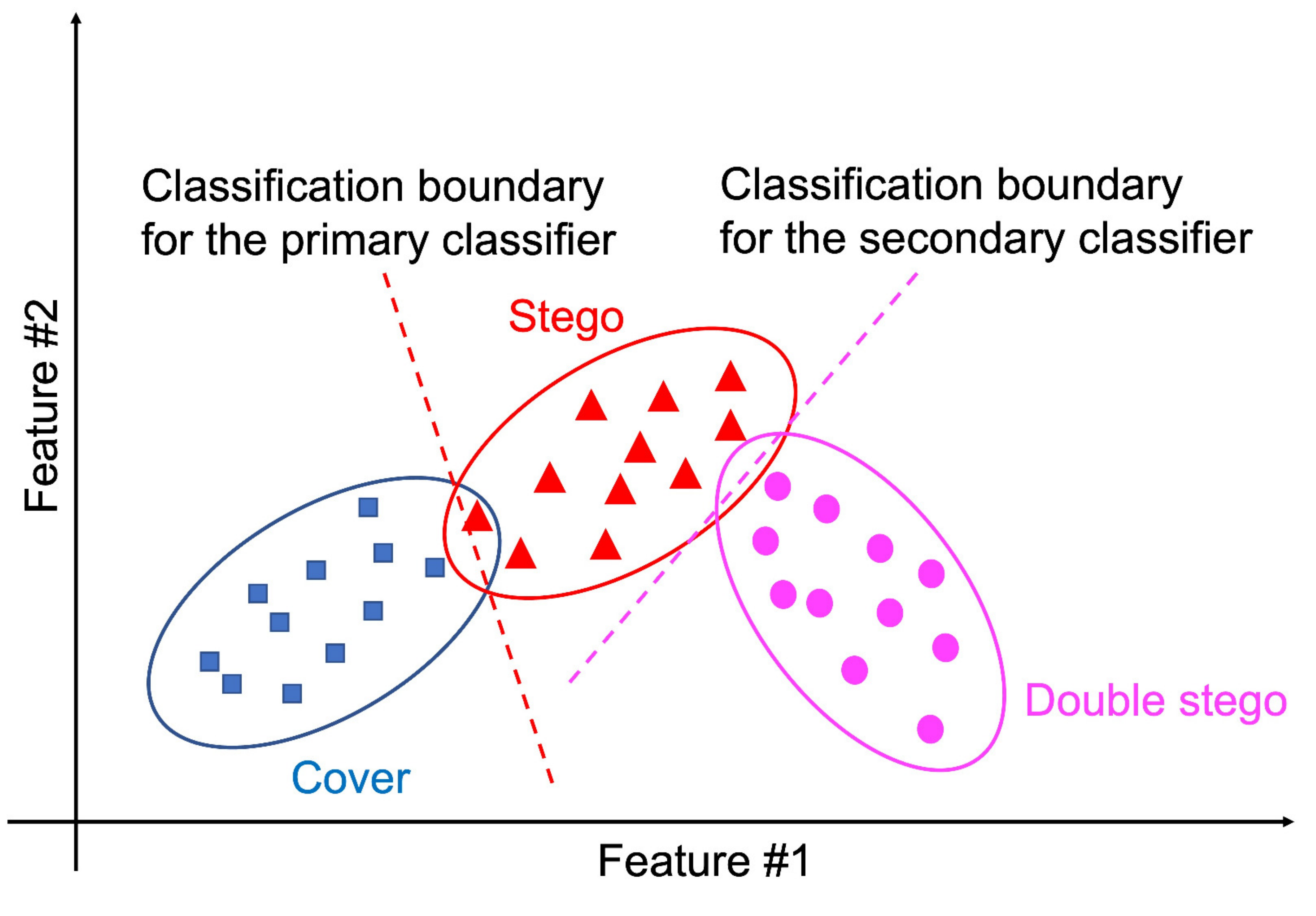}} \qquad
  \subfloat[\textcolor{black}{Classification with non-directional features only.}]{\label{fig:hypo3} \includegraphics[width=.3\textwidth]{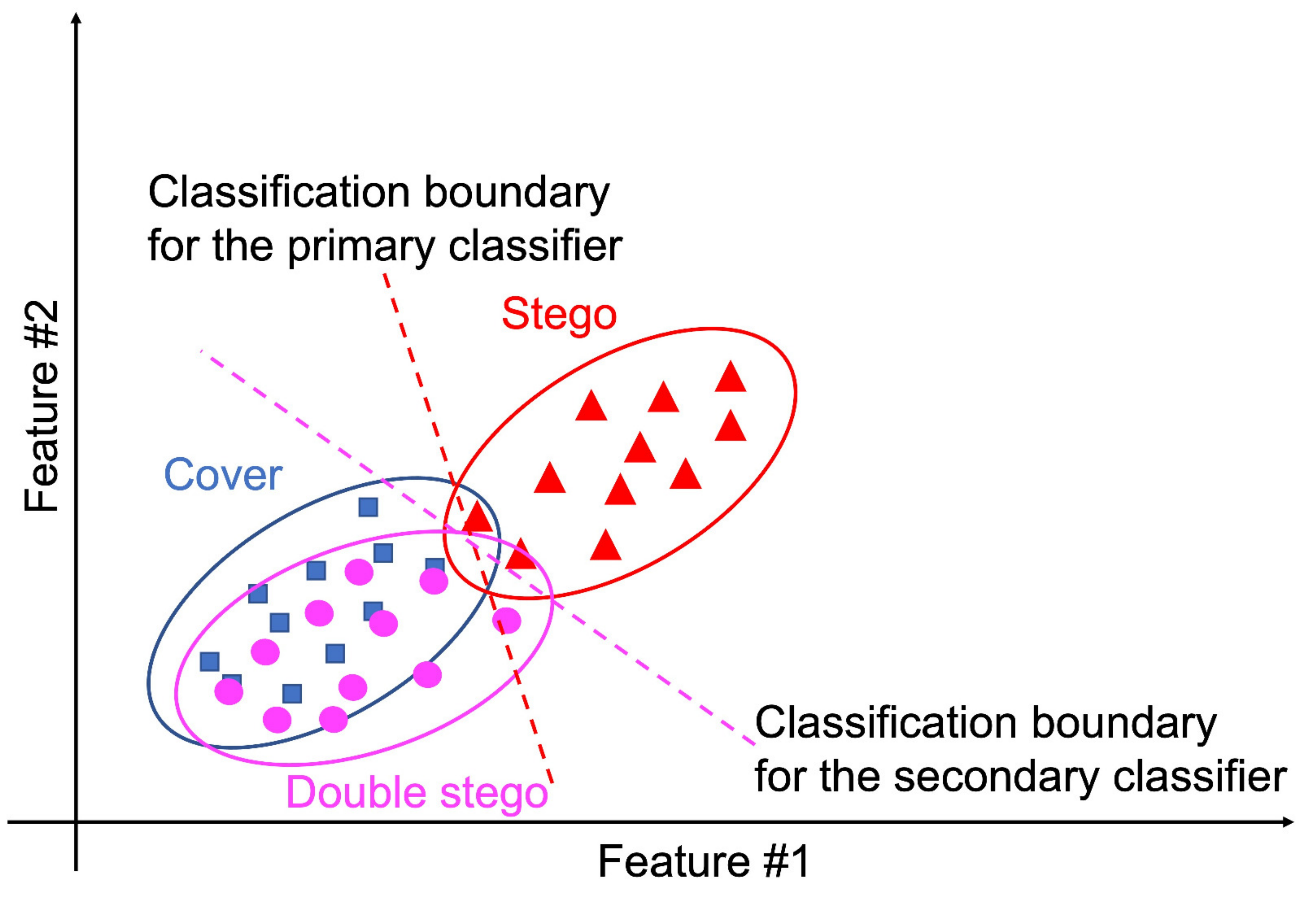}} \qquad
  \caption{\textcolor{black}{Three scenarios of feature directionality in the working hypothesis.}}
  \label{fig:hypo}
\end{figure*}

\subsection{Directional features}
\label{sec:directional}
In this work, we exploit the classifiers $\pred^{\mathsf A}_{{\stego}_p,\mathcal{F}}$ and $\pred^{\mathsf B}_{{\stego}_p,\mathcal{F}}$ not only in the obvious way, i.e the former to classify cover and stego images and the latter to classify stego and ``double stego'' images, but in a more sophisticated framework. This framework requires classifying a cover image in $\mathcal C$ using $\pred^{\mathsf B}_{{\stego}_p,\mathcal{F}}$, which is not trained with cover samples, and 
classifying a ``double stego'' image in $\mathcal D$ using $\pred^{\mathsf A}_{{\stego}_p,\mathcal{F}}$, which is not trained with ``double stego'' samples. 

A \textbf{sufficient condition} for those classifications to be possible is that \textbf{some relevant features be directional wrt the embedding function}. 
A feature $f_i$ is called directional wrt embedding if it satisfies one of the following two conditions:
 \begin{enumerate}[i)]
     \item if $f_i$ increases after the first data embedding, this feature shall further increase with a second embedding, or
     \item if $f_i$ decreases after the first data embedding, it shall further decrease after a second data embedding.
 \end{enumerate}
If the second embedding reverted the sign of the changes of the first embedding \textbf{for all features}, it may lead to mistaking a ``double stego'' image by a cover one, or conversely. On the other hand, if there is a relevant set of features satisfying the directionality condition, the classifiers will work properly, since there will be a set of features that will be significantly different from cover and ``double stego'' images.
%If this directionality condition is fulfilled by a collection of features and for vast the majority all images.
\color{black}
This idea is formalized as a working hypothesis below.

\vspace{3mm}
\hrule
\begin{hypo}
When classifying cover, stego and ``double stego'' images, for a successful classification that does not mistake cover and ``double stego'' images, a sufficient condition is to have some directional features (for most of the images).

This hypothesis is illustrated in Fig. \ref{fig:hypo}, where three different situations are shown. For simplicity, the graphical illustration is limited to two dimensions, corresponding to two features. In the best scenario (Fig. \ref{fig:hypo1}), all features are directional, and it is straightforward for a linear classifier to discriminate the three different types of images without confusing cover and ``double stego'' images.  In the worst scenario (Fig. \ref{fig:hypo3}), we can see that when all features reverse the sign change after a subsequent embedding, the sets of cover and ``double stego'' images overlap in the feature space, and a linear classifier will be in trouble to separate them, making the applications presented in Section \ref{sec:practical} unfeasible. The most common situation is none of the those two extreme cases, but a mixture of both directional and non-directional features, which is the case illustrated in Fig. \ref{fig:hypo2}. In this scenario, the feature represented in the abscissae is directional and preserves the sign change after one and two subsequent embeddings. However, the feature represented in the ordinates reverses the sign change in the second embedding and, thus, is not directional. We can see that, even in this case, a linear classifier will work, since the sets for cover, stego and ``double stego'' images are disjoint in the feature space, and no confusion between cover and ``double stego'' images occurs. Several experiments have been carried out  is to validate this hypothesis with real images and features, whose results are provided and discussed in the Appendices (supplemental material).

This graphical representation shows why the directionality of some features is a \textbf{sufficient condition} for successful classification with linear classifiers. However, it is not a necessary condition, since some non-linear classifiers may find more complex variations in features (for example, the Euclidean norm of a subset of features) making a successful classification possible even with no apparent directional features.
\end{hypo}
\hrule
\vspace{3mm}
\color{black}

To formalize the directionality condition, we first define the operator $\Delta$ as the variation of the feature $f_i$ after data embedding, i.e:
\[
\Delta f_i(X)=f_i\left(\stego^{\#}_p(X)\right)-f_i(X),
\]
for a cover image $X\in \mathcal C$. We further define a second-order variation $\Delta^2$ in a similar way: 
\[
\Delta^2 f_i(X)=f_i\left(\stego^{\#}_p\left(\stego^{\#}_p(X)\right)\right)-f_i\left(\stego^{\#}_p(X)\right).
\]

The directionality condition can now be written as follows:
\begin{equation}
\label{eqn:direct1}
\Delta f_i(X)\times \Delta^2 f_i(X)>0.
\end{equation}
This expression represents the features $f_i$ for which the changes occur in the same direction after one and two data embeddings. If the variation $\Delta f_i(X)$ is negative, the second-order variation $\Delta^2f_i(X)$ shall also be negative for Expression (\ref{eqn:direct1}) to hold, and the same goes for positive variations. For directional features, Expression (\ref{eqn:direct1}) should be satisfied for most images. Checking all possible cases is usually not feasible due to the large input size (the domain of all cover images). A relaxed property can be proposed based on the expected value of the features, i.e., we define:
\[
\begin{split}
E\left[\Delta f_i(X)\right]&=
E\left[f_i\left(\stego^{\#}_p(X)\right)\right]-f_i(X),\\
E\left[\Delta^2 f_i(X)\right]&=
E\left[f_i\left(\stego^{\#}_p\left(\stego^{\#}_p(X)\right)\right)\right]-
E\left[f_i\left(\stego^{\#}_p(X)\right)\right],
\end{split}
\]
where $E[\cdot]$ denotes the expected value. Hence, a relaxed condition for the directionality of $f_i$ can be expressed as follows:
\begin{equation}
\label{eqn:direct2}
E\left[\Delta f_i(X)\right]\times E\left[\Delta^2 f_i(X)\right]>0.
\end{equation}
Expression (\ref{eqn:direct1}) is stronger than Expression (\ref{eqn:direct2}), but the latter suffices to guarantee directionality in most cases. We refer to Expressions (\ref{eqn:direct1}) and (\ref{eqn:direct2}) as the ``hard directionality'' and ``soft directionality'' conditions, respectively.

Now, the question is whether such kind of features exists. Is directionality a common property for feature changes in subsequent embedding? Sections \ref{sec:simple} and \ref{sec:theoretical} focus on showing that there exists a simple feature model that fulfills the directionality property for most relevant features, and provides some sufficient (and sometimes necessary) conditions for this to occur. Once this property is established for a particular and simple model, it is reasonable to accept that the same property holds by other more complex feature definitions, as shown empirically in Section \ref{sec:validation}.

\section{Simple feature model for image steganalysis}
\label{sec:simple}
This section proposes a simple feature model for image steganalysis and analyzes the effect of the embedding process on the expected value of the features.

\subsection{Prediction and residual image}
For an image $X=(x_{i,j})$, we denote a prediction of a pixel as $\hat x_{i,j}$, which is usually obtained from the values of neighboring pixels. For
example, we can have horizontal ($\hat x_{i,j} = x_{i-1,j}$), vertical
($\hat x_{i,j} = x_{i,j-1}$) and diagonal ($\hat x_{i,j} = x_{i-1,j-1}$) predictions,
among others.

In general, we consider that the predicted image is obtained through a prediction function $\Psi$ such that $\hat X=\Psi(X)=(\hat x_{i,j})$. Note that $\hat X$ is typically smaller than the size of $X$, since there are some pixels for which the prediction does not exist. For example, with the horizontal prediction, the first column of pixels cannot be predicted and hence, the size of $\hat X$ is $(w-1)\times h$. Similarly, the size of $\hat X$ is $w\times(h-1)$ for a vertical prediction  and $(w-1)\times(h-1)$ for a diagonal prediction.

Let the residual image $R=(r_{i,j})$ be formed by the residuals (or differences) between the true and the predicted values of the pixels, i.e. $r_{i,j}=x_{i,j}-\hat x_{i,j}$. The size of $R$ is the same as that of $\hat X$.

\subsection{Histogram of the residual image}
\label{sec:histogram}
Let $H=\left(h_k\right)$, for $k=v_{\min}, ..., v_{\max}$, be the histogram of $R$, i.e. $h_k$ stands for the number of elements for which $r_{i,j}=k$. Since both $x_{i,j}$ and $\hat x_{i,j}$ are in the range $[0,\dots,2^n-1]$, we define $v_{\min}=-(2^n-1)$ and  $v_{\max}=2^n-1$ such that the we have a histogram bin for each possible value of the residuals. We also define $N$ as the total number of elements (residuals) in $R$, and it follows that
\[N=\sum_{k=v_{\min}}^{v_{\max}}h_k.\]
For example, for a horizontal prediction, $N=(w-1)\times h$.

%{\color{black}No entenc aquesta part. N es la suma del valor de totes les barres de l'histograma? Donat que l'jljklñhistograma es simetric, N sera gairebe zero. Per a que serveix?}

%{\color{black} N es la suma de l'alcada de les barres (de la coordenada Y). Ha de ser igual al nombre de ``pixels'' de la imatge residual.}

\subsection{Basic model of feature extraction}
\label{sec:basic}
Many feature extraction functions have been defined in the steganalysis literature. Here, we define a basic steganalytic system that uses the histogram of the residual image as features $V=\mathcal{F}(Z)=(h_k)$, for $k=v_{\min},\dots,v_{\max}$.

%However, it allows to carry out the theoretical analysis detailed in the next section.
This feature set is possibly very limited to lead to accurate classification results, but it serves for the theoretical analysis carried out in the next few sections.
In the sequel, we analyze how the embedding method ${\stego}_p$ modifies the histogram of the residual image. 

Let $H'=(h'_k)$ be the histogram of the residual image of $X'=\stego^{\#}_p(X)\in \mathcal S$, for $X\in\mathcal C$, i.e. $H'$ refers to the histogram of the residual image \textbf{of a stego image}. When a pixel $x'_{i,j}$ differs from $x_{i,j}$, the corresponding residual may also be affected, resulting in a change in the residual histogram. For simplicity, we assume that the prediction is carried out using another single pixel (e.g. horizontal, vertical or diagonal prediction). If a combination of different pixels were used, the reasoning would be similar, but the discussion would be far more cumbersome. 

In the discussion below, we make use of the operator $\Delta$ to refer to the variation of a pixel after data embedding, i.e.
$\Delta x_{i,j}=x'_{i,j}-x_{i,j}$. As remarked in Section \ref{sec:lowprob}, it is assumed that $\abs{\Delta x_{i,j}}\leq1$.

\subsection{Analysis for non-adaptive steganography}
\label{sec:non-adaptive}
First, we consider the case when the selection of pixels in steganography is carried out as independent events, i.e. the probability for the selection of a pixel is not affected by the selection of its neighboring pixels. This leads to the following assumption:

\begin{assm}
\label{assm:independent}
Pixel changes are independent events. 
\end{assm}

This assumption is not true for adaptive steganography, since those methods produce more pixel changes in textured areas and edges, which correspond to regions that are more difficult to model. A discussion about an adaptive steganography case is provided in Section \ref{sec:adaptive}.

\begin{assm}
\label{assm:bounded}
The difference between two consecutive bins in the residual image histogram is bounded by $\delta$ (a relatively small number):
\[
\abs{h_{k+1}-h_k}\leq \delta,\forall k\in[v_{\min},v_{\max}].
\]
\end{assm}

\begin{remark}
This assumption means that the histogram of the residual image is ``soft'' and that the variations between neighboring bins are small. This is typically the case for most natural images.
\end{remark}

\begin{lemma}
\label{lemma:expected}
If we assume independent (Assumption \ref{assm:independent}) random $\pm1$ changes of the pixel values with probability $\beta=\alpha/2$   for $\alpha\in(0,1]$, under Assumption \ref{assm:bounded}, then the expected value of the histogram bin $h'_k$ can be approximated as follows:
\begin{equation}
\label{eqn:expected}
    E[h'_k]\approx (1-\alpha)h_k+\frac{\alpha}{2}\left(h_{k-1}+h_{k+1}\right).
\end{equation}
\end{lemma}

\begin{proof}
We first define the probability that a pixel be modified in $X'=\stego^{\#}_P(X)$ (i.e. $x_{i,j}\neq x'_{i,j}$) as $\probmod$ %$P(y_{i,j}\neq x_{i,j})$. 
The complementary probability (pixel unchanged) is thus \[\probeq=1-\probmod.\]
As discussed above, we have $\probmod=\beta$ and $\probeq=1-\beta$.

\begin{table}[ht]
\caption{Combined probabilities of changes in $x_{i,j}$ and $\hat x_{i,j}$.}
\label{tab:prob1}
\centering 
\begin{tabular}{||c||c|c||} 
\hhline{~|t:==:t|} 
\multicolumn{1}{c||}{} & $\Delta\hat x_{i,j}=0$ & $\Delta\hat x_{i,j}\neq 0$ \\ \hhline{|t:=::==:|}
$\Delta x_{i,j}=0$ & $(1-\beta)^2$ & $\beta(1-\beta)$ \\
\hhline{||-||-|-||} 
$\Delta x_{i,j}\neq0$ &$\beta(1-\beta)$ & $\beta^2$ \\
\hhline{|b:=:b:==:b|}
\end{tabular}
\end{table}

Regarding combined pixel changes for a pixel $x_{i,j}$ and its prediction $\hat x_{i,j}$, we have four different possibilities, whose probabilities are detailed in Table \ref{tab:prob1}.
Under Assumption \ref{assm:independent}, there are four possible cases wrt the changes of a pixel $x_{i,j}$ and its prediction $\hat x_{i,j}$:
\begin{enumerate}
    \item Neither the pixel nor its prediction are changed. This occurs with probability $(1-\beta)^2$. The expected number of pixels that satisfy this condition for a residual histogram value $h_k$ are $(1-\beta)^2h_k$.
    \item The pixel $x_{i,j}$ is modified ($\Delta x_{i,j}\neq0$), but the prediction $\hat x_{i,j}$ is unchanged. This occurs with probability $\beta(1-\beta)$. In this case, if the residual histogram value associated to $x_{i,j}$ is $h_k$, a $+1$ change would ``move'' the pixel to the bin $h'_{k+1}$ and a $-1$ change would ``move'' it to the bin $h'_{k-1}$. Both possibilities occur with a probability $\beta(1-\beta)/2$.
    \item The prediction $\hat x_{i,j}$ is modified ($\Delta\hat x_{i,j}\neq0$), but the pixel $x_{i,j}$ is unchanged. Again, this occurs with probability $\beta(1-\beta)$. This leads to a ``movement'' from $h_k$ to either $h_{k-1}$ or $h_{k+1}$, both with probability $\beta(1-\beta)/2$.
    \item Finally, both the prediction $\hat x_{i,j}$ and the pixel $x_{i,j}$ can be modified, which occurs with probability $\beta^2$. In this case, we have four possibilities (each with the same probability $\beta^2/4$): 
    \begin{enumerate}
 \item Both $x_{i,j}$ and $\hat x_{i,j}$ change $+1$. In this case, the corresponding residual remains at $h'_k$.
 \item $x_{i,j}$ is modified $+1$ whereas $\hat x_{i,j}$ is modified $-1$. In this case, the residual ``moves'' to the bin $h'_{k+2}$.
 \item $x_{i,j}$ is modified $-1$ whereas $\hat x_{i,j}$ is modified $+1$. In this case, the residual ``moves'' to the bin $h'_{k-2}$.
 \item Both $x_{i,j}$ and $\hat x_{i,j}$ change $-1$. In this case, the corresponding residual remains at $h'_k$.
    \end{enumerate}  
\end{enumerate}

Considering all four cases, the expected value of $h'_k$ can be obtained as follows:
\begin{multline*}
E[h'_k]=
\left(1-\beta\right)^2h_k+\beta(1-\beta)\left(h_{k-1}+h_{k+1}\right)
\\+\frac{\beta^2}{2}h_k+\frac{\beta^2}{4}\left(h_{k-2}+h_{k+2}\right),
\end{multline*}
and, hence:
\begin{equation}
\label{eqn:expect2}    
\begin{split}
E[h'_k]=&
\left(1-2\beta+\frac{3}{2}\beta^2\right)h_k+\left(\beta-\beta^2\right)\left(h_{k-1}+h_{k+1}\right)
\\ & + \frac{\beta^2}{4}\left(h_{k-2}+h_{k+2}\right)
\\ =&\left(1-2\beta\right)h_k+\beta\left(h_{k-1}+h_{k+1}\right)+\varepsilon
\\ =&\left(1-\alpha\right)h_k+\frac{\alpha}{2}\left(h_{k-1}+h_{k+1}\right)+\varepsilon,
\end{split}
\end{equation}
with
\[
\varepsilon=\frac{3}{2}\beta^2 h_k -\beta^2 h_{k-1} - \beta^2 h_{k+1} + \frac{\beta^2}{4}h_{k-2}+ \frac{\beta^2}{4}h_{k-2}. 
\]
Now we only need to show that $\varepsilon\approx 0$ to complete the proof. To this aim,
as already discussed, we make use of the fact that $\alpha$ is typically close to zero and, thus, $\beta=\alpha/2$ is even closer to zero. Consequently, $\beta^2$ is small compared to $\beta$.
We can now rewrite $\varepsilon$ as follows:
\begin{equation}
\label{eqn:epsilon}
    \varepsilon=\beta^2\left(\frac{1}{4}h_{k-2}-h_{k-1}+\frac{3}{2}h_k-h_{k+1}+\frac{1}{4}h_{k+2}\right)
\end{equation}
and, thus:
\begin{multline*}
\varepsilon=\beta^2\left(-\frac{1}{4}\left(h_{k-1}-h_{k-2}\right)+\frac{3}{4}\left(h_{k}-h_{k-1}\right)\right.\\-\left.\frac{3}{4}\left(h_{k+1}-h_{k}\right)+\frac{1}{4}\left(h_{k+2}-h_{k+1}\right) \right).
\end{multline*}
Hence,
\begin{multline*}
\abs{\varepsilon}\leq\beta^2\left(\frac{1}{4}\abs{h_{k-1}-h_{k-2}}+\frac{3}{4}\abs{h_{k}-h_{k-1}}\right.\\+\left.\frac{3}{4}\abs{h_{k+1}-h_{k}}+\frac{1}{4}\abs{h_{k+2}-h_{k+1}}\right),
\end{multline*}
which yields
\[
\abs{\varepsilon}\leq \beta^2\left(\frac{1}{4}\delta+\frac{3}{4}\delta+\frac{3}{4}\delta+\frac{1}{4}\delta\right) = 2\beta^2\delta.
\]
Finally, since both $\beta^2$ and $\delta$ are small quantities, we have $\abs{\varepsilon}\approx 0$, which completes the proof.
\end{proof}

\begin{remark}
Strictly speaking, $\varepsilon\approx 0$ means that it is much smaller than the other dominant terms in Expression (\ref{eqn:expected}), i.e. $$\abs{\varepsilon}\ll\abs{(1-\alpha)h_k+\frac{\alpha}{2}\left(h_{k-1}+h_{k+1}\right)}.$$ 
The error $\varepsilon$ can  be a relatively large number, but it is much smaller than the other two terms.
\end{remark}

\begin{remark}
There are positive and negative terms in the second factor of Expression (\ref{eqn:epsilon}) that tend to cancel each other yielding a number much lower than the upper bound ($2\beta^2 \delta$) used in the proof of Lemma \ref{lemma:expected}. 
Informally, if a group of consecutive bins in a residual image histogram are typically similar: $h_{k-2}\approx h_{k-1} \approx h_k \approx h_{k+1} \approx h_{k+2} \approx \bar h_k$, we can see that $\varepsilon\approx 0$:
\[
\begin{split}
  \varepsilon&=\frac{3}{2}\beta^2 h_k -\beta^2 h_{k-1} - \beta^2 h_{k+1} + \frac{\beta^2}{4}h_{k-2}+ \frac{\beta^2}{4}h_{k-2}\\
  &\approx \frac{3}{2}\beta^2 \bar h_k -\beta^2 \bar h_k - \beta^2 \bar h_k + \frac{\beta^2}{4}\bar h_k+ \frac{\beta^2}{4}\bar h_k,
\end{split}
\]
and hence
\[
  \varepsilon\approx \beta^2 \bar h_k \left( \frac{3}{2}-1-1+\frac{1}{4}+\frac{1}{4}\right)=0.
\]
In fact, $\beta^2$ is a small value multiplying another factor that is very close to zero. That is the reason why $\varepsilon$ is really small compared to the dominant terms in Expression(\ref{eqn:expect2}).
\end{remark}

\subsection{Analysis for adaptive steganography}
\label{sec:adaptive}

Now, we analyze the variations of the histogram of the residual image after embedding when Assumption \ref{assm:independent} does not hold, i.e. when pixel changes are not independent events.

\begin{lemma}
\label{lemma:adapt}
If we assume dependent random $\pm1$ changes of the pixel values, the expected value of the histogram bin $h'_k$ can be approximated as follows:
%\begin{equation}
%\label{eqn:expect3}    
$$E[h'_k]\approx (1-\alpha)h_k+\frac{\alpha}{2}\left(h_{k-1}+h_{k+1}\right).$$
%\end{equation}
\end{lemma}

\begin{remark}
This is the same expression obtained in Lemma \ref{lemma:expected} for non-adaptive steganography.
\end{remark}

\textcolor{black}{For brevity, the proof of Lemma \ref{lemma:adapt} is omitted here and can be found in the Appendices (supplemental material)}.

\section{Theoretical analysis of subsequent embedding for the simple model}
\label{sec:theoretical}
This section is focused on analyzing the variation of the bins of the histogram of the residual image after two successive embeddings. More precisely, we establish sufficient conditions such that the soft directionality property of Expression (\ref{eqn:direct2}) be satisfied for the simple model.

\subsection{Directionality of histogram variations after subsequent embedding} 

Here, we introduce some notations. Let $\Delta$ be the variation of the $k$-th histogram bin of the residual image after one embedding, i.e, if $H=(h_k)$ is the histogram of the residual image of $X$ and $h'_k$ is the histogram of the residual image of $X'={\stego}_p^{\#}(X)$, then \textcolor{black}{$\Delta h_k=h'_k-h_k$}. Similarly, if \textcolor{black}{$X''={\stego}_p^{\#}(X')$} is the ``double stego'' image, whose histogram of the residual image is given by $H''=(h''_k)$, we have $\Delta^2 h_k= \Delta h'_k=h''_k-h'_k$. %More generally, the operator $\Delta^n$ refers to the variation of the histogram bins after $n$ subsequent embeddings. In the sequel, we use the cases $n=1$ and $n=2$ only.

\begin{thm}
\label{thm:negative}
%f the value of $h'_k$ is the approximate expected value given in 

With the expected value of $h'_k$ given in  
Expression (\ref{eqn:expected}), the ``expected variation'' $E\left[\Delta h_k\right]=E\left[h'_k\right]-h_k$ is negative [positive] if and only if the value $h_k$ is above [below] the straight line that connects $h_{k-1}$ and $h_{k+1}$.
\end{thm}

\begin{proof}
First, we use Expression (\ref{eqn:expected}) to write:
\[
E\left[h'_k\right]=(1-\alpha)h_k +\frac{\alpha}{2}\left(h_{k-1}+h_{k+1}\right).
\]
Now, we can compute the ``expected variation'' between $h'_k$ and $h_k$ as follows:
\begin{equation}
\label{eqn:diff}
\begin{split}
E\left[\Delta h_k\right] &= E\left[h'_k\right]-h_k\\
&=-\alpha h_k  +\frac{\alpha}{2}\left(h_{k-1}+h_{k+1}\right),
\end{split}
\end{equation}
and, hence $E\left[h'_k\right]-h_k<0$ if and only if:
\[
%-\alpha h_k +\frac{\alpha}{2}\left(h_{k-1}+h_{k+1}\right)<0 \iff
\frac{\alpha}{2}\left(h_{k-1}+h_{k+1}\right)<\alpha h_k.\]
Finally, since $\alpha$ is positive:
\begin{equation}
\label{eqn:cond1}
E\left[\Delta h_k\right]%=E\left[h'_k\right]-h_k
<0 \iff h_k > \frac{h_{k-1}+h_{k+1}}{2}.   
\end{equation}
This means that if $h_k$ is above the straight line that connects $h_{k-1}$ and $h_{k+1}$, the sign of the expected variation $E\left[\Delta h_k\right]$ is negative and, conversely, if $h_k$ lies below the straight line that connects $h_{k-1}$ and $h_{k+1}$, the sign of the expected variation is positive:
\[
E\left[\Delta h_k\right]%=h'_k-h_k
>0 \iff h_k < \frac{h_{k-1}+h_{k+1}}{2}.   
\]
\end{proof}
In the sequel, we work with the negative case, i.e. the condition of Expression (\ref{eqn:cond1}), since it is the common one for the central part of the histogram (around its peak value). However, the same analysis can be carried out for the positive case simply by exchanging the ``$<$'' and ``$>$'' signs.

\begin{remark}
\label{rem:diff}
Note that Expression (\ref{eqn:diff}) also indicates that $E\left[\Delta h_k\right]$ will be larger in magnitude when $\alpha$ is larger, since $\abs{E\left[\Delta h_k\right]}$ is directly proportional to $\alpha$. This means that the number of pixels selected by the embedding algorithm affects the expected variation $E\left[\Delta h_k\right]$. The more pixels are selected by the embedding algorithm ($\alpha$), the more $\pm1$ changes ($\beta$) will occur, leading to larger variations in the features, as expected. For instance, if the embedding bit rate is doubled, the probability of selecting a given pixel will also be approximately doubled and, hence, the expected variation $E\left[\Delta h_k\right]$ will also be, approximately, doubled. 
\end{remark}

Now, we analyze the variation of the histogram bin after a subsequent embedding.

%, i.e., suppose we embed a new message using the same steganographic scheme ($\stego$) into the image $Y$ and obtain a new image $Z=\left(z_{i,j}\right)$ for $(i,j)=(1,1), \dots, (h,w)$. Let $H''=(h_k'')$ be the histogram of the residual image of $Z$. 

\begin{assm}
\label{assm:alpha}
We assume $2-4\alpha\geq0$. This occurs for all $0<\alpha\leq\frac{1}{2}$. This assumption will hold for typical embedding algorithms, since $\alpha$ is the ratio of pixels selected by the steganographic method that will be typically very low for undetectability. 
\end{assm}

%\newpage

\begin{thm}
\label{thm:nega2}
Under Assumption \ref{assm:alpha}, with the expected valued of $h'_k$ given in  Expression (\ref{eqn:expected}), the ``expected second-order variation'' 
%if the values of $h'_k$ and $h''_k$ are the approximate expected values given in  Expression (\ref{eqn:expected}), then the ``expected variation'' 
$E\left[\Delta^2 h_k\right]=E\left[h''_k\right]-E\left[h'_k\right]$ is negative [positive] if %and only if 
\begin{enumerate}[i)]
\color{black}
\item \label{cond1} the value $h_{k}$ is above [below] the straight line that connects $h_{k-1}$ and $h_{k+1}$, 
\color{black}
\\ %for 
%\item \label{cond2} the value $h_{k-1}$ is above [below] the straight line that connects $h_{k-2}$ and $h_k$, \\ %for $\ell=k-1$ and $\ell=k+1$, and 
%\item \label{cond3} the value $h_{k+1}$ is above [below] the straight line that connects $h_k$ and $h_{k+2}$, and \\ %for $\ell=k-1$ and $\ell=k+1$, and 
\item \label{cond2}  the value $h_k$ is above [below] the straight line that connects $h_{k-2}$ and $h_{k+2}$.
\end{enumerate}
\end{thm}

\begin{proof}
Again, we prove the negative case only, since the positive one is completely analogous. To prove this theorem, we can borrow Expression (\ref{eqn:diff}) and extend it for a subsequent embedding as follows:
\begin{equation}
\label{eqn:diff2}
\begin{split}
    E\left[\Delta^2 h_k\right]&=E\left[h''_k\right]-E\left[h'_k\right]\\
    &=-\alpha E[h'_k]+\frac{\alpha}{2}\left(E[h'_{k-1}]+E[h'_{k+1}]\right).
\end{split}
\end{equation}
Now, we can apply Theorem \ref{thm:negative} using Expression (\ref{eqn:cond1}) to write:
\[
E\left[\Delta^2 h_k\right]<0 \iff E\left[h'_k\right] > \frac{E\left[h'_{k-1}\right]+E\left[h'_{k+1}\right]}{2}.
\]
If we replace $E\left[h'_k\right]$, $E\left[h'_{k-1}\right]$ and $E\left[h'_{k+1}\right]$ using the expected value obtained in Lemma \ref{lemma:expected} (or Lemma \ref{lemma:adapt}), this condition can be rewritten as follows: 
\begin{multline*}
(1-\alpha)h_k+\frac{\alpha}{2}\left(h_{k-1}+h_{k+1}\right) > \\ \frac{1}{2}\left( 
(1-\alpha)h_{k-1}+\frac{\alpha}{2}\left(h_{k-2}+h_k\right) \right. \\ + \left. (1-\alpha)h_{k+1}+\frac{\alpha}{2}\left(h_k+h_{k+2}\right)
\right).
\end{multline*}
Then, grouping terms, we can write:
\begin{multline*}
    \frac{2-3\alpha}{2}h_k > (1-2\alpha) \left(\frac{1}{2}\left(h_{k-1}+h_{k+1}\right)\right) \\
    +\frac{\alpha}{2}\left(\frac{1}{2}\left(h_{k-2}+h_{k+2}\right)\right).
\end{multline*}
Now, under Assumption\footnote{In fact, this assumption is introduced to guarantee that both $2-4\alpha$ and $2-3\alpha$ are positive in the expressions below.}  \ref{assm:alpha}, we have $2-3\alpha>0$, and the following equivalent condition is obtained:
\begin{multline*}
h_k > \frac{2-4\alpha}{2-3\alpha}\left(\frac{1}{2}\left(h_{k-1}+h_{k+1}\right)\right) + \\ \frac{\alpha}{2-3\alpha}\left(\frac{1}{2}\left(h_{k-2}+h_{k+2}\right)\right).
\end{multline*}
Then, we define 
$$\varphi=\frac{2-4\alpha}{2-3\alpha},\:\: 1-\varphi=\frac{\alpha}{2-3\alpha}.$$ Again, under Assumption \ref{assm:alpha}, it can be checked that $0\leq\varphi, (1-\varphi)\leq 1$. Using the definition for $\varphi$, the condition can be written as follows:
\begin{multline}
\label{eqn:cond2}
h_k > \varphi\left(\frac{1}{2}\left(h_{k-1}+h_{k+1}\right)\right) + \\ (1-\varphi)\left(\frac{1}{2}\left(h_{k-2}+h_{k+2}\right)\right).
\end{multline}
Hence, we only need to prove that Expression (\ref{eqn:cond2}) holds for all $0\leq\varphi\leq1$ from the conditions established in the theorem.

%Now, using the conditions \ref{cond1} and \ref{cond2} of the theorem, we have
%\begin{multline*}
%    \frac{1}{2}\left(h_{k-1}+h_{k+1}\right) > \\
%    \frac{1}{2}\left(\frac{1}{2}\left(h_{k-2}+h_{k}\right)
%    +\frac{1}{2}\left( h_{k}+h_{k+2}\right)
%    \right),
%\end{multline*}
%and hence:
%\[
%    \frac{1}{2}\left(h_{k-1}+h_{k+1}\right) >
%    \frac{1}{4}\left( h_{k-2}+2h_k+h_{k+2}
%    \right).
%\]

\color{black}
A \textbf{sufficient} condition for the inequality of Expression (\ref{eqn:cond2}) to hold can be obtained replacing $h_{k-1}+h_{k+1}$ with an upper bound of it, which is given directly in Condition \ref{cond1} of the theorem:
$$\frac{1}{2}\left(h_{k-1}+h_{k+1}\right)<h_k.$$

Thus, we can write such a sufficient condition for Expression (\ref{eqn:cond2}) as follows:
$$
h_k > \varphi h_k+(1-\varphi)\frac{1}{2}\left(h_{k-2}+h_{k+2}\right),$$
or
\[
(1-\varphi)h_k >(1-\varphi)\frac{1}{2}\left(h_{k-2}+h_{k+2}\right),
\]
and, thus:
\[
h_k>\frac{1}{2}\left(h_{k-2}+h_{k+2}\right),
\]
which is exactly Condition \ref{cond2} in the statement of the theorem. This completes the proof.
\end{proof}
\color{black}

\begin{remark}
\label{rem:diff2}
Similarly to Remark \ref{rem:diff}, it can be easily seen, from Expression (\ref{eqn:diff2}), that $\abs{E\left[\Delta^2 h_k\right]}$ is directly proportional to $\alpha$. Hence, the larger the embedding bit rate, the larger variation is expected in the feature after a subsequent embedding is obtained.
\end{remark}

\subsection{Interpretation using a continuous variable}
Sometimes, it is convenient to consider a histogram as an approximation of some probability density function (pdf). To this aim, we define the normalized histogram $\widehat {H}=(\hat h_k)$ as follows:
$$\hat h_k=\frac{1}{N\uptau} h_k,$$
where $\uptau$ is some ``sampling interval''. With this definition, it follows that $$\sum_{k=v_{\min}}^{v_{\max}} \hat h_k=\frac{1}{\uptau}.$$
Therefore, $\hat h_k$ can be viewed as empirical values of samples of the pdf\footnote{Strictly speaking, such a pdf would require a continuous definition of the residuals, which is impossible for digital images.} of the values of $r_{i,j}$.

\begin{assm}
\label{assm:function}
We assume that the normalized histogram bins $\hat h_k$ of a residual image are samples of some continuous probability density function $f(s)$ for $s\in \mathbb{R}$, i.e. $\hat h_k= f(k\uptau)$ for some ``sampling interval'' $\uptau$. 
\end{assm}

\begin{remark}
Under Assumption \ref{assm:function}, we have:
%$$
%\int_{-\infty}^{\infty} f(s)ds \approx \uptau\sum_{k=-\infty}^{\infty} % f(k\uptau).
%$$
%Hence, we have
\begin{multline*}
\int_{-\infty}^{\infty} f(s)ds \approx \uptau\sum_{k=-\infty}^{\infty} f(k\uptau) = \\ \uptau\sum_{k=-\infty}^{\infty} \hat h_k =
\uptau\sum_{k=v_{\min}}^{v_{\max}} \hat h_k =\uptau\frac{1}{\uptau}=1,
\end{multline*}
which is consistent with the concept of pdf.
\end{remark}

\begin{thm}
\label{thm:concave}
Under Assumption \ref{assm:function}, Theorem \ref{thm:negative} holds if $f(s)$ is a strictly downward concave\footnote{We use ``concave'' to refer to a downward concave function and ``convex'' to refer to a downward convex one. Thus, we omit the term ``downward'' below.}
function in $s\in[(k-1)\uptau,(k+1)\uptau]$.
\end{thm}

\begin{proof}
By definition, if $f(s)$ is strictly concave in $s\in[(k-1)\uptau,(k+1)\uptau]$, we have
\begin{multline*}
f\left(t(k-1)\uptau+(1-t)(k+1)\uptau\right)>\\
tf((k-1)\uptau)+(1-t)f((k+1)\uptau),
\end{multline*}
for all $t\in[0,1]$. Now, for $t=1/2$:
\begin{gather*}
f(k\uptau)>\frac{f((k-1)\uptau) + f((k+1)\uptau)}{2} \iff \hat h_k>\frac{\hat h_{k-1} + \hat h_{k+1}}{2}.
\end{gather*}
Therefore, since $\hat h_k=h_k/(N\uptau)$ for some $N>0$ and $\uptau>0$, it follows that
$$\hat h_k>\frac{\hat h_{k-1} + \hat h_{k+1}}{2} \iff
h_k>\frac{h_{k-1} + h_{k+1}}{2},$$
which completes the proof.
\end{proof}

\begin{coro}
\label{coro:diff}
If $f(s)$ is a twice differentiable function, under Assumption \ref{assm:function}, Theorem \ref{thm:negative} holds if 
$$\frac{d^2}{ds}f(s)<0,\forall  s\in[(k-1)\uptau,(k+1)\uptau].$$
\end{coro}

\begin{proof}
For a twice differentiable function $f(s)$, the condition $$\frac{d^2}{ds}f(s)<0,$$
is equivalent to concavity, and the Corollary directly follows from Theorem \ref{thm:concave}.
\end{proof}

The following theorem and corollary are the counterparts of Theorem  \ref{thm:concave} and Corollary \ref{coro:diff}, and their proofs are completely analogous.

\begin{thm}
\label{thm:convex}
Under Assumption \ref{assm:function}, Theorem \ref{thm:negative} holds, with a positive variation $E\left[\Delta h_k\right]=E\left[h'_k\right]-h_k>0$,  if $f(s)$ is a strictly convex function in $s\in[(k-1)\uptau,(k+1)\uptau]$.
\end{thm}

\begin{coro}
\label{coro:diffpos}
If $f(s)$ is a twice differentiable function, under Assumption \ref{assm:function}, Theorem \ref{thm:negative} holds,  with a positive variation $E\left[\Delta h_k\right]=E\left[h'_k\right]-h_k>0$, if  $$\frac{d^2}{ds}f(s)>0,\forall  s\in[(k-1)\uptau,(k+1)\uptau].$$
\end{coro}

\begin{remark}
Note that Theorem \ref{thm:concave} (and \ref{thm:convex}) and Corollary \ref{coro:diff} (and \ref{coro:diffpos}) establish sufficient but not necessary conditions. In fact, the values of $f(s)$ outside the samples $s=-v_{\min}\uptau,\textcolor{black}{(-v_{\min}+1)\uptau},\dots,v_{\max}\uptau$ are not relevant for Theorem \ref{thm:negative} and, hence, the particular shape of the function $f(s)$ ``between samples'' is not significant.
\end{remark}

\begin{thm}
\label{thm:conca2}
Under Assumptions \ref{assm:alpha} and \ref{assm:function}, Theorem \ref{thm:nega2} holds if $f(s)$ is a strictly concave function in $s\in[(k-2)\uptau,(k+2)\uptau]$.
\end{thm}

\color{black}
\begin{proof}
By definition, if $f(s)$ is a strictly concave in $s\in[(k-2)\uptau,(k+2)\uptau]$, it is necessarily strictly concave in $s\in[(k-1)\uptau,(k+1)\uptau]$ (which is a subinterval of the former interval) and, as proved in Theorem \ref{thm:concave}, this means that
$$h_k>\frac{h_{k-1} + h_{k+1}}{2},$$
which is the first condition of the statement of Theorem \ref{thm:nega2}. The second condition is also obtained from the definition of concavity in $s\in[(k-2)\uptau,(k+2)\uptau]$:
\begin{multline*}
f\left(t(k-2)\uptau+(1-t)(k+2)\uptau\right)>\\
tf((k-2)\uptau)+(1-t)f((k+2)\uptau),
\end{multline*}
for $t=1/2$.
\end{proof}
\color{black}

\begin{coro}
\label{coro:diff2}
If $f(s)$ is a twice differentiable function, under Assumptions \ref{assm:alpha} and \ref{assm:function}, Theorem \ref{thm:nega2} holds if 
$$\frac{d^2}{ds}f(s)<0,\forall  s\in[(k-2)\uptau,(k+2)\uptau].$$
\end{coro}

\begin{proof}
The proof is analogous to that of Corollary \ref{coro:diff}.
\end{proof}

Again, we have the corresponding counterpart of Theorem  \ref{thm:conca2} and Corollary \ref{coro:diff2} for the positive variation case, whose proofs are completely analogous.

\begin{thm}
\label{thm:convex2}
Under Assumptions \ref{assm:alpha} and \ref{assm:function}, Theorem \ref{thm:nega2} holds, with a positive variation
$E\left[\Delta^2h_k\right]=E\left[h''_k\right]-E\left[h'_k\right]>0$
%$\Delta^2h_k=h''_k-h'_k>0$, 
if $f(s)$ is a strictly convex function in $s\in[(k-2)\uptau,(k+2)\uptau]$.
\end{thm}

\begin{coro}
%\label{coro:diffpos}
If $f(s)$ is a twice differentiable function, under Assumptions \ref{assm:alpha} and \ref{assm:function}, Theorem \ref{thm:nega2} holds, with a positive variation $E\left[\Delta^2h_k\right]=E\left[h''_k\right]-E\left[h'_k\right]>0$, if $$\frac{d^2}{ds}f(s)>0,\forall  s\in[(k-2)\uptau,(k+2)\uptau].$$
\end{coro}

\subsection{Concluding remarks about the simple model}

This model provides a powerful tool to predict the signs of the expected values $E\left[\Delta h_k\right]$ and  $E\left[\Delta^2h_k\right]$. The theorems provided in the previous sections establish sufficient  conditions for Expression (\ref{eqn:direct2}) to hold.

%If the histograms $H$ and $H'$ are used as feature vectors by a steganalyst, as suggested in Section \ref{sec:basic}, the signs of $\Delta h_k$ and  $\Delta^2h_k$ represent the variation of the features after one and two embeddings, respectively. If both values have the same sign, this can be exploited by classification algorithms to separate ``double'' stego images from stego images, since the features of ``double'' stego images would be further from the values of the features of cover images. On the other hand, if the signs of $\Delta h_k$ and  $\Delta^2h_k$ are opposite, then, intuitively, that feature would move ``back'' after a second embedding, closer to the feature of a cover image. Note that, for a classifier to work, it is only require that some of the features preserve this ``directionality'' in subsequent embedding. The theoretical discussion provided in the previous sections establish some conditions for the simple model to satisfy this directionality. 

Theorems \ref{thm:negative} and \ref{thm:nega2} are applicable in any case, and Theorems \ref{thm:concave}, \ref{thm:convex}, \ref{thm:conca2} and \ref{thm:convex2} provide sufficient conditions when a real-valued function $f(s)$ can approximate the histogram curve. According to those theorems, if $f(s)$ is concave in a given interval, the expected variation of the histogram bins within that interval is negative after one and two embeddings. Conversely, if the $f(s)$ is convex in some interval, the expected variation of the value of the histogram bins after one or two embeddings is positive. There is uncertainty about the sign of those variations for the bins around the inflection points. 

The conclusion is that the concavity/convexity properties of the approximate model given by the function $f(s)$ determines the expected sign variations of the histogram bin values and, more importantly, that such a sign tends to be preserved when a second embedding occurs. Thus, this proves that, for a large number of histogram bins, as far as the assumptions are satisfied, the direction of the changes in the histogram values is preserved for one and two subsequent embeddings. This justifies that a Machine Learning algorithm can exist to exploit this directionality. The features of the simple model exhibit the soft directionality property that makes it possible for some classification algorithm to discriminate between cover, stego and ``double stego'' images, because, for directional features, double stego images tend to be disturbed from a stego one in the same direction that a stego image is disturbed from a cover one. Hence, the directional features for a ``double stego'' image will differ from those of a cover image even more than those of a stego image.

\textcolor{black}{A generalization of the theoretical framework to actual (not simple) feature models, and especially SPAM \cite{SPAM} and ``minmax'' \cite{Fridrich:2011:minmax} features that are included in Rich Models \cite{Fridrich:2012:RM}, is provided in the Appendices (supplemental material). In addition, some experimental results of the practical application proposed in Section \ref{sec:prediction} obtained with the EfficientNet B0 \cite{Tan:2019} CNN model are also presented in the Appendices.}

\begin{figure}[ht]
  \centering
  \includegraphics[width=.7\columnwidth]{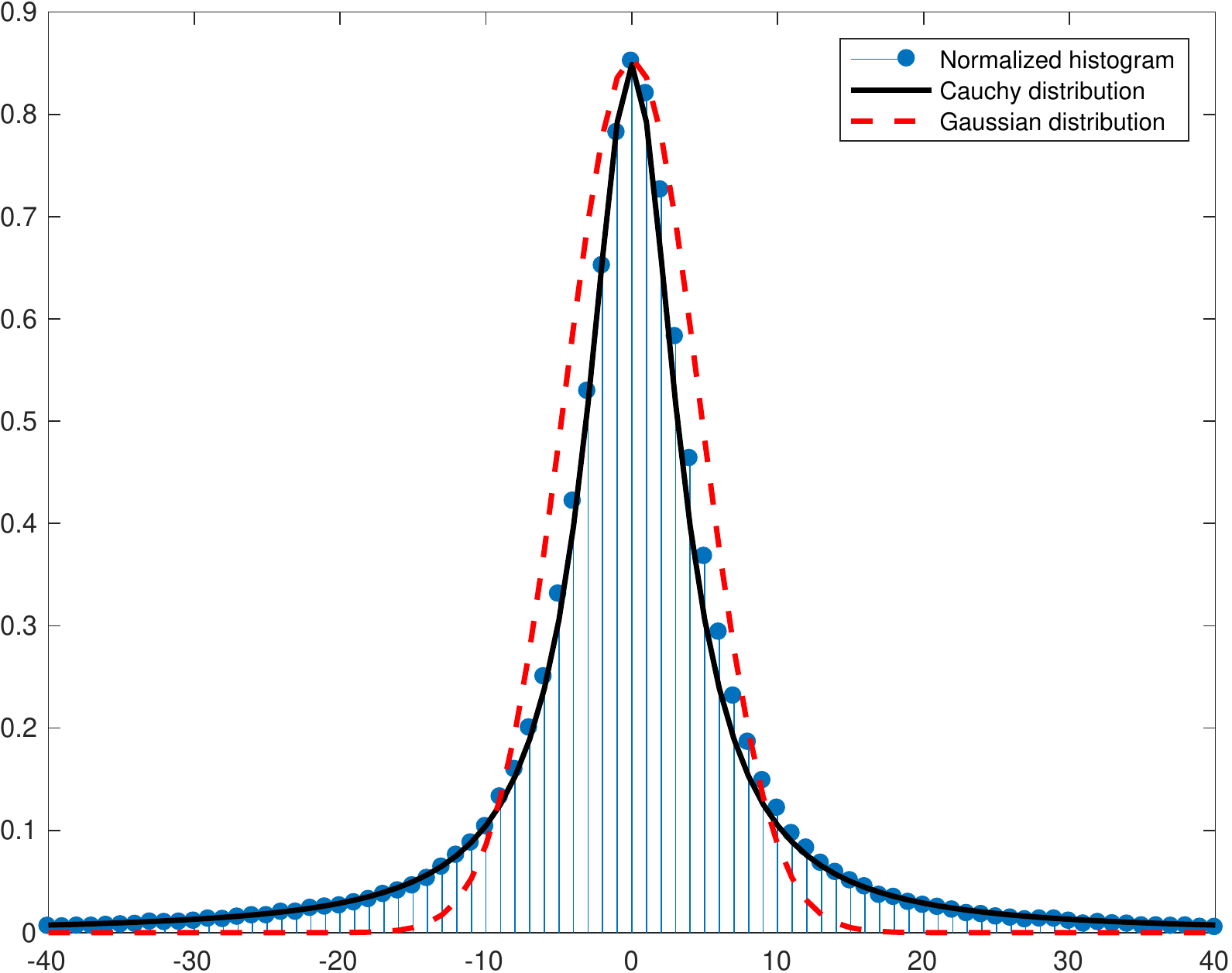}
  \caption{Normalized histogram compared for a Gaussian and a Cauchy distribution.}
  \label{fig:cauchy}
\end{figure}

\begin{figure}[ht]
  \centering
   \includegraphics[width=.7\columnwidth]{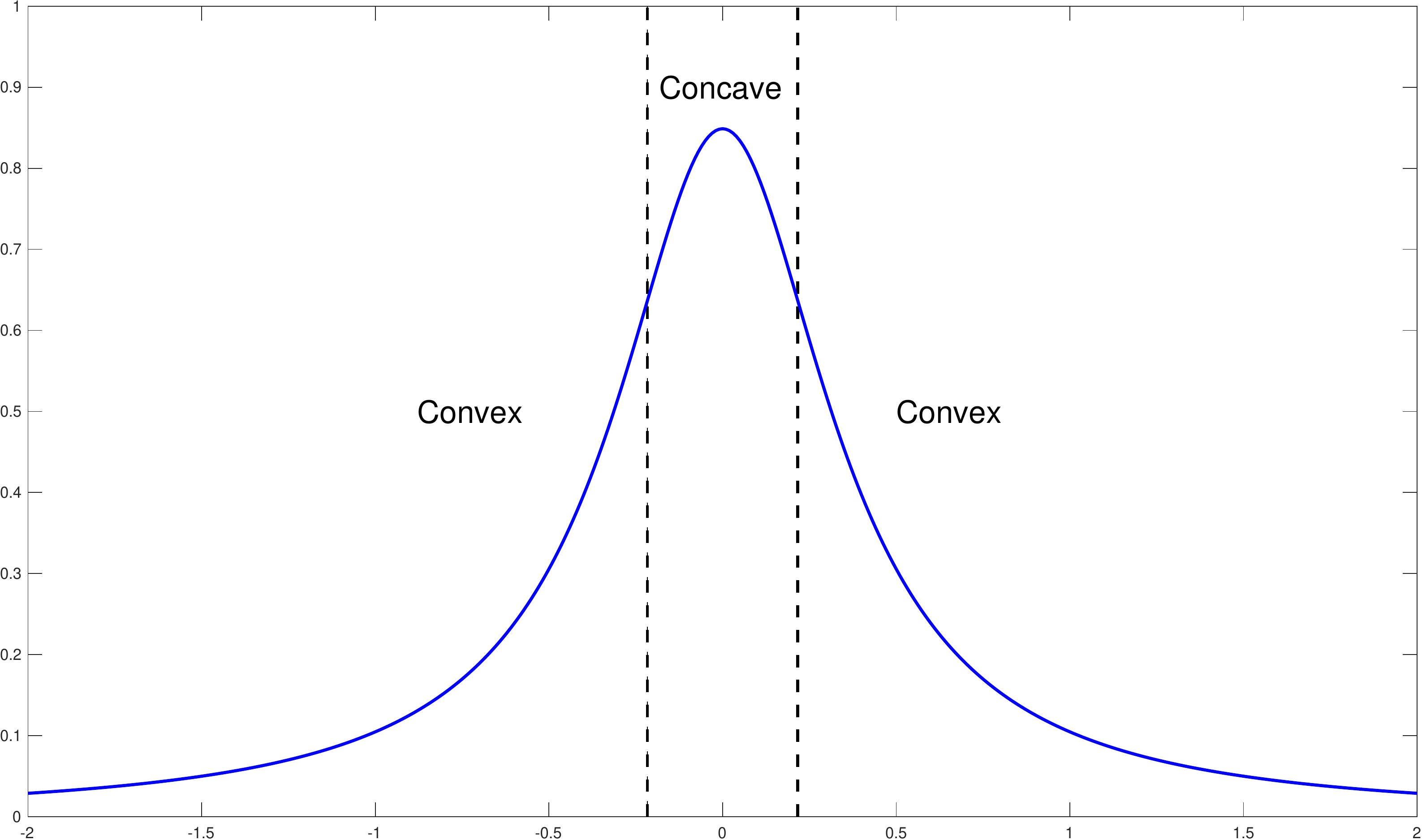}
  \caption{Concavity and convexity intervals of a Cauchy distribution.}
  \label{fig:concave}
\end{figure}

\section{Model validation}
\label{sec:validation}

\begin{figure*}[ht]
  \centering
  \subfloat[Expected (model) variations of the bins of the residual image: $E\lbrack\Delta h_k\rbrack$ (top) and  $E\lbrack\Delta^2h_k\rbrack$ (bottom).] {\label{fig:variations} \includegraphics[width=.7\columnwidth]{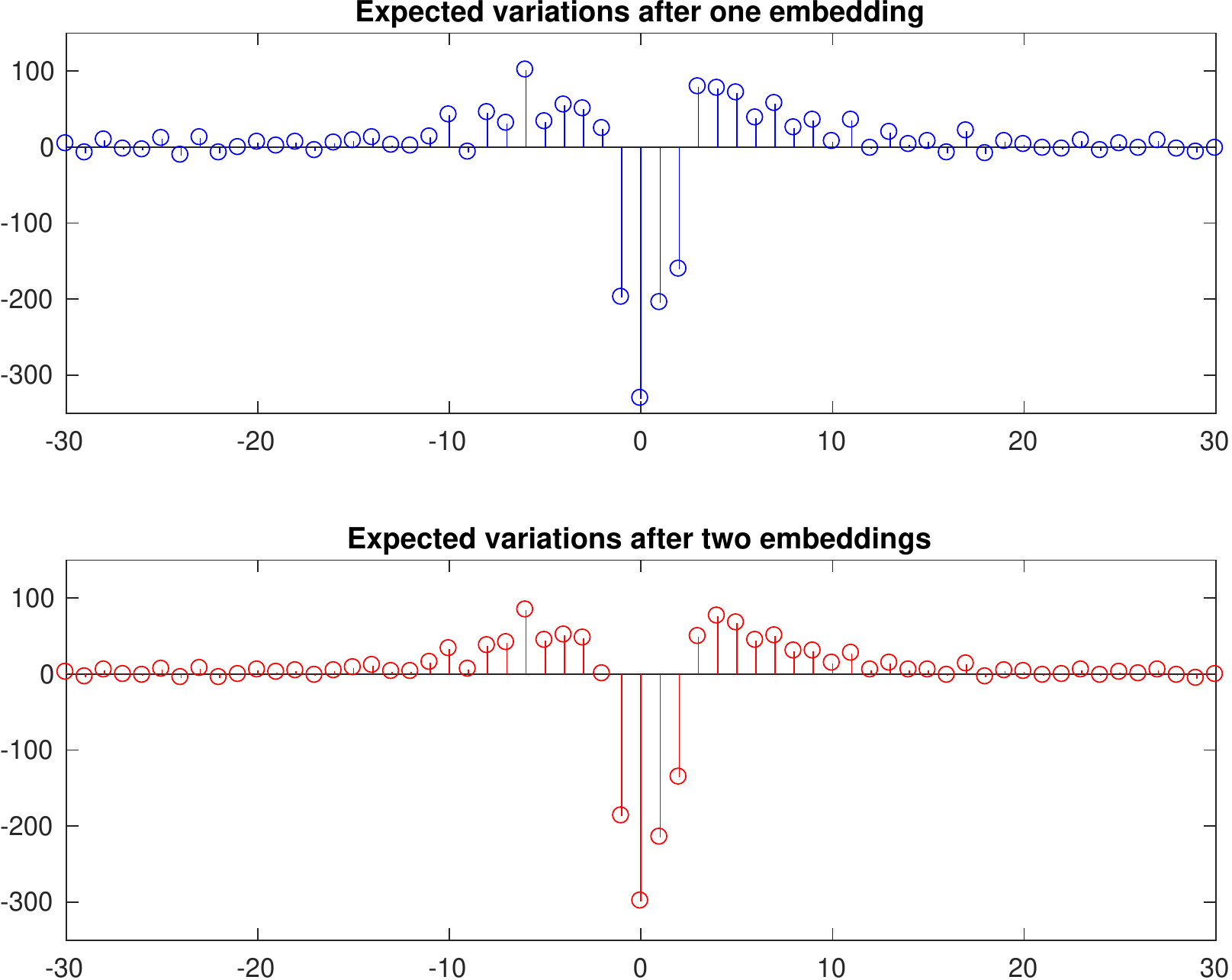}} \qquad
  \subfloat[Average (true) variations of the bins of the residual image: $\Delta h_k$ (top) and $\Delta^2h_k$ (bottom) for 1,000 experiments.]{\label{fig:var2} \includegraphics[width=.7\columnwidth]{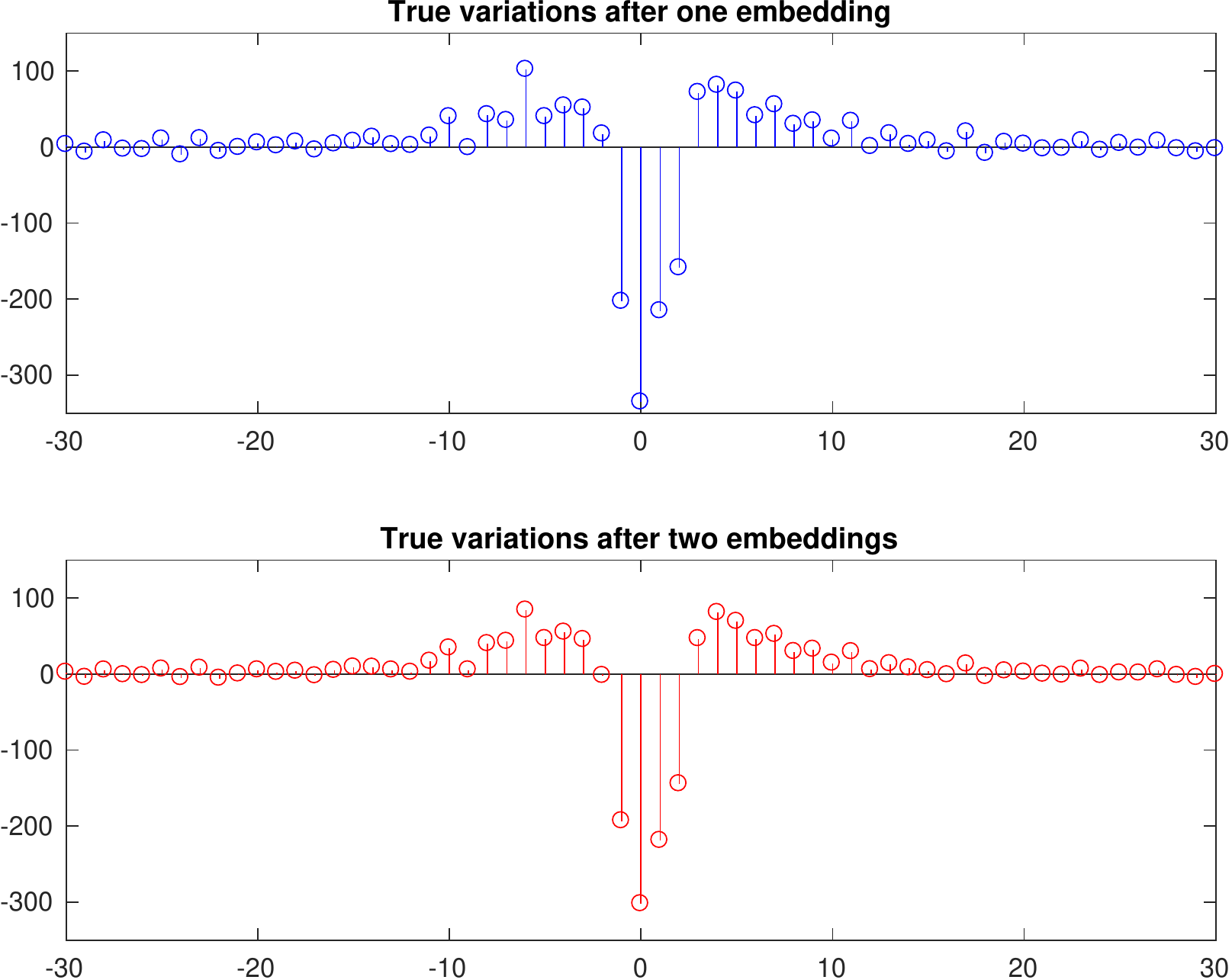}}
  \caption{Expected (model) and true variations of the bins of the residual image after one and two embeddings.}
  \label{fig:allvariations} 
\end{figure*}

This section presents the validation of the model presented in Section \ref{sec:theoretical} for real images. First, we show that Assumption \ref{assm:function} holds for standard images. Fig. \ref{fig:cauchy} shows the histogram of the residual image for Lena with horizontal prediction. The true values of the normalized bins of the residuals, represented by bullets, are compared with a normal distribution (dashed line) and a Cauchy distribution (solid line). It can be observed that the matching between the Cauchy distribution function and the histogram bins of the residual image is very accurate (and much better than that of a normal distribution). This behavior has been confirmed with different test images. Hence, for many natural images, the values of the histogram bins can be approximated by samples of a Cauchy distribution.

%\begin{figure}[ht]
%  \centering
  %\subfloat[Histogram]{\label{fig:cauchy}
%  \ifpdf \includegraphics[width=\columnwidth]{concave.png}
%  \else \includegraphics[width=0\columnwidth]{concave.pdf}
%  \fi
%  \caption{Concavity and convexity intervals of the pdf of a Cauchy distribution.}
%  \label{fig:concave}
%\end{figure}

If we assume that the pdf of the Cauchy distribution is centered around 0 (as is the most typical case for histograms of residual images), we can write:
$$
f(s;\gamma)=\frac{1}{\pi}\frac{\gamma}{s^2+\gamma^2},
$$
for some $\gamma>0$. This expression makes it possible to obtain the concavity and convexity intervals just by computing the second derivative and analyzing its sign intervals, and the result is shown in Fig. \ref{fig:concave}. As it can be observed, there is a narrow interval near the peak where the function is concave, and it is convex elsewhere. 

\begin{remark}
The particular function (Cauchy, Gaussian, or other) of the pdf is not relevant for the results. The only relevant issue is the presence of concave and convex intervals, which is a common feature of many possible pdf functions.
\end{remark}

\subsection{Validation of the simple model}
According to this convexity/concavity analysis, the expected differences for the few bins in the concave area should be negative, and positive (decaying to zero) elsewhere. This is clearly confirmed in Fig. \ref{fig:variations}. We can see that, for both the expected variations $E\left[\Delta h_k\right]$ and the second-order expected variations $E\left[\Delta^2h_k\right]$, the sign is negative in the central bins, where the function $f(s)$ is concave, and the sign is positive elsewhere, where $f(s)$ is convex. Of course, there are some minor inaccuracies due to two facts: 1) the term $\varepsilon$ that is dropped in Lemma \ref{lemma:expected}, and 2) the matching between the true histogram bin values and the Cauchy pdf samples is not perfect in Fig. \ref{fig:cauchy}.

Another test has been carried out to check whether the true value of the variations is consistent with the expected value (Fig. \ref{fig:variations}). To do so, we have performed 1,000 experiments with a single embedding and two subsequent embeddings, using LSB matching with $\alpha=0.25$ (the same value used for Fig. \ref{fig:variations}) to the same Lena image used above. For each embedding, a different secret key has been used (in the case of two subsequent embeddings, two different keys were used). After these 1,000 experiments, the average of the variations $\Delta h_k$ and $\Delta^2h_k$ were computed and the results are shown in Fig. \ref{fig:var2}. As it can be observed, the differences between the model of Fig. \ref{fig:variations} and the true values of Fig. \ref{fig:var2} are minimal. This shows that the theoretical model presented in Section \ref{sec:theoretical} is accurate. Note, also, that the concavity and convexity properties of the approximate Cauchy distribution illustrated in Fig. \ref{fig:cauchy} predicts the sign of the expected variations $E\left[\Delta h_k\right]$ and $E\left[\Delta^2h_k\right]$, and, more importantly, the signs of the expected first and second variations are the same for almost all the histogram bins. This confirms that the embedding process produces a disturbance in the feature space that is directional. A second embedding disturbs the features in the same direction as the first embedding, pushing the features' values further from those of a non-embedded image.

\begin{figure}[ht]
  \centering
  %\subfloat[Histogram]{\label{fig:cauchy}
  \includegraphics[width=.7\columnwidth]{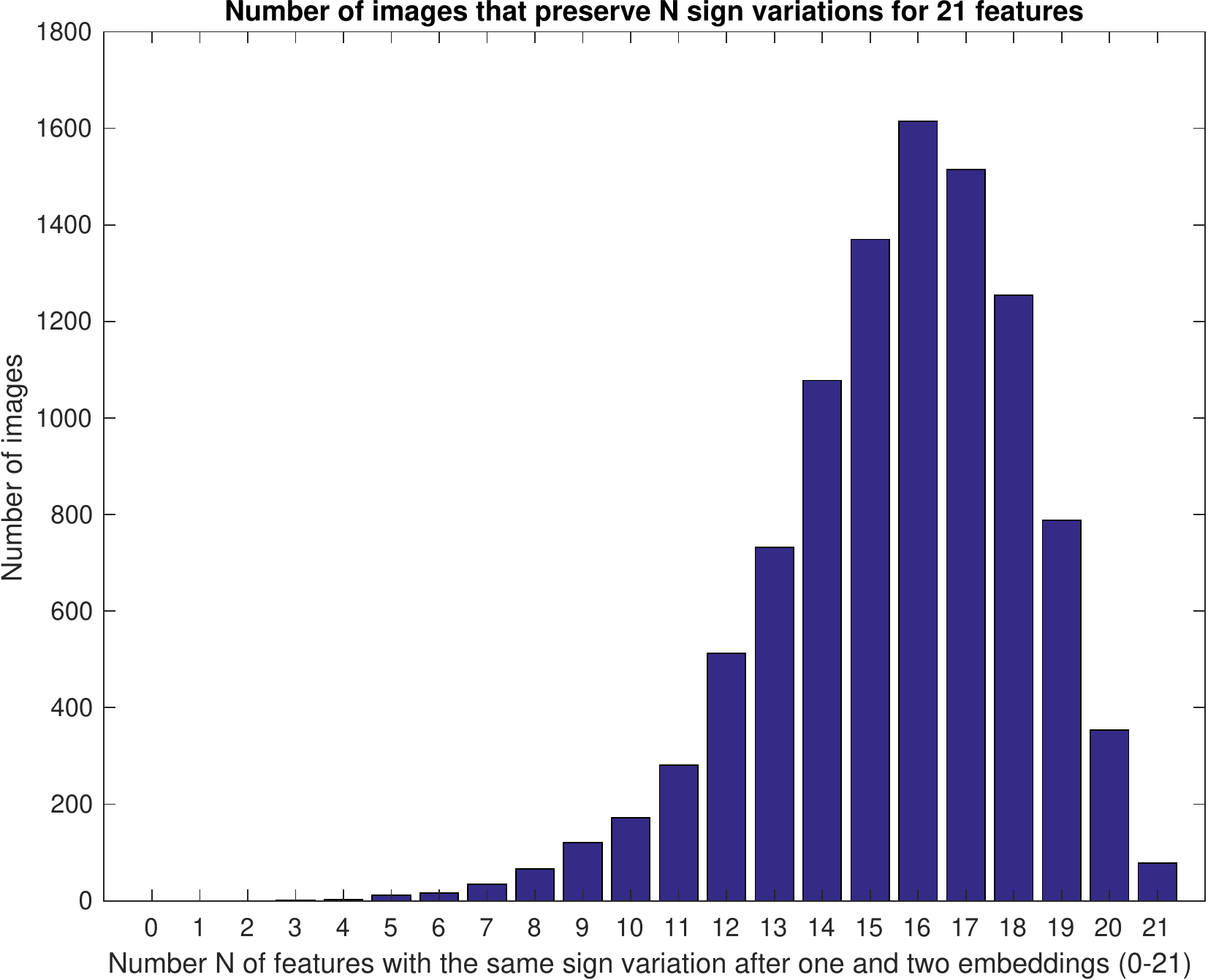}
  \caption{Number of the bins of the residual image that preserve the sign variation in two subsequent embeddings.}
  \label{fig:preserve}
\end{figure}

The following experiment has been carried out to illustrate that the results are not particular of a single image (Lena) but extend to the majority of images. We have analyzed the sign of the variations $\Delta h_k$ and $\Delta^2h_k$ after two subsequent embeddings for the 10,000 images of the BOSS database \cite{BOSS}. In this case, we have not computed an estimation of the expected value (by averaging different experiments) since the number of images is large enough to provide, on average, results close to the expected value. In particular, we have focused on the 21 middle bins of the histogram, i.e. $h_k$, $h'_k$ and $h''_k$ for $k\in[-10,10]$ to guarantee values that are significant enough (not too close to zero). For each bin and image, we consider a success when the signs of both variations are identical and a failure otherwise. Hence, the best result for an image is that the variation of the 21 bins preserves the same sign, whereas the worst case is 0. The results for the 10,000 images are shown in Fig. \ref{fig:preserve}. It can be noticed that the directionality is preserved for the majority of the images and bins. The average ``preservation value'' for the 10,000 images is $15.59$ (out of 21) and the mode is 16. This shows that the majority of the bins are distorted in the same direction after one and two embeddings, which is consistent with the theoretical model and the results illustrated above. Note that if directionality was ``random'', failures and successes would occur with the same probability, leading to a distribution centered between 10 and 11, which is not the case.

\subsection{Directionality for true feature models}
\label{sec:true}
\begin{figure}[ht]
  \centering
   \includegraphics[width=.8\columnwidth]{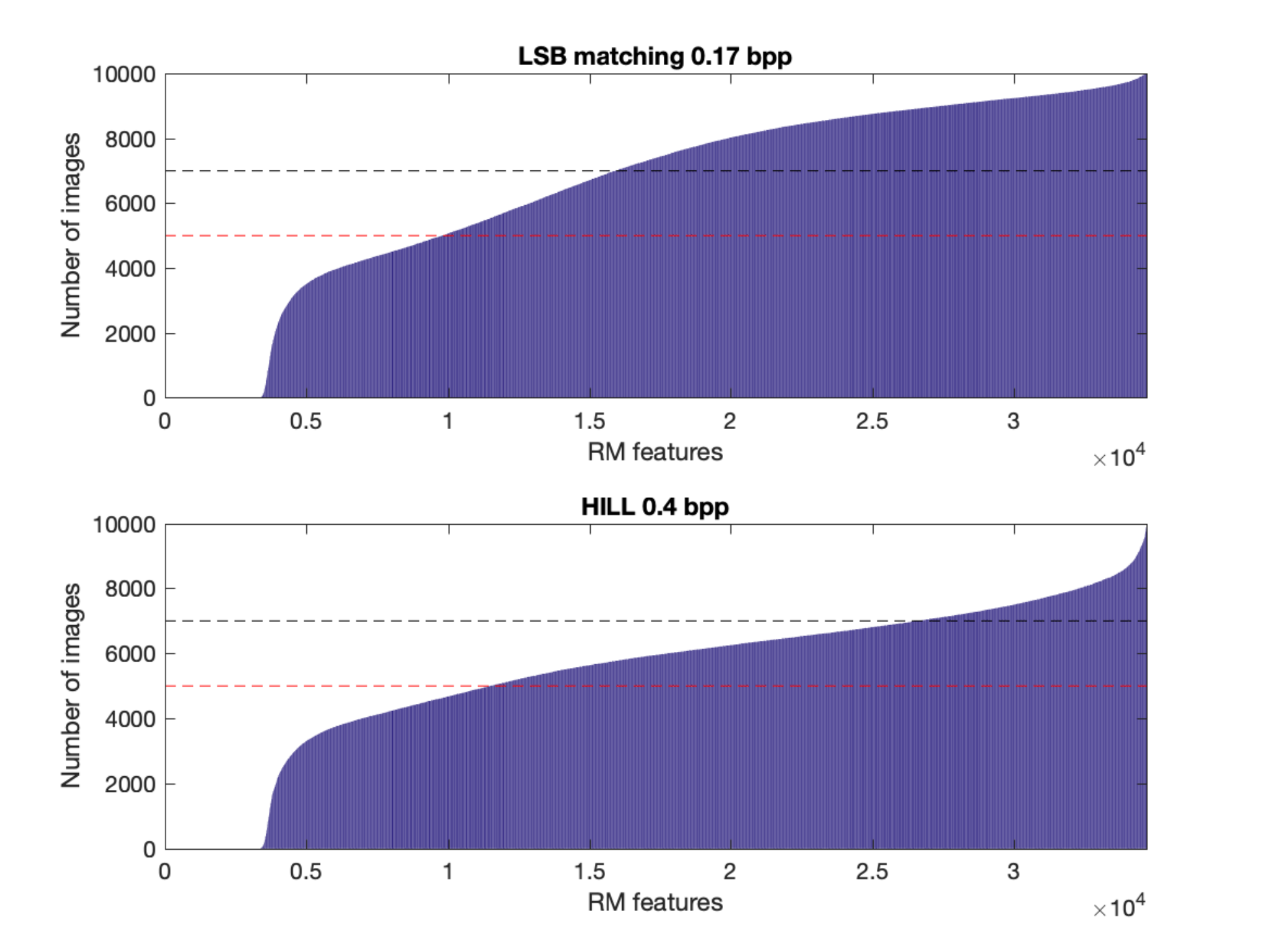}
   \caption{Directionality of RM features for LSB Matching at $0.17$ bpp (top) and HILL steganography at $0.40$ bpp (bottom)}
  \label{fig:truemodel}
\end{figure}

Although the directionality property is validated for the simple feature model in the previous section, the extent to which this property also holds for actual feature models is analyzed next. We have checked the directionality property for the Rich Model features \cite{Fridrich:2012:RM}, a set of 34,671 features combining different steganalytic models. The testing has been carried out both for LSB Matching with an embedding bit rate of $0.17$ bpp and HILL steganography at $0.40$ bpp\footnote{As remarked in Section \ref{sec:lowprob}, the chosen embedding bit rates for LSB matching and HILL lead to a similar value of $\alpha=0.17$.}, using the 10,000 images of the BOSS database. The number of images for which the features are directional is shown in Fig. \ref{fig:truemodel}. For LSB Matching (top), 24,894 features (72\%) are directional for more than half of the images and 18,759 features (54\%) are directional for more than 70\% of the images. This shows that directionality is a common property for RM features with LSB matching. Regarding HILL steganography (bottom), 23,181 features (67\%) are directional for more than half of the images and 8,099 features (23\%) are directional for more than 70\% of the images. Dashed lines show the thresholds for 50\% and 70\% of the images, and the features are sorted according to the number of images that exhibit directionality.

The results of this experiment lead to the following two observations:
\begin{enumerate}[i)]
    \item Directionality is not a rare property for true steganalytic models. Directional features exist for a great deal of images making it possible to classify images in the sets $\mathcal C$, $\mathcal S$ and $\mathcal D$.
    \item Directionality depends on the embedding algorithm. It appears that simple non-adaptive steganographic schemes, such as LSB matching, are more directional than modern adaptive and highly undetectable steganography, such as HILL. However, even for adaptive schemes, there are enough directional features for the primary and secondary classifiers to work. 
\end{enumerate}

\begin{figure}[ht]
\label{fig:directional}
   \centering
 \subfloat[Two directional features for LSB matching.]{\includegraphics[width=.6\columnwidth]{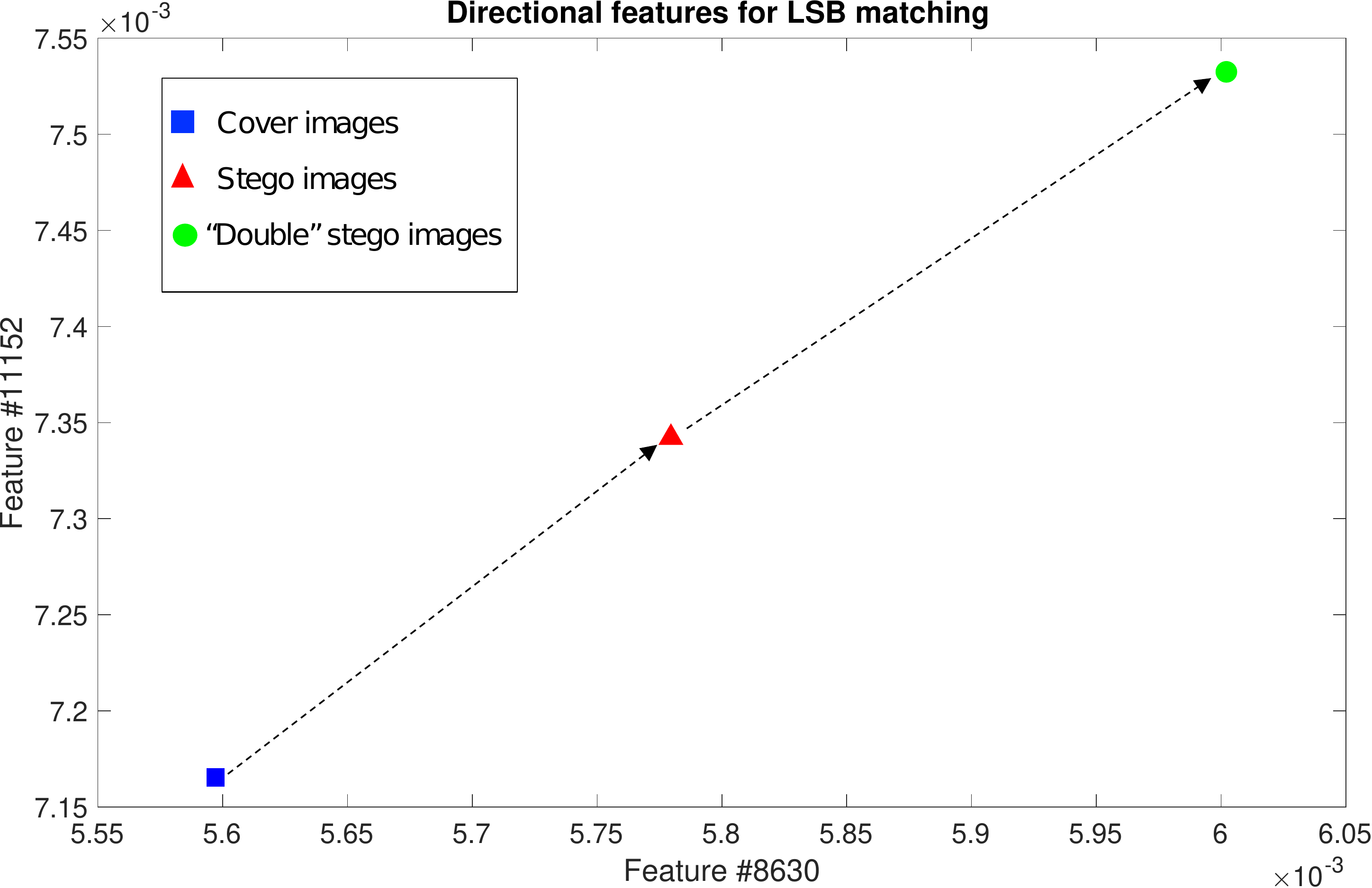}} \\
 \subfloat[Two directional features for HILL steganography.]{\includegraphics[width=.6\columnwidth]{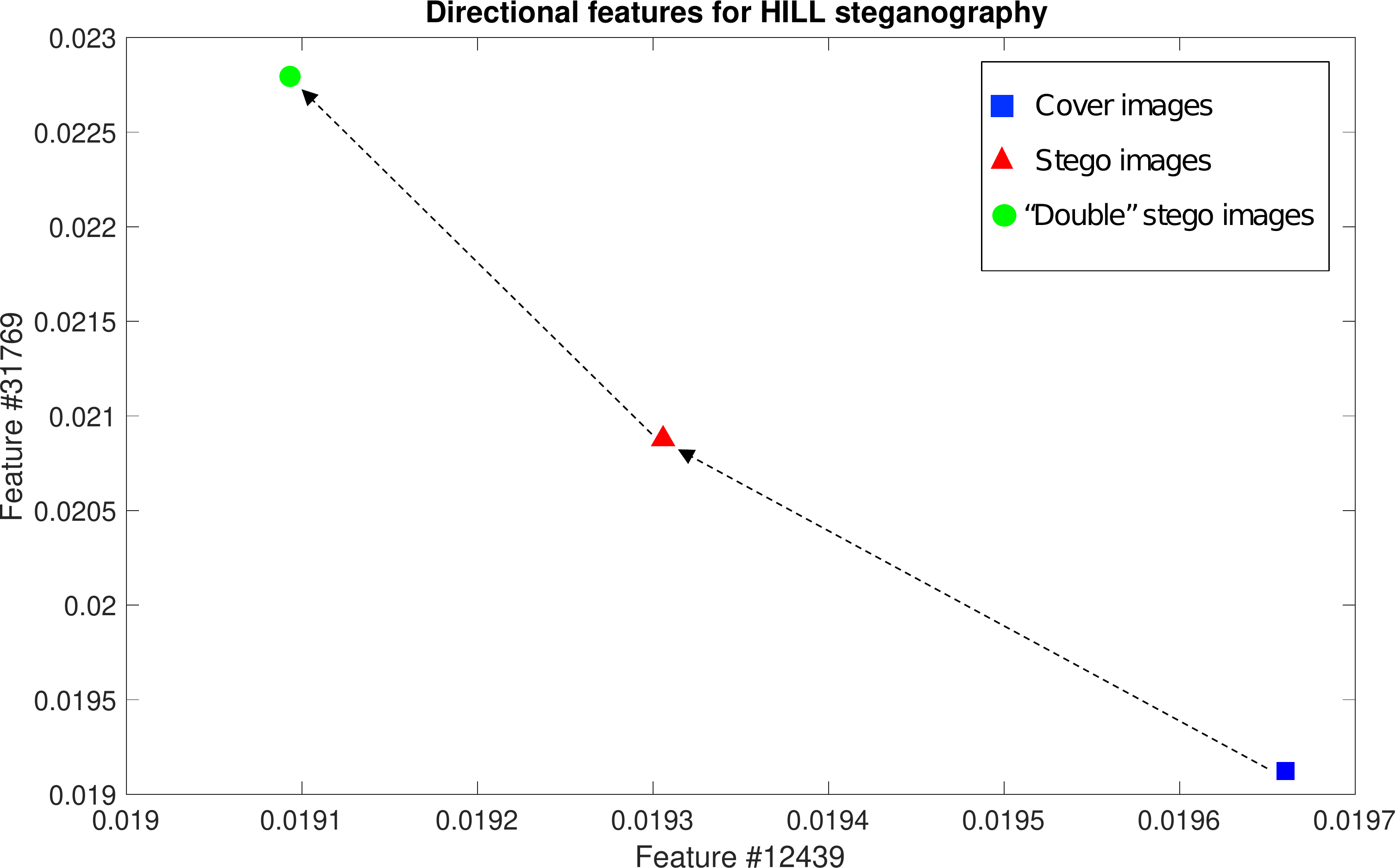}}
  \caption{Example of directional features for LSB matching and HILL steganography.}
  \label{fig:arrows}
\end{figure}

Finally, directionality can be expressed more graphically as depicted in Fig. \ref{fig:arrows}. For this figure, two directional features have been chosen both for LSB matching at $0.17$ bpp (top) and HILL at $0.4$ bpp (bottom). The features are selected from Rich Models such that 80\% of the BOSS images preserve directionality after two subsequent embeddings. The selected features are \#8630 and \#11152 for LSB matching, and \#12439 and \#31769 for HILL steganography. The pictures show the centroid for the selected features for the whole database of 10,000 images. A square is used for the centroid of cover images, a triangle for stego images and a circle for ``double'' stego images. It can be seen that the selected features exhibit directionality, since the vector from the stego to ``double'' stego centroids is similar (and clearly in the same quadrant) to the vector that goes from the cover to the stego centroid. This is precisely the behavior of directional features that makes it possible to apply the practical methods described in the next section.

\section{Practical applications}
\label{sec:practical}
This section presents some practical applications of subsequent embedding based on directional features.

\subsection{Prediction of the classification error}
\label{sec:prediction}

This section describes how subsequent embedding can be used to predict the accuracy of standard steganalysis without knowledge of the true classes (cover/stego) of the testing images. This application is an extension of the work presented in \cite{Lerch-Hostalot:2019}, which is referred to as inconsistency detection (NC detection) in the sequel. The method is based on the classifiers $\pred^\mathsf{A}_{{\stego}_p,\mathcal{F}}$ and $\pred^\mathsf{B}_{{\stego}_p,\mathcal{F}}$ in the following way:
\begin{enumerate}
    \item We have a training set\footnote{The superscript or subscript ``${\mathsf L}$'' is inspired by ``learning'' (in the sense of ``training''), whereas ``${\mathsf T}$'' is borrowed from the word ``testing''.} of images $\mathsf{A^{L}}=\{A_1,\dots A_{N_\mathsf L}\}\subset\mathcal{I}=\mathcal{C} \cup \mathcal{S}$. Without loss of generality, we assume that the first few images in $\mathsf{A^{L}}$ are cover and the remaining ones are stego, i.e. $\{A_1,\dots,A_{M_{\mathsf L}}\}\subset\mathcal{C}$ and $\{A_{M_{\mathsf L}+1},\dots,A_{N_{\mathsf L}}\}\subset\mathcal{S}$, with $1\leq M_{\mathsf L}\leq N_{\mathsf L}$. The training set $A^{\mathsf L}$ is used to train the primary classifier $\pred^\mathsf{A}_{{\stego}_p,\mathcal{F}}$.
    
    \item Next, we create the secondary training set by carrying out a subsequent (random) embedding in all the images of $\mathsf{A^L}$, as $\mathsf{B^L}=\stego^{\#}_p(\mathsf{A^L})=\{B_1,\dots B_{N_\mathsf L}\}$ such that $B_k=\stego^{\#}_p(A_k)$.
    Hence, $\{B_1,\dots,B_{M_{\mathsf L}}\}\subset\mathcal{S}$ and $\{B_{M_{\mathsf L}+1},\dots,B_{N_{\mathsf L}}\}\subset\mathcal{D}$. This secondary training set is used to train the secondary classifier $\pred^\mathsf{B}_{{\stego}_p,\mathcal{F}}$.
   
    \item The objective of steganalysis is to classify a testing set $\mathsf{A^T}=\{A'_1,\dots,A'_{N_{\mathsf T}}\}\subset\mathcal{I}=\mathcal{C} \cup \mathcal{S}$. The result of the classification of $\mathsf{A^T}$ using $\pred^\mathsf{A}_{{\stego}_p,\mathcal{F}}$ is two sets of indexes, namely i) $I_0=\{k_1,\dots,k_{M_{\mathsf T}}\}$ for presumably cover images: $\pred^\mathsf{A}_{{\stego}_p,\mathcal{F}}(A'_{k_i})=0$  for  $i=k_1,\dots,k_{M_{\mathsf T}}$; and ii) $I_1=\{k_{m_{\mathsf T}+1},\dots,k_{N_{\mathsf T}}\}$ for presumably stego images: $\pred^\mathsf{A}_{{\stego}_p,\mathcal{F}}(A'_{k_i})=1$  for  $i=k_{M_{\mathsf T}+1},\dots,k_{N_{\mathsf T}}$, with $1\leq M_{\mathsf T} \leq N_{\mathsf T}$.
   
    \item Now, we create the secondary testing set with a subsequent (random) embedding as $\mathsf{B^T}=\stego^{\#}_p(\mathsf{A^T})=\{B'_1\dots,B'_{N_{\mathsf T}}\}$, with $B'_k=\stego^{\#}_p(A'_k)$. Then, we classify $\mathsf B^T$ using $\pred^{\mathsf B}_{{\stego}_p,\mathcal F}$, obtaining the indexes $I'_0=\{k'_1,\dots,k'_{M'_{\mathsf T}}\}$ for images classified as stego and $I'_1=\{k'_{M'_{\mathsf T}+1},\dots,k'_{N_{\mathsf T}}\}$ for images classified as ``double stego'', with $1\leq M'_{\mathsf T} \leq N_{\mathsf T}$.
    
    \item \textbf{Consistency Filter --- Type 1:} Each index $1,\dots,N_{\mathsf T}$ now belongs to either $I_0$ or $I_1$, and to either $I'_0$ or $I'_1$. If some index $i\in I_0$, this means that $A_i$ is classified as cover by $\pred^{\mathsf A}_{{\stego}_p,\mathcal F}$. Consequently, $B_i=\stego^{\#}_p(A_i)$ should be classified as stego (not as ``double stego'') by $\pred^{\mathsf B}_{{\stego}_p,\mathcal F}$, and we expect $i\in I'_0$. Conversely, if $i\in I_1$, we expect $i\in I'_1$. Then, images that are consistently classified by both classifiers are in $(I_0 \cap I'_0)\cup (I_1 \cap I'_1)$. On the other hand, images that are not consistently classified by the primary and secondary classifiers are the ones with indices in $I_{\mathsf{NC},1}=(I_0 - I'_0) \cup (I_1 - I'_1)$.
   
    \item The testing images $A'_i$ for $i\in I_{\mathsf{NC},1}$ would better not be classified either as cover or stego, since the classification obtained with the secondary classifier is not consistent with that of the primary classifier. Hence, we propose replacing their classification value from 0 or 1 to another category, namely inconsistent (``NC'' from ``non-consistent'').
    
    \item \textbf{Consistency Filter --- Type 2:} If there are enough directional features in the model $\mathcal F$, on classifying the images in $\mathsf A^{\mathsf T}$ using the secondary classifier, all of them should be classified as stego or $\pred^{\mathsf B}_{{\stego}_p,\mathcal F}(A_k)=0$, for all $k=1,\dots,N_{\mathcal T}$. The indices that do not satisfy this condition are kept in the set $I''_1=\{k''_{M''_{\mathsf T}+1},\dots,k''_{N_{\mathsf T}}\}$. This means that $\pred^{\mathsf B}_{{\stego}_p,\mathcal F}(A_i)=1$ for $i\in I''_1$. Similarly, if we classify the images of $\mathsf{B^T}$ using the primary classifier $\pred^{\mathsf A}_{{\stego}_p,\mathcal F}$, with enough directional features, we do not expect any of the images be classified as cover, since all of them are stego or ``double stego''. Hence, we can collect another set of indexes with an inconsistent classification result: $I'''_0=\{k'''_1,\dots,k'''_{M'''_{\mathsf T}}\}$ such that $\pred^{\mathsf A}_{{\stego}_p,\mathcal F}(B_i)=0$, for $i\in I'''_0$. Both Type 2 inconsistencies can be combined in a single set as $I_{\mathsf{NC},2}=I''_1 \cup  I'''_0$.
    
   \item Again, the images $A_i$ for $i\in I_{\mathsf{NC},2}$ would better not be classified as either cover or stego, but as inconsistent (``NC'').
   
   \item Finally, we have four sets of indices: $I^{\mathcal C}$ for images consistently classified as cover, $I^{\mathcal S}$ for images consistently classified as stego, $I_{{\mathsf {NC}}}^{\mathcal C}$ for images classified as cover by $\pred^{\mathsf A}_{{\stego}_p,\mathcal F}$ but with other inconsistent classifications, and $I_{{\mathsf {NC}}}^{\mathcal S}$ for images classified as stego by $\pred^{\mathsf A}_{{\stego}_p,\mathcal F}$ but with other inconsistent classifications.
   
\end{enumerate}

\begin{table}[ht]
    \caption{Classification cases and final output. The ``Type'' column refers to the type of Consistency Filter that applies in that case, if any.}
    \label{tab:cases}
    \centering
    \resizebox{\columnwidth}{!}{
    \begin{tabular}{||c|c|c|c||c|c||}
    \hhline{|t:====:t:==:t|}
$\pred^{\mathsf B}_{{\stego}_p,\mathcal F}(A'_k)$ & $\pred^{\mathsf A}_{{\stego}_p,\mathcal F}(B'_k)$ &
$\pred^{\mathsf A}_{{\stego}_p,\mathcal F}(A'_k)$ & $\pred^{\mathsf B}_{{\stego}_p,,\mathcal F}(B'_k)$ &
Type & Output \\ \hhline{|:====::==:|}
0 & \textcolor{black}{$0^*$} & 0|1 & 0|1 & 2 & NC \\ \hhline{|:====::==:|}
\textbf{0} & \textbf{1} & \textbf{0} & \textbf{0} & \textbf{---} & \textbf{Cover} \\ \hhline{||----||--||}
0 & 1 & \textcolor{black}{$0^*$} & \textcolor{black}{$1^*$} & 1 & NC \\ \hhline{||----||--||}
0 & 1 & \textcolor{black}{$1^*$} & \textcolor{black}{$0^*$} & 1 & NC \\ \hhline{||----||--||}
\textbf{0} & \textbf{1} & \textbf{1}& \textbf{1} & \textbf{---} & \textbf{Stego} \\ \hhline{|:====::==:|}
\textcolor{black}{$1^*$} & \textcolor{black}{$0^*$} & 0|1 & 0|1 & 2 & NC \\ \hhline{|:====::==:|}
\textcolor{black}{$1^*$} & 1 & 0|1 & 0|1 & 2 & NC \\ \hhline{|b:====:b:==:b|}
    \end{tabular}}
\end{table}

This method requires each testing image $A'_k$ be classified once with $\pred^{\mathsf A}_{{\stego}_p,\mathcal F}$ and once with $\pred^{\mathsf B}_{{\stego}_p,\mathcal F}$, and the corresponding secondary testing image \textcolor{black}{$B'_k=\stego^{\#}_p(A'_k)$} also be classified once with $\pred^{\mathsf A}_{{\stego}_p,\mathcal F}$ and once with $\pred^{\mathsf B}_{{\stego}_p,\mathcal F}$. For each image, there are a total of $2^4=16$ outcomes of these four  classifications, and most of the combinations are not consistent. The 16 possible cases are summarized in Table \ref{tab:cases}, where ``0|1'' represents that the final result is the same with either value provided by that classification and, hence, the second, seventh and eighth rows represent four cases each. Note that only two (boldfaced rows) out of the sixteen possible cases lead to a consistent classification of the image $A'_k$ as cover or stego. All other 14 combinations are inconsistent. The values causing the inconsistencies are marked with a ``*'' sign. In addition, it is worth pointing out that some cases for which the Type 2 filter applies may also lead the Type 1 filter to output an inconsistency, but this is simplified in the table for brevity.

The number of inconsistencies: $N_{\mathsf{NC}}=\abs{I_{\mathsf{NC},1}\cup I_{\mathsf{NC},2}}$ can be used to predict the accuracy of a standard classifier that only uses $\pred^{\mathsf A}_{{\stego}_p,\mathcal F}(A'_k)$ as the classification of the image. First, we define the error of the classifier as follows:
$\mathrm{Err}=\frac{\FP+\FN}{N_{\mathsf T}}$,
where $\FP$ and $\FN$ stand for the number of false positives and false negatives, respectively. On the other hand, the classification error computed after removing inconsistencies is denoted as $\overline{\mathrm{Err}}$.

Let us assume that the true (unknown) number of cover images in $\mathsf{A^T}$ is $M_{\mathsf T}$, which, in general, is different from $M'_{\mathsf T}$ (the number of images classified as cover by the primary classifier). We also assume that the images declared as inconsistent are classified by $\pred^{\mathsf A}_{{\stego}_p,\mathcal F}$ in a way equivalent to random guessing, since it is clear that the four different outputs of the classifiers are not well aligned. 

Now, we discuss how we can predict $\FP$ and $\FN$ from the inconsistencies. Inconsistencies can be split into two types $N_{\mathsf{NC}}^{\mathcal C} = \abs{I_{\mathsf{NC}}^{\mathcal C}}$, if the corresponding image is classified as cover by $\pred^{\mathsf A}_{{\stego}_p,\mathcal F}$, and $N_{\mathsf{NC}}^{\mathcal S}=\abs{I_{\mathsf{NC}}^{\mathcal S}}$, if the corresponding image is classified as stego by $\pred^{\mathsf A}_{{\stego}_p,\mathcal F}$. Hence, we have $N_{\mathsf{NC}}=N_{\mathsf{NC}}^{\mathcal C}+N_{\mathsf{NC}}^{\mathcal S}$.
In both cases, we can assume that the primary classifier has classified the image randomly, either as cover or as stego. For the images counted in $N_{\mathsf{NC}}^{\mathcal C}$, the probability of guessing their class correctly is $p_{\mathcal C}=M_{\mathsf T}/N_{\mathsf T}$, and the probability of misclassification is $1-p_{\mathcal C}$. Similarly, for the images counted in $N_{\mathsf{NC}}^{\mathcal S}$, the probability of correct classification is $1-p_{\mathcal C}$, whereas the probability of misclassification is $p_{\mathcal C}$. Hence, if we assume that images with a consistent classification produce no errors, the false negatives and false positives obtained with the primary classifier (without removing inconsistencies) can be estimated as follows:
$\widehat{\FN}=(1-p_{\mathcal C})N_{\mathsf{NC}}^{\mathcal C}$ and $\widehat{\FP}=p_{\mathcal C}N_{\mathsf{NC}}^{\mathcal S}$, and an estimate of the classification error is the following:
\begin{equation}
\label{eqn:pred}
\widehat{\mathrm{Err}}_{p_{\mathcal C}}=\frac{p_{\mathcal C}N_{\mathsf{NC}}^{\mathcal S}+(1-p_{\mathcal C})N_{\mathsf{NC}}^{\mathcal C}}{N_{\mathsf T}}.    
\end{equation}
Such a prediction requires knowing the true ratio of cover ($p_{\mathcal C}$) and stego ($1-p_{\mathcal C}$) images in the testing set, which will be unknown. However, we can define an interval for the predicted classification error for $p_{\mathcal C},(1-p_{\mathcal C})\in[0,1]$: $\widehat{\mathrm{Err}}_0=N_{\mathsf{NC}}^{\mathcal C}/N_{\mathsf T}$ (all stego case), $\widehat{\mathrm{Err}}_1=N_{\mathsf{NC}}^{\mathcal S}/N_{\mathsf T}$ (all cover case), and the special case of an equal number of cover and stego images $p_{\mathcal C}=1-p_{\mathcal C}=0.5$:
\begin{equation}
\label{eqn:pred2}
\widehat{\mathrm{Err}}_{0.5}=\frac{N_{\mathsf{NC}}^{\mathcal S}+N_{\mathsf{NC}}^{\mathcal C}}{2N_{\mathsf T}}=\frac{N_{\mathsf{NC}}}{2N_{\mathsf T}}.    
\end{equation}
Note that Expression (\ref{eqn:pred2}) is exactly the same provided in \cite{Lerch-Hostalot:2019} for an equal number of cover and stego images, whose generalization is Expression (\ref{eqn:pred}).

\begin{figure}[ht]
\centering
 \subfloat[Overestimated bit rate.]{\label{fig:over}\includegraphics[width=0.7\columnwidth]{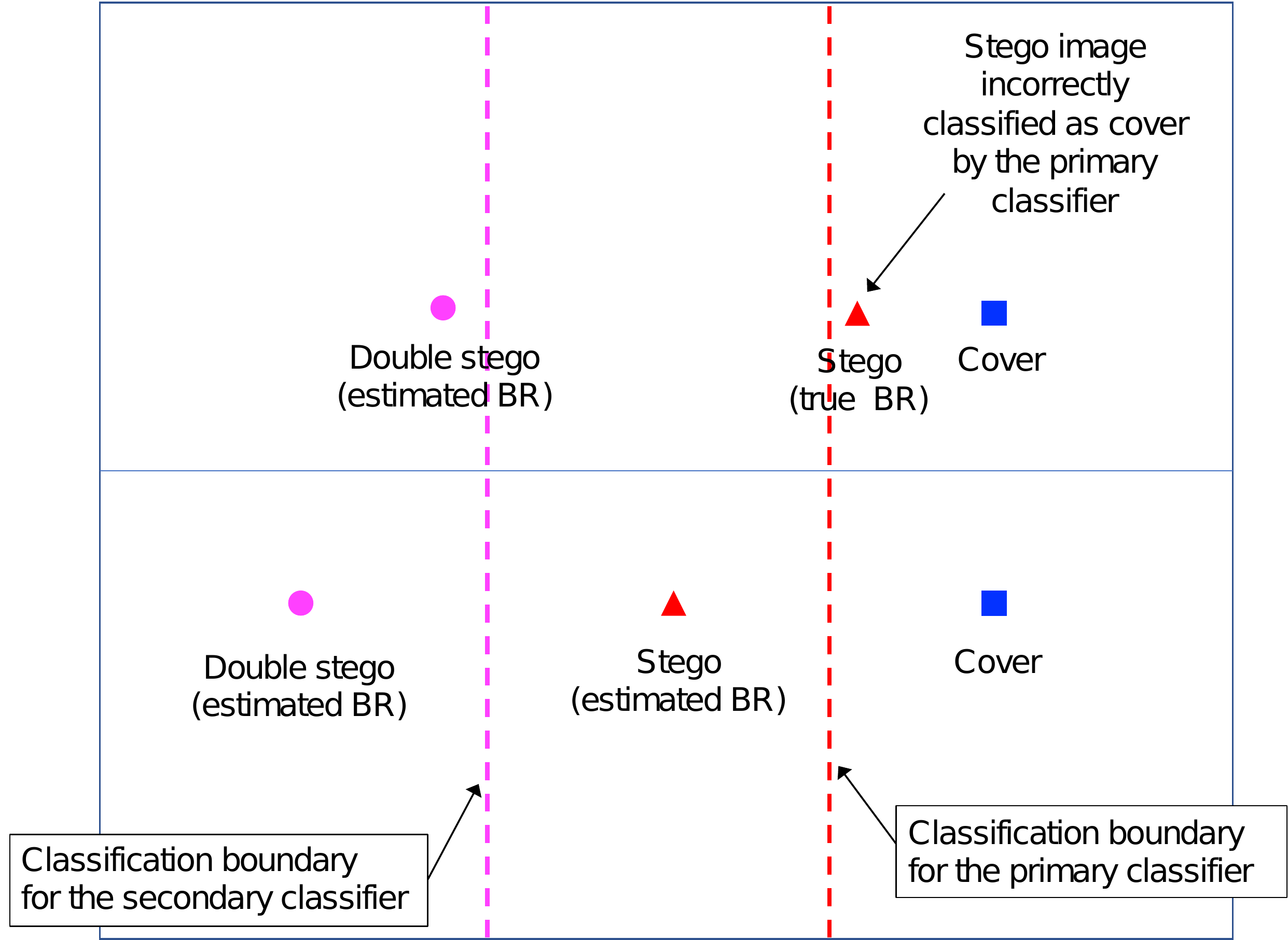}} \\
 \subfloat[Underestimated bit 
 rate.]{\label{fig:under}\includegraphics[width=0.7\columnwidth]{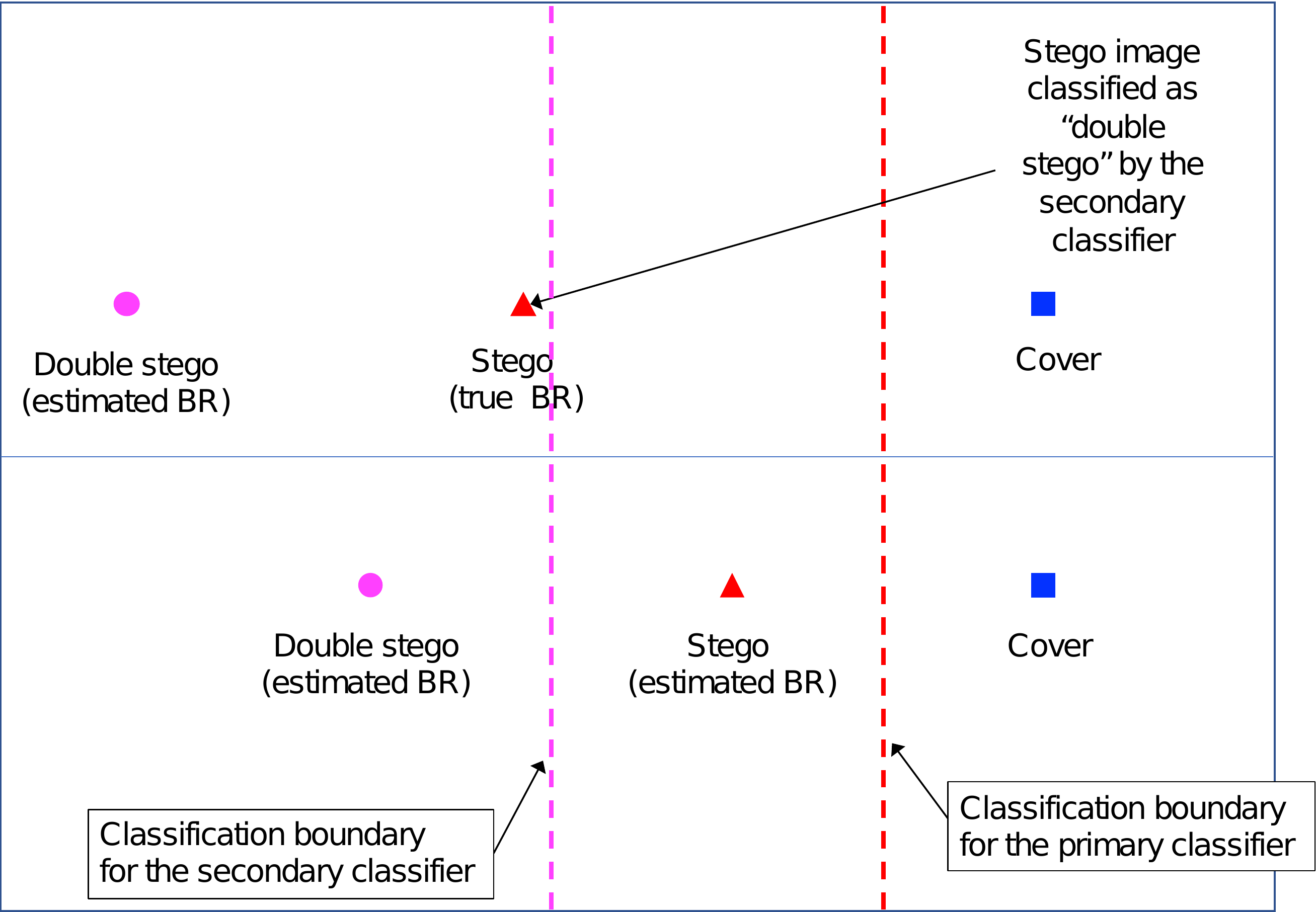}}
  \caption{Classification for overestimation and underestimation of the message length.}
  \label{fig:UML}
\end{figure}

\begin{table*}
\centering
\caption{Experimental results with stego source mismatch (SSM). ``C/S'': \#cover images/\#stego images in the testing set.}
\label{tab:SSM}
\begin{tabular}{||c|c|r|r|r|r|r|r|r|r|r|r|r|r||}
\hhline{|t:==============:t|}

\multicolumn{1}{||c}{\textbf{C/S}} & \multicolumn{1}{|c}{$\bm{p_{\mathsf C}}$} &
\multicolumn{1}{|c}{\textbf{Err}}& 
\multicolumn{1}{|c}{\textbf{TP}} & \multicolumn{1}{|c}{\textbf{TN}} & \multicolumn{1}{|c}{\textbf{FP}} & \multicolumn{1}{|c}{\textbf{FN}} &
\multicolumn{1}{|c}{$\bm{\overline{\mathrm{Err}}}$} & 
\multicolumn{1}{|c}{$\bm{N_{\mathsf{NC}}}$} & 
\multicolumn{1}{|c}{$\bm{N_{\mathsf{NC}}^{\mathcal C}}$} &
\multicolumn{1}{|c}{$\bm{N_{\mathsf{NC}}^{\mathcal S}}$}  &
\multicolumn{1}{|c}{$\bm{\widehat{\mathrm{Err}}_{0.5}}$} & \multicolumn{1}{|c}{$\bm{\widehat p_{\mathsf C}}$} & \multicolumn{1}{|c||}{{$\bm{\widehat{\mathrm{Err}}_{\widehat p_{\mathsf C}}}$}} 
\\ \hhline{|:==============:|}
500/0 & 1 & 0.2840  & 0   & 223 & 41  & 0 & 0.118 & 236 & 135 & 101 & 0.236 & 0.845   & 0.213 \\ \hhline{||--------------||}
500/250 & $2/3$ & 0.2453  & 108 & 223 & 41  & 9 & 0.131  & 369 & 168 & 201 & 0.246 & 0.609   & 0.251 \\ \hhline{||--------------||}
500/500 & $1/2$ & 0.2170 & 218 & 223 & 41  & 18 & 0.155  & 500 & 192 & 308 & 0.250 & 0.482   & 0.248 \\ \hhline{||--------------||}
250/500 & $1/3$ & 0.1960 & 218 & 110 & 23  & 18 & 0.111 & 381 & 125 & 256 & 0.254 & 0.347  & 0.227  \\ \hhline{||--------------||}
0/500 & 0  & 0.1500 & 218 & 0   & 0   & 18 & 0.076 & 264 & 57  & 207 & 0.264 & 0.076 & 0.137 \\ \hhline{|b:==============:b|}          
\end{tabular}
\end{table*}

The midpoint value, $\widehat{\mathrm{Err}}_{0.5}$, is particularly relevant, since it predicts how the primary classifier would perform for a balanced testing set formed by an equal number of cover and stego images. If the true ratio of the testing set were all cover ($p_{\mathcal C}=1$), a ``dummy'' primary classifier that returned $0$ always would make no classification errors but, in the balanced case, the same classifier would have an accuracy of 50\%, equivalent to random guessing. Similarly, a ``dummy'' classifier that returned 1 always would have a perfect accuracy in the all stego case ($p_{\mathcal C}=0$), but its performance in the balanced case would be equivalent to random guessing. In general, the midpoint prediction $\widehat{\mathrm{Err}}_{0.5}$ is a much better indicator of the performance of the primary classifier. When $\widehat{\mathrm{Err}}_{0.5}=0.5$, we can assume that the primary classifier is not working for that particular testing set. Several reasons may prevent a classifier from working for a testing set, such as CSM, SSM or an inappropriate feature model. 

Apart from $\widehat{\mathrm{Err}}_{0.5}$, we can try to use an estimate of $p_{\mathcal C}$ in Expression (\ref{eqn:pred}). As already discussed, the true value of this ratio will be unknown, but we can use the images classified without inconsistencies to approximate this ratio as 
\[\widehat p_{\mathcal C}=\frac{N^{\mathcal C}}{N^{\mathcal C}+N^{\mathcal S}}, \]
where $N^{\mathcal C}=\abs{I^{\mathcal C}}$ and $N^{\mathcal S}=\abs{I^{\mathcal S}}$, i.e. $\widehat p_{\mathcal C}$ denotes the ratio of images classified as cover among all consistently classified images. With this estimate, we can have a prediction $\widehat{\mathrm{Err}}_{\widehat p_{\mathcal C}}$ of the classification error using $\widehat p_{\mathcal C}$ in Expression (\ref{eqn:pred}).

This possibility, however, is only advisable if the primary classifier is not random guessing, such that the predicted ratio $\widehat{p}_{\mathcal C}$ is good enough. Hence, $\widehat{\mathrm{Err}}_{0.5}$ should be clearly below $0.5$ (possibly below $0.25$) for $\widehat{\mathrm{Err}}_{\widehat p_{\mathcal C}}$ to be accurate. The next section illustrates how to use this approach for SSM.

\subsection{Stego Source Mismatch}
\label{sec:ssm}
This section analyses the results with the proposed approach when the tested and the true embedding algorithms differ.  The experiments have been carried out for the BOSS database selecting 5,000 images for training, using both the cover and a stego version of each image. For the testing set, different images from BOSS were chosen and only one version of each image (either cover or stego) was included. The training and subsequent embeddings in testing have been carried out using HILL steganography with an embedding bit rate of $0.4$ bpp, but the true embedding algorithm of the stego images of the testing set was UNIWARD at $0.4$ bpp. I.e., we are trying to detect UNIWARD steganography using HILL as the targeted scheme, which is a clear example of SSM. 

The experiments have been carried out for different ratios of cover images, ranging from all cover to all stego, i.e. $p_{\mathsf C}\in\{1,2/3,1/2,1/3,0\}$. 
The results, shown in Table \ref{tab:SSM}, illustrate the usefulness of the proposed approach. It can be seen that, irrespective of the ratio of cover images in the testing set, the prediction $\widehat{\mathrm{Err}}_{0.5}$ is always around $0.25$, meaning that the classifier is random guessing approximately 50\% of the testing images when we have a balanced testing set with an equal number of cover and stego images. Note, also, that the predicted ratio of cover images ($\widehat p_{\mathsf C}$)  and the corresponding prediction ($\widehat{\mathrm{Err}}_{\widehat p_{\mathsf C}}$) are quite accurate, compared to the column ``Err'', which shows the actual classification error obtained with the standard classifier $\pred^{\mathsf A}_{{\stego}_p,\mathcal F}$ without removing any inconsistencies. In addition, the classification errors after removing inconsistencies (column ``$\overline{\mathrm{Err}}$'') are always better than those obtained without applying the consistency filters. Finally, looking at the results of the last row, with a predicted error $\widehat{\mathrm{Err}}_{\widehat p_{\mathsf C}}=0.137$, we may decide to use the classification of $\pred^{\mathsf A}_{{\stego}_p,\mathcal F}$ as a valid one, including inconsistent samples. 

In summary, the proposed method provides additional indicators, $N_{\mathsf{NC}}$ 
$N_{\mathsf{NC}}^{\mathcal C}$ and $N_{\mathsf{NC}}^{\mathcal S}$, and other values derived from those ($\widehat{\mathrm{Err}}_{0.5}$, $\widehat p_{\mathsf C}$ and $\widehat{\mathrm{Err}}_{\widehat p_{\mathsf C}}$), that can help the steganalyst to decide without requiring side information about the testing set.

\subsection{Unknown message length}
As described in the sections above, the application of subsequent embedding requires the knowledge of ${\stego}_p$, i.e. the embedding algorithm and the parameters $p$, which include the approximate embedding bit rate, also known as message length. However, the message length is only known by the steganographer and the performance of the classification can be seriously affected if the bit rate is underestimated or overestimated. Here, we present a method to find out the correct message length. 

The idea behind the proposed method is depicted in Fig. \ref{fig:UML}. For simplicity, a 1-dimensional feature space is illustrated, but the approach extends to the multiple dimension case. In the pictures, the horizontal axis represents the value of the feature, a square shows the (centroid) position of cover images, a triangle that of stego images, and a circle that of 
``double stego'' images. The primary classifier is trained with an estimated bit rate and produces the a classification line that splits images as cover or stego.  In both diagrams, the bottom part represents the classification (boundary) lines obtained with the estimated or hypothesized embedding bit rate during training, whereas the top part represents an image from the testing set. Two different situations are considered:
overestimated bit rate and underestimated bit rate.

The \textcolor{black}{top} picture (Fig. \ref{fig:over}) shows what may occur if the message length is overestimated. In such a case, a stego image ($A'_k$) from the testing set might be misclassified as cover with the primary classifier\footnote{Note that the distance between cover and stego images, and between stego and ``double stego'' images, increases with the embedding bit rate, as discussed in Remarks \ref{rem:diff} and \ref{rem:diff2}.}. In that case, since the corresponding ``double stego'' image ($B'_k$) of the secondary testing set is created using the estimated bit rate (not the true one), it may be classified as ``double stego'' by the secondary classifier, leading to an inconsistency. This would produce a significant number of inconsistencies for images classified as cover with $\pred^{\mathsf A}_{{\stego}_p,\mathcal F}$, expectedly larger than the number of inconsistencies for images classified as stego.  This leads to the condition $N_{\mathsf{NC}}^{\mathcal C}>N_{\mathsf{NC}}^{\mathcal S}$.

Similarly, the \textcolor{black}{bottom} picture (Fig. \ref{fig:under}) shows the case of underestimating the message length. In that situation, if an image ($A'_k$) of the primary set is stego, when classified with the secondary classifier, it may be identified as ``double stego'', producing an inconsistency. Now, inconsistencies would be more likely for images classified as stego, leading to reverse condition compared to the prior case, i.e. $N_{\mathsf{NC}}^{\mathcal S}>N_{\mathsf{NC}}^{\mathcal C}$.

\RestyleAlgo{boxruled}
\begin{algorithm}
\begin{minipage}[adjusting]{0.9\columnwidth}
\label{algo:uml}
{\footnotesize
\begin{enumerate}
    \item Select a large embedding bit rate for the first experiment.
    \item \label{step:class} Apply the classifiers $\pred^{\mathsf A}_{{\stego}_p,\mathcal F}$ and $\pred^{\mathsf B}_{{\stego}_p,\mathcal F}$ to obtain the test image classes and inconsistencies, as detailed in Section \ref{sec:prediction}. 
    \item Check whether $N_{\mathsf{NC}}^{\mathcal C}>N_{\mathsf{NC}}^{\mathcal S}$. If so, decrease the embedding bit rate and go back to step \ref{step:class}). Otherwise, return the current embedding bit rate.
\end{enumerate}}
\caption{Unknown message length}
\end{minipage}
\end{algorithm}

\begin{table}[ht!]
\centering
\caption{Experimental results with an unknown message length for LSB matching and HILL steganography. ``C/S'': \#cover images/\#stego images in the testing set.}
\label{tab:UML}
\subfloat[][LSB matching steganography.]{
\setlength\tabcolsep{5pt}
\begin{tabular}{||c|c|r|r|r|r|r||}
\hhline{|t:=======:t|}
\textbf{Est./True} & \textbf{C/S} & \textbf{Error} & $\bm{\widehat{\mathrm{Err}}_{0.5}}$ &  $\bm{N_{\mathsf{NC}}}$ & $\bm{N_{\mathsf{NC}}^{\mathcal C}}$ & $\bm{N_{\mathsf{NC}}^{\mathcal S}}$  \\ \hhline{|:=======:|}
0.35/0.20 & 500/250 & 0.112 & 0.064 & 96    & 80 & 16 \\ \hhline{||-------||}
0.30/0.20     & 500/250 & 0.075     & 0.043 & 65    & 51 & 14 \\ \hhline{||-------||}
0.25/0.20     & 500/250 & 0.057     & 0.042 & 63    & 40 & 23 \\ \hhline{||-------||}
0.20/0.20     & 500/250 & 0.056     & 0.048 & 72    & 39 & 33 \\ \hhline{||-------||}
\color{black} \textbf{0.15/0.20}   & \color{black} \textbf{500/250} & \color{black} \textbf{0.056}   & \color{black} \textbf{0.075} & \color{black} \textbf{112} & \color{black} \textbf{44} & \color{black} \textbf{68} \\ \hhline{||-------||}
0.10/0.20     & 500/250 & 0.068     & 0.205 & 307   & 53 & 254 \\ \hhline{|:=======:|}
0.35/0.20     & 500/500 & 0.162     & 0.089 & 177   & 157      & 20 \\ \hhline{||-------||}
0.30/0.20     & 500/500 & 0.080     & 0.046 & 92    & 73 & 19 \\ \hhline{||-------||}
0.25/0.20     & 500/500 & 0.052     & 0.039 & 77    & 47 & 30 \\ \hhline{||-------||}
\textbf{0.20/0.20}   & \textbf{500/500} & \textbf{0.048}   & \textbf{0.045} & \textbf{90}    & \textbf{41} & \textbf{49} \\ \hhline{||-------||}
0.15/0.20     & 500/500 & 0.048     & 0.086 & 171   & 47 & 124 \\ \hhline{||-------||}
0.10/0.20     & 500/500 & 0.056     & 0.272 & 543   & 56 & 487 \\ \hhline{|:=======:|}
0.35/0.20     & 250/500 & 0.205     & 0.105 & 158   & 144      & 14 \\ \hhline{||-------||}
0.30/0.20     & 250/500 & 0.096     & 0.050 & 75    & 60 & 15 \\ \hhline{||-------||}
0.25/0.20     & 250/500 & 0.057     & 0.037 & 56    & 31 & 25 \\ \hhline{||-------||}
\textbf{0.20/0.20}   & \textbf{250/500} & \textbf{0.051}   & \textbf{0.043} & \textbf{64}    & \textbf{20} & \textbf{44} \\ \hhline{||-------||}
0.15/0.20     & 250/500 & 0.045     & 0.097 & 146   & 26 & 120 \\ \hhline{||-------||}
0.10/0.20     & 250/500 & 0.045     & 0.336 & 504   & 27 & 477 \\ \hhline{|:=======:|}
0.35/0.20     & 0/500   & 0.296     & 0.149 & 149   & 139      & 10 \\ \hhline{||-------||}
0.30/0.20     & 0/500   & 0.128     & 0.063 & 63    & 53 & 10 \\ \hhline{||-------||}
0.25/0.20     & 0/500   & 0.064     & 0.036 & 36    & 20 & 16 \\ \hhline{||-------||}
\textbf{0.20/0.20}   & \textbf{0/500} & \textbf{0.044}   & \textbf{0.040} & \textbf{40}    & \textbf{8} & \textbf{32} \\ \hhline{||-------||}
0.15/0.20     & 0/500   & 0.034     & 0.122 & 122   & 9 & 113 \\ \hhline{||-------||}
0.10/0.20     & 0/500   & 0.022     & 0.473 & 473   & 6 & 467     
\\ \hhline{|b:=======:b|}
\end{tabular}} \\

\subfloat[][HILL steganography.]{
\setlength\tabcolsep{5pt}
\begin{tabular}{||c|c|r|r|r|r|r||}
\hhline{|t:=======:t|}
\textbf{Est./True} & \textbf{C/S} & \textbf{Error} & $\bm{\widehat{\mathrm{Err}}_{0.5}}$ & $\bm{N_{\mathsf{NC}}}$ & $\bm{N_{\mathsf{NC}}^{\mathcal C}}$ & $\bm{N_{\mathsf{NC}}^{\mathcal S}}$  \\ \hhline{|:=======:|}
0.60/0.40 & 500/250      & 0.257 & 0.171 & 257   & 150      & 107 \\
\hhline{||-------||}
0.50/0.40 & 500/250      & 0.255 & 0.198 & 297   & 168      & 129 \\
\hhline{||-------||}
\textbf{0.40/0.40} & \textbf{500/250} & \textbf{0.257} & \textbf{0.241} & \textbf{361} & \textbf{169} & \textbf{192} \\
\hhline{||-------||}
0.30/0.40 & 500/250      & 0.273 & 0.321 & 482   & 215      & 267 \\
\hhline{||-------||}
0.20/0.40 & 500/250      & 0.332 & 0.389 & 583   & 223      & 360 
\\ \hhline{|:=======:|}
0.60/0.40 & 500/500      & 0.282 & 0.172 & 343   & 183      & 160 \\
\hhline{||-------||}
0.50/0.40 & 500/500      & 0.264 & 0.200 & 399   & 203      & 196 \\
\hhline{||-------||}
\textbf{0.40/0.40} & \textbf{500/500} & \textbf{0.244} & \textbf{0.241} & \textbf{482} & \textbf{197} & \textbf{285} \\
\hhline{||-------||}
0.30/0.40 & 500/500      & 0.245 & 0.334 & 667   & 244      & 423 \\
\hhline{||-------||}
0.20/0.40 & 500/500      & 0.285 & 0.405 & 810   & 250      & 560 
\\ \hhline{|:=======:|}
\color{black} \textbf{0.60/0.40} & \textbf{\color{black} 250/500}      & \textbf{\color{black}0.317} & \textbf{\color{black}0.176}      & \textbf{\color{black}264}   & \textbf{\color{black}131}      & \textbf{\color{black}133} \\
\hhline{||-------||}
0.50/0.40 & 250/500      & 0.280 & 0.200 & 300   & 140      & 160 \\
\hhline{||-------||}
0.40/0.40 & 250/500 & 0.232 & 0.242 & 363 & 130 & 233 \\
\hhline{||-------||}
0.30/0.40 & 250/500      & 0.223 & 0.349 & 524   & 159      & 365 \\
\hhline{||-------||}
0.20/0.40 & 250/500      & 0.249 & 0.416 & 624   & 147      & 477 
\\ \hhline{|:=======:|}
0.60/0.40 & 0/500 & 0.376 & 0.183 & 183   & 84      & 0 \\
\hhline{||-------||}
0.50/0.40 & 0/500 & 0.312 & 0.189 & 189   & 95 & 94 \\
\hhline{||-------||}
\textbf{0.40/0.40} & \textbf{0/500} & \textbf{0.204} & \textbf{0.246} & \textbf{246} & \textbf{62} & \textbf{184} \\
\hhline{||-------||}
0.30/0.40 & 0/500 & 0.174 & 0.400 & 400   & 71 & 329 \\
\hhline{||-------||}
0.20/0.40 & 0/500 & 0.164 & 0.488 & 488   & 76 & 412 
\\ \hhline{|b:=======:b|}
\end{tabular}}
\end{table}

This analysis can be easily converted into an iterative algorithm to determine the true bit rate, with Algorithm \ref{algo:uml}. This simple, but effective, algorithm has been tested for LSBM matching and HILL steganography. The experiments have been carried out for the BOSS database selecting 5,000 images for training, using both the cover and a stego version of each image. Different images from BOSS were chosen for the testing set and only one version of each image (either cover or stego) was included. An Ensemble Classifier with RM features has been used for the primary and secondary classification. The true embedding bit rate is $0.20$ bpp for LSB matching and $0.4$ bpp for HILL, and the prospective bit rates are: $0.35$, $0.3$, $0.25$, $0.2$, $0.15$ and $0.1$ bpp for LSB matching and  $0.6$, $0.5$, $0.4$, $0.3$ and $0.2$ bpp for HILL. The results are shown in Table \ref{tab:UML}, where the first column details the tested and the true steganographic methods, which are LSBM or HILL in all cases, but with different embedding bit rates. The second column shows the number of cover and stego images in the testing set, whereas the remaining columns are self-explanatory. The true ratio of cover images in the testing set varies from $2/3$ in the first block (the first six or five rows) to 0 in the last block (the last six or five rows). Note that the ``all cover'' case would make no sense in this scenario, since there would be no true bit rate to find out. It can be observed that Algorithm \ref{algo:uml} finds the correct embedding bit rate in almost all cases, remarked as a boldfaced row for each block, except for the first block of LSBM matching, where the estimated bit rate is $0.15$ bpp, whereas the true bit rate was $0.2$ bpp, and the third block for HILL steganography ($0.6$ detected for a true bit rate of $0.4$ bpp). Apart from those two cases, the first result with a lower number of inconsistencies for images classified as cover with $\pred^{\mathsf A}_{{\stego}_p,\mathcal F}$ is the one that provides the right estimate of the true embedding bit rate. Even in the extreme case that all images are stego (last block), the first experiment with $N_{\mathsf{NC}}^{\mathcal S}>N_{\mathsf{NC}}^{\mathcal C}$ provides the correct estimate of the message length for both HILL and LSB matching.

\begin{table*}[ht]
\centering
\caption{Real-world scenario, comparison between a standard classifier with embedding bit rate estimated at $0.30$ bpp, detection of inconsistencies with bit rate estimated at $0.30$ bpp and Algorithm \ref{algo:real}. ``$N$'': Number of images in the subset, ``Acc.'': classification accuracy without inconsistencies.}
\label{tab:real}
\resizebox{.9\textwidth}{!}{
\begin{tabular}{||c|r||r|r|r||r|r|r|r||r|r|r|r||}
\hhline{~~|t:===:t:====:t:====:t|} 
 \multicolumn{2}{c||}{} & \multicolumn{3}{c||}{\textbf{Standard} $\pred^{\mathsf A}_{{\stego}_p,\mathcal F}$, $\mathbf{\textbf{BR}=0.30}$} & \multicolumn{4}{c||}{\textbf{NC detection}, $\mathbf{\textbf{BR}=0.30}$} & \multicolumn{4}{c||}{\textbf{Algorithm \ref{algo:real}}} \\
 \hhline{|t:==::===::====::====:|}
 \textbf{Subset} & $N$ & \textbf{TP+TN} & \textbf{FP+FN} & \textbf{Acc.}  &  
 \textbf{TP+TN} & \textbf{FP+FN} & $\bm{N_{\mathsf{NC}}}$ & \textbf{Acc.} & 
  \textbf{TP+TN} & \textbf{FP+FN} & $\bm{N_{\mathsf{NC}}}$ & \textbf{Acc.} \\
  \hhline{|:==::===::====::====:|}
  Stego $0.20$ & 25 & 9 & 16 & $0.360$ & 7 & 10 & 8 & $0.412$ & 10 & 12 & 3 & $\mathbf{0.455}$ \\ \hhline{||--||---||----||----||}
  Stego $0.30$ & 25 & 21 & 4 & $0.840$ & 16 & 1 & 8 & $\mathbf{0.941}$ & 20 & 3 & 2 & $0.870$ \\ \hhline{||--||---||----||----||}
  Stego $0.40$ & 25 & 20 & 5 & $0.800$ & 11 & 1 & 13 & $\mathbf{0.917}$ & 18 & 2 & 5 & $0.900$ \\ \hhline{||--||---||----||----||}
  Stego $0.50$ & 25 & 23 & 2 & $0.920$ & 4 & 0 & 21 & $\mathbf{1.000}$ & 19 & 0 & 6 & $\mathbf{1.000}$ \\ \hhline{||--||---||----||----||}
  Stego $0.60$ & 25 & 22 & 3 & $\mathbf{0.880}$ & 1 & 2 & 22 & $0.333$ & 13 & 2 & 10 & $0.867$ \\ \hhline{||--||---||----||----||}
  All stego & 125 & 95 & 30 & $0.760$ & 39 & 14 & 72 & $0.736$ & 80 & 19 & 26 & $\mathbf{0.806}$ \\ \hhline{||--||---||----||----||}  
  All cover & 125 & 80 & 45 & $0.640$ & 33 & 10 & 82 & $\mathbf{0.767}$ & 54 & 28 & 43 & $0.659$ \\ \hhline{||--||---||----||----||}
  All  & 250 & 175 & 75 & $0.700$ & 72 & 24 & 154 & $\mathbf{0.760}$ & 134 & 47 & 69 & $0.740$ \\ 
  \hhline{|b:==:b:===:b:====:b:====:b|}
\end{tabular}}
\end{table*}

\subsection{Real-world scenario}
\label{sec:real}
In a real-world scenario, when a steganalyst receives a batch of images, we do not expect only cover and stego images, but also, stego images with different message sizes. This is indeed challenging for the proposed approach, since the approximate bit rate is assumed to be known when applying the transformation $\stego^{\#}_p(\cdot)$ to obtain the secondary sets from the primary ones. Even in that case, inconsistencies can be detected to improve the classification accuracy, as illustrated in the following example. Let us consider a batch of 250 images taken from the BOSS database, half of them (125) are cover and 125 are embedded with HILL steganography with embedding bit rates in $\{0.2,0.3,0.4,0.5, 0.6\}$. There are 25 stego images for each bit rate, and all the images are different. We assume that the steganalyst knows that HILL is used and that the bit rate is between $0.2$ and $0.6$. The method described in Section \ref{sec:prediction} cannot be applied directly, since we have a mixture of message lengths. However, the steganalyst can apply the method for all the bit rates in the set \textcolor{black}{provided above}. This requires that each image must be classified $4\times 5=20$ times (four classifications per each bit rate as detailed in Table \ref{tab:cases}). For each embedding rate, there are four possible outcomes of the classification: ``Cover'', ``Stego'', ``Non-consistent type 1'' (``NC1'') and ``Non-consistent type 2'' (``NC2'').

\RestyleAlgo{boxruled}
\begin{algorithm}[ht]
\label{algo:real}
\begin{minipage}[adjusting]{0.9\columnwidth}
{\footnotesize
\begin{enumerate}
   \item Order the classification results for all images from highest to lowest targeted bit rate.
   \item Given the image $A'_k$ in the testing set, if there are less than two classification results different from ``NC2'', classify the image as inconsistent.
    \item Otherwise, if the classification results are regular, moving from cover or inconsistent to stego or inconsistent, and at least one result is ``Stego'', classify $A_k'$ as ``Stego''.
    \item If there is no regular behavior, take the number of classifications as ``Cover'' and ``Stego'' and make a vote, by returning the most repeated outcome. In case of an equal number of ``Cover'' and ``Stego'' votes, classify $A'_k$ as inconsistent.
\end{enumerate}}
\end{minipage}
\caption{Multiple bit rate scenario}
\end{algorithm}

The results can be sorted from highest to lowest targeted bit rate (from $0.6$ to $0.2$ bpp). In principle, cover images should be identified as cover with all the classifiers, although some inconsistencies may appear. Regarding stego images, those with a true embedding bit rate lower than $0.6$ bpp might produce an inconsistency for the higher estimated bit rates as shown in Fig. \ref{fig:UML}, or the image may be misclassified as cover. Then, when the targeted bit rate gets closer to the true one, the inconsistency may disappear and the image would be correctly identified as stego. Another inconsistency may arise for very low targeted bit rates due to underestimation. If this regular behavior appears (inconsistencies/cover for higher targeted bit rates and inconsistencies/stego for lower targeted bit rates) the image can be classified as stego. 

Inspired by this idea, Algorithm \ref{algo:real} is proposed to be applied in this real-world classification problem. In this scenario, we have tested three different solutions: 1) a standard RM-based classifier trained with a bit rate of $0.3$ bpp, 2) the inconsistency detection system presented in Section \ref{sec:prediction} with an estimated bit rate of $0.3$ bpp, and 3) Algorithm \ref{algo:real}. The estimated bit rate of $0.3$ bpp used in 1) and 2) has been chosen to be in the center of all true bit rates, which are between $0.2$ and $0.6$ bpp as described above. 
The results obtained for this real-world scenario are provided in Table \ref{tab:real}. The five rows labeled as ``Stego 0.20''-``Stego 0.60'' show the classification outcome obtained for stego images in the testing set, ordered by true embedding rate (from lowest to highest). The next row is the cumulative result of all stego images (125 in total). The row labeled as ``All cover'' presents the results for the 125 cover images of the testing set. Finally, the total results are given in the last row. Note that for stego images, TN and FP are always 0. Similarly, for cover images, $\text{FN} = \text{TP} = 0$.

In terms of accuracy, the standard classifier is the best one only for stego images with the highest bit rate ($0.6$ bpp). The inconsistency detection method provides the best accuracy for the central values of the embedding bit rate (from $0.3$ to $0.5$) and for cover images, but there is a considerable price to pay in the form of inconsistencies. Indeed, the NC detection with an estimated bit rate of $0.3$ bpp yields 154 inconsistencies, meaning that only $38.6\%$ of the images are classified. Finally, the method proposed in Algorithm \ref{algo:real} is a trade-off solution between both extremes. This method provides the best accuracy results for extreme embedding bit rates ($0.2$ and $0.5$ bpp),the best aggregate accuracy results for all stego images (correct classification for more than 80\% of the classified stego images), and better results for cover images than those of the standard classifier. The overall accuracy of Algorithm \ref{algo:real} is better than that of the standard classifier and not far from the NC detection method and, compared to NC detection, the number of inconsistencies is significantly reduced, since $72.2\%$ images are classified (almost twice compared to the NC detection approach ($38.6\%$ of classified images). The conclusion is that the detection of inconsistencies based on the directionality of features can be used in real-world scenarios to help the steganalyst.

\section{Conclusions}
\label{sec:conclusion}

Subsequent embedding was proposed as a tool to help in image steganalysis in previous works. The idea behind this technique is to embed new random data in the training and the testing sets to obtain transformed sets with stego and ``double stego'' images. The transformed set can build new classifiers that make it possible to improve the classification results in different ways.

However, the subsequent embedding strategy requires some conditions to work. This technique assumes that additional data embeddings will distort the features of the images in the same direction in the feature space, something that was conjectured but not proved before. This paper provides a theoretical basis for subsequent embedding, proving that there exists a simple set of features that satisfy the directionality of features under certain conditions that are analyzed mathematically. The theoretical (simplified) model is also validated in several experiments with different images. Once the directionality principle has been proved for the simplified feature model, the same property is verified for some state-of-the-art feature models that are too complex for such a theoretical analysis. 

Moving from theory to practice, the applicability of the ideas introduced in the paper is illustrated in four different cases. Namely, 1) estimating the classification error for the testing set with standard steganalysis without knowledge of the true classes of the images, 2) stego source mismatch, 3) unknown message length, and 4) a real-world scenario in which cover and stego images with different embedding bit rates are given to the steganalyst. In those four cases, the proposed approach is shown to succeed in helping the steganalyst to improve the classification results.

Regarding future work, we plan to extend the subsequent embedding technique to more ambitious scenarios, paving the way for moving steganalysis from lab to real-world conditions. 

\section*{Acknowledgments}
We gratefully acknowledge the support of NVIDIA Corporation with the donation of an NVIDIA TITAN Xp GPU card that has been used in this work. This work was partly funded by the Spanish Government through grant RTI2018-095094-B-C22 ``CONSENT''. The authors also acknowledge the funding obtained from the EIG {CON\-CERT}-Japan call to the project entitled ``Detection of fake newS on SocIal MedIa pLAtfoRms'' (DISSIMILAR) through grant PCI2020-120689-2 (Ministry of Science and Innovation, Spain).
\textcolor{black}{We are most grateful to our colleague Daniel Blanche-T., who has carried out a thorough proofreading of the paper, making it possible to detect and correct several writing mistakes.}

\bibliographystyle{IEEEtranS}
\bibliography{main}

\vspace{-1.25cm}
\begin{IEEEbiography}[{\includegraphics[width=1in,height=1.2in,clip,keepaspectratio]{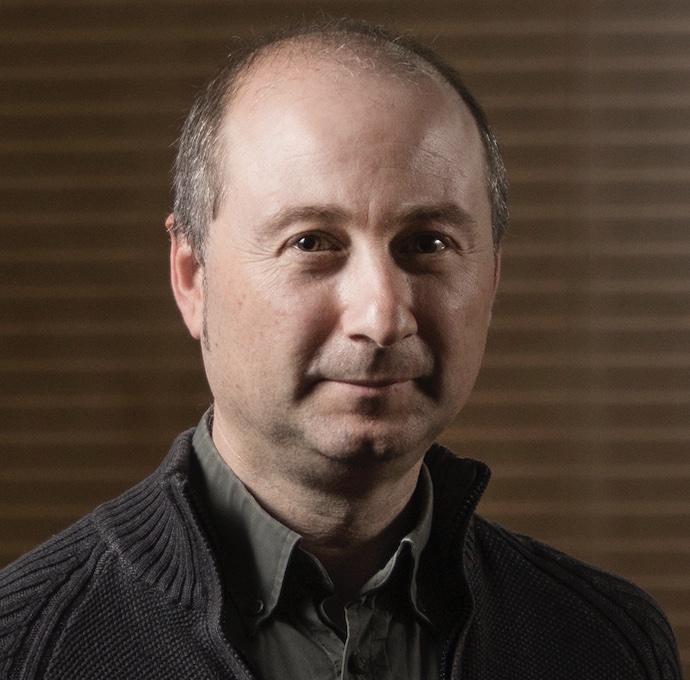}}]{David Meg\'ias}
is Full Professor
at the Universitat Oberta de Catalunya
(UOC), Barcelona, Spain, and also the current
Director of the Internet Interdisciplinary Institute (IN3) at UOC. He has authored more than 100 research papers in
international conferences and journals. He has participated
in different national research projects
both as a contributor and as a Principal Investigator.
He has also experience in international
projects, such as the European Network of Excellence
of Cryptology of the 6th Framework Program of the European Commission.
His research interests include security, privacy, data hiding, protection of
multimedia contents, privacy in decentralized networks and information
security. He is a member of IEEE.
%\begin{IEEEbiography}
\end{IEEEbiography}

\vspace{-1.25cm}

\begin{IEEEbiography}
    [{\includegraphics[width=1in,height=1.25in,clip,keepaspectratio]{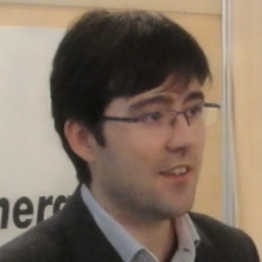}}]{Daniel Lerch-Hostalot} is course tutor and research associate of Data Hiding and Cryptography at the Universitat Oberta de Catalunya (UOC). He has received his PhD in Network and Information Technologies from Universitat Oberta de Catalunya (UOC), in 2017. His research interests include steganography, steganalysis, image processing and machine learning.
\end{IEEEbiography}

%%% Aquests comptadors estan comprovats (David) %%%
\setcounter{table}{5}
\setcounter{figure}{8}
\setcounter{lemma}{1}
\setcounter{remark}{12}
\setcounter{equation}{11}

% ----> Arxiv page
%\thispagestyle{empty}
%\begin{minipage}[t][5cm][t]{1.\textwidth}
%
%{\Huge This is a preprint of the paper: }
%{\Large
%
%~
%
% David Meg\'{\i}as, Daniel Lerch-Hostalot, ``Subsequent embedding in image %steganalysis: Theoretical framework and practical applications''.
% 
%~
%
% This work has been submitted to the IEEE for possible publication. Copyright may be transferred without notice, after which this version may no longer be accessible.
%}

%\end{minipage}
%\newpage
%\setcounter{page}{0}
% <---- Arxiv page

\title{Subsequent embedding in \textcolor{black}{targeted} image steganalysis: Theoretical framework and practical applications -- Supplemental material}

\author{David Meg\'{\i}as,~\IEEEmembership{Member,~IEEE,} and Daniel Lerch-Hostalot
\IEEEcompsocitemizethanks{\IEEEcompsocthanksitem D. Meg\'{\i}as and D. Lerch-Hostatlot are with the Internet Interdisciplinary Institute (IN3), Universitat Oberta de Catalunya (UOC), CYBERCAT-Center for Cybersecurity Research of Catalonia, Castelldefels (Barcelona), Catalonia, Spain.\protect\\
E-mail: \{dmegias,dlerch\}@uoc.edu}}

	%\markboth{IEEE TRANSACTIONS ON DEPENDABLE AND SECURE COMPUTING, VOL. X, NO. X, JANUARY 2021}%
	%{Authors: title}
	
	\maketitle

\IEEEpeerreviewmaketitle{}

\appendices

\begin{table*}
\centering
\textcolor{black}{\caption{Literature review and addressed challenges.}
\label{tab:literature}}
\begin{tabular}{||m{6.5cm}||c|c|c|c||}
\hhline{|t:=:t:====:t|}
\textbf{Reference} & \multicolumn{1}{m{1cm}|}{\textbf{CSM}} & \multicolumn{1}{m{2.5cm}|}{\textbf{Unknown message length}} & \multicolumn{1}{m{1cm}|}{\textbf{SSM}} & \multicolumn{1}{m{2.5cm}||}{\textbf{Error prediction}}\\  \hhline{|:=::====:|}
Westfeld and Pfitzmann (2000) \cite{westfeld} & --- & \checkmark & --- & --- \\ \hhline{||-||----||} 
Fridrich, Goljan, Hogea, and Soukal (2003) \cite{Fridrich2003} & --- & \checkmark & --- & --- \\ \hhline{||-||----||} 
Pevn\'{y} (2011) \cite{Pevny:2011} & --- & \checkmark & --- & --- \\ \hhline{||-||----||} 
Kodovsk\'{y}, Sedighi, and Fridrich (2014) \cite{Fridrich:2014:csm_mitig} & \checkmark & --- & --- & --- \\ \hhline{||-||----||} 
Pasquet, Bringay, and Chaumont (2014) \cite{Pasquet:2014} & \checkmark & --- & --- & --- \\ \hhline{||-||----||} 
Xu, Dong, Wang, and Tan (2015) \cite{Xu:2015:csm} & \checkmark & --- & --- & --- \\ \hhline{||-||----||} 
Boroumand, Chen, and Fridrich (2019) \cite{Boroumand:2019:SRNet} & \checkmark & \checkmark & \checkmark & --- \\ \hhline{||-||----||} 
Yousfi, Butora, and Fridrich (2020) \cite{Yousfi:2019:breaking_alaska1} & \checkmark & \checkmark & \checkmark & --- \\ \hhline{||-||----||} 
Giboulot, Cogranne, Borghys, and Bas (2020) \cite{Giboulot:2020} & \checkmark & --- & --- & --- \\ \hhline{||-||----||} 
Yousfi and Fridrich (2020) \cite{Yousfi:2020:onehot} & \checkmark & \checkmark & \checkmark & --- \\ \hhline{||-||----||} 
Yousfi, Butora, Khvedchenya, and Fridrich (2020) \cite{Yousfi:2020:alaska2} & \checkmark & \checkmark & --- & --- \\ \hhline{||-||----||} 
Ruiz, Yedroudj, Chaumont, Comby, and Subsol (2021) \cite{ruiz:2021} &\checkmark & --- & --- & --- \\ \hhline{||-||----||} 
\textbf{Proposed (subsequent embedding)} & \textbf{\checkmark} & \textbf{\checkmark} & \textbf{\checkmark} & \textbf{\checkmark} \\ \hhline{||-||----||} 
\hhline{|b:=:b:====:b|}      
\end{tabular}
\end{table*}

\section{Contributions of the paper and related work}
\label{sec:contribution}
This appendix summarizes the contributions of this paper, analyzes related works of the literature, and compares the proposed method with prior art.

\subsection{Contributions of the paper}
This article makes different contributions to steganalysis. All these contributions are based on the idea of embedding additional information --referred to as ``subsequent embedding'' in this paper-- in the image being analyzed. Such an idea is rooted in our previous articles \cite{Lerch-Hostalot:2015} and \cite{Lerch-Hostalot:2019}.

From a theoretical point of view, the main contribution of this paper consists in a consistent mathematical framework that supports the use of subsequent embedding for targeted steganalysis. Such a theoretical framework was not provided in \cite{Lerch-Hostalot:2015} and \cite{Lerch-Hostalot:2019}. In \cite{Lerch-Hostalot:2015}, it is shown that subsequent embedding makes it possible to develop an unsupervised steganalysis method, whereas in \cite{Lerch-Hostalot:2019}, an approach to estimate the classification error of steganalysis is presented. In this paper, we endorse the subsequent embedding approach with a solid theoretical framework and provide new applications and experiments in more complex scenarios, much closer to those of the real world.

From a practical point of view, the most important outcome of this article is the possibility of applying steganalysis in scenarios in which a classifier might not work as expected, including CSM, an unknown length of the hidden message (unknown message length or unknown embedding bit rate) and SSM. With previous (standard) steganalysis tools, the steganalyst would not know whether his/her classifier is working well or not. However, with the inconsistency analysis introduced in Section \ref{sec:practical}, the steganalyst will have additional information about the quality of the classification, which will ultimately allow him/her to determine whether the classification results are sufficiently reliable or not. We would like to remark that such an estimation it is not available with standard steganalysis. 

It is worth pointing out that, in the subsequent embedding framework, the underlying classifier is the same as in standard steganalysis. Thus, when we compare our results using Rich Models or Deep Learning, we are actually using exactly the same classifier as in standard steganalysis and, thus, its classification results (for the primary classifier) are the same. What makes the difference with the proposed approach is the analysis of the classification inconsistencies obtained through an additional (secondary) classifier. The additional information obtained from subsequent embedding and the secondary classifier makes it possible to improve the classification results, either leaving inconsistent samples unclassified, or estimating the classification error of the primary (standard) classifier.   

The proposed method, based on subsequent embedding, may not provide the ultimate solution to the CSM, unknown message length, and SSM problems, but we expect that it will constitute a further step in the right direction, paving the way to the application of steganalysis in the conditions that occur in the real world.

\subsection{Related work}
As discussed in Section 1, several approaches have been proposed to try to overcome the CSM problem. Most of the methods proposed in the literature to face CSM have not been widely accepted and, currently, the most popular approach is to build a sufficiently large and complete image database to use it to train a large neural network. Although this solution may be successful in the future, this is something that has not been achieved yet. Apart from CSM, an unknown message length and SSM are other sources of uncertainty that degrade the classification results of steganalysis in real-world scenarios. 

In the remainder of this section, we provide a detailed review of image steganalysis works that address at least one of the relevant challenges of modern steganalysis, namely, CSM, unknown message length and SSM. We also remark the main contributions of the proposed subsequent embedding approach compared to those works.
A summary of the surveyed works and the  addressed challenges is shown in Table \ref{tab:literature}, where the different papers are listed in chronological order, from least to most recent. The details about those schemes are the summarized below:
\begin{itemize}
    \item Westfeld and Pfitzmann \cite{westfeld} present a collection of visual and statistical attacks against four different steganographic tools (EzStego, Jsteg, Steganos, and S-Tools). The attacks described in the paper are aimed at detecting whether images carry secret information embedded using those tools. The steganalytic attacks are classified as visual, which rely on a human observer, and statistical, which are  easier to automate. The detection of the selected steganographic schemes is quite successful with the proposed attacks even in the case of an unknown message length. The paper describes the Chi-square attack that was very popular a few years ago. However, this technique is not effective against modern steganography, and has been replaced by Machine Learning-based methods.
    
    \item Fridrich, Goljan, Hogea, and Soukal \cite{Fridrich2003} directly tackle the unknown message length problem for a collection of popular steganographic schemes, namely, the F5 and OutGuess algorithms for JPEG, the EzStego algorithm with random straddling for palette images, and  LSB embedding for uncompressed image formats. The results presented in the paper are very remarkable and the estimation of the message length is quite accurate for the tested images. However, modern steganography is much more difficult to detect and the proposed methodology is not designed against the current advanced methods (especially adaptive steganography).
    
    \item Pevn{\'{y}} \cite{Pevny:2011} shows that facing an unknown message length significantly decreases the accuracy of steganalysis. To deal with the problem, the paper presents two approaches. The first one consists in using quantitative steganalysis. The second one is based on transforming a one-sided hypothesis test into a simple hypothesis test, assuming a uniformly distributed payload. Both approaches are tested experimentally against LSB matching and nsF5. The paper shows that the accuracy of steganalysis highly depends on the knowledge about the payload.
    
    \item Kodovsk\'{y}, Sedighi, and Fridrich  \cite{Fridrich:2014:csm_mitig} discuss the impact of CSM on steganalysis and work with some techniques to mitigate its effects. To that purpose, they experiment with the nsF5 algorithm, different JPEG qualities, and different image sources. The results show that the steganalytic techniques that work with some data sources do not work properly with others. In other words, steganalysis largely depends  on the scenario and the knowledge that the steganalyst has about the source, and whether those steganalytic techniques are adapted to it. 
    
    \item Pasquet, Bringay, and Chaumont \cite{Pasquet:2014} consider a scenario in which the message length and the steganographic method are known, but the image source is not. To face the CSM problem, they use a small image database and an ensemble classifier with feature selection. Additionally, they present the \textit{islet approach}, which allows improving the performance of the classifier.
    
    \item Xu, Dong, Wang, and Tan  \cite{Xu:2015:csm} propose a method to build a sufficiently representative image database, since small databases often represent only a small part of the actual data distribution. In addition, the authors  propose a weighted Ensemble Classifier that can be adapted to the testing set. In this way, the authors, using a large number of images downloaded from the Internet, manage to build a small database without any significant loss in terms of performance.
    
    \item Boroumand, Chen, and Fridrich \cite{Boroumand:2019:SRNet} propose a deep residual network for steganalysis, called SRNet. This convolutional network was the first one to allow end-to-end training from randomly initialized kernels regardless of the analyzed domain. This network is verified experimentally for both JPEG and spatial domains.  At the time of its publication, this scheme achieved the best state-of-the-art results. Furthermore, this architecture is general enough to address the CSM, SSM and unknown message length issues to some extent, as shown later on in the ALASKA competition.
    
    \item Yousfi, Butora, and Fridrich \cite{Yousfi:2019:breaking_alaska1} describe their winning solution to the first ALASKA steganalysis competition. Their approach used SRNet-based classifiers to build several multi-class tile detectors trained on various combinations of input channels. The classifiers were trained by grouping the images by quality factor. The authors also introduce a JPEG reverse compatibility attack for images with qualities 99 and 100.
    
    \item Giboulot, Cogranne, Borghys, and Bas \cite{Giboulot:2020} carry out an in-depth analysis of the effects of CSM and the parameters that produce it. The paper includes a large number of experiments in multiple scenarios using different feature vectors that make it possible to analyze their impact in CSM. The authors identify the processing pipeline as the main cause of CSM.
    
    \item Yousfi and Fridrich \cite{Yousfi:2020:onehot} propose a different approach to steganalysis in the JPEG domain. Given the evidence that feature-based detectors perform better than CNNs, since the latter fail to compute simple statistics of DCT coefficients, the authors propose the use of OneHot networks. This new architecture is based on the clipped one-hot encoding, which allows the network to learn to compute higher-order statistics for DCT coefficients. This type of network was part of Guanshuo Xu's winning solution of the ALASKA 2 steganalysis competition.
    
    \item In the ALASKA 2 competition,  it became clear that some network architectures, such as EfficientNet, MixNet or ResNet, designed and pre-trained for computer vision, significantly outperform networks designed for steganalysis, such as SRNet. Yousfi, Butora, Khvedchenya and Fridrich \cite{Yousfi:2020:alaska2} analyze these architectures and show how the suppression of pooling/stride in the first few layers improve the learning process of the architecture. The authors conjecture that these networks work so well because the ImageNet pretraining has exposed them to a large number of different images.
    
    \item Ruiz, Yedroudj, Chaumont, Comby, and Subsol \cite{ruiz:2021} present Large Scale Steganalysis Database (LSSD), a database of two million JPEG images. The article describes the pipeline used to create the database. This can be regarded as another step towards the construction of a sufficiently diverse and large enough database, a necessary tool to investigate how to deal with the CSM problem in the real world.
\end{itemize}

The next section presents a discussion of the works summarized above and remarks the differences between them and the proposed approach. 

\subsection{Discussion and comparative analysis}
In the articles outlined in the previous section --some of which were published in the context of the ALASKA competitions-- the problem of the CSM is dealt with by creating a large and heterogeneous database. This is, in fact, what the ALASKA Image Database attempts to reproduce. However, even if a database is large and heterogeneous, other databases may exist with a different statistical distribution of the images (or their features). In the future, it may be possible to circumvent the CSM problem using huge databases, although this is still an open problem.

Currently, most of the existing approaches do not consider a testing database that differs from the training one to study the effects of the CSM. As already mentioned, the most popular approach consists in trying to build a large enough database to bypass the problem, but it is uncertain if this is a convenient strategy in the real world, even if the proposed solutions have obtained good results for the ALASKA databases. On the other hand, the method proposed in this paper allows detecting the existence of CSM by using the error prediction technique. Multiple experiments obtained with the subsequent embedding technique for differing training and testing databases can be found in our previous article \cite{Lerch-Hostalot:2019}. This type of analysis is uncommon in the literature. 

Regarding SSM, the analyzed articles face the problem by considering a mixture of different steganographic methods to train the classifier, but do not try to detect schemes that were not used in the training phase.  On the other hand, with the proposed subsequent embedding strategy, we also consider the latter (more complex) scenario, which is presented in Section \ref{sec:ssm}. The experiments presented in 
in Section \ref{sec:ssm} illustrate how a steganographic scheme (UNIWARD) can be detected using a different algorithm (HILL) for training the classifier. In any case, since many steganographic schemes use similar techniques, training with multiple steganographic algorithms seems to be a valid option to deal with the SSM problem. We also consider this approach in Section \ref{sec:mulembed} (Appendix \ref{sec:towards}).

As the unknown message length problem is concerned, it is not too difficult to build a training database with multiple payloads. Hence, this problem is possibly less challenging than the previous ones. Although this strategy can significantly increase the number of images needed for training, this is not typically a problem with modern CNN-based detectors. As it can be seen in the analyzed articles, all the methods that work with the ALASKA databases, especially those used in ALASKA 2, solve this problem satisfactorily.

Finally, to the best of our knowledge, there is no prior art proposing a mechanism to estimate the prediction of the classification error that is obtained for a given testing set. The proposed method, thus, brings also novelty in this aspect, which is a very valuable tool to deal with CSM (an other sources of uncertainty) in the real world. To conclude this discussion, it is worth pointing out that the techniques proposed in this paper can be combined with any standard steganalytic classifier to obtain a prediction of the classification error. Hence, the proposed methodology provides the steganalyzer with additional information to complement the standard classification results and identify sources of uncertainty, at a very reduced cost.

\section{Proof of Lemma \ref{lemma:adapt}}
\label{sec:proof}

\begin{lemma}
If we assume dependent random $\pm1$ changes of the pixel values, the expected value of the histogram bin $h'_k$ can be approximated as follows:
\[
    E[h'_k]\approx (1-\alpha)h_k+\frac{\alpha}{2}\left(h_{k-1}+h_{k+1}\right).
\]
\end{lemma}

\begin{proof}
We first obtain the joint probabilities for the pixel changes in $x_{i,j}$ and $\hat x_{i,j}$ similarly as discussed for the non-adaptive case. Let $\probmodhat=\beta=\alpha/2$ be the probability of modifying the pixel prediction, and the corresponding complementary probability be $\probeqhat=1-\beta$. First, the joint probability of modifying both the pixel and its prediction is:
\begin{multline*}
\prob{(\Delta x_{i,j}\neq0)\cap (\Delta \hat x_{i,j}\neq0)} = \\ \probmodhat\prob{\Delta x_{i,j}\neq0 \left| \Delta \hat x_{i,j}\neq0\right.}=\beta\beta',
\end{multline*}
where $\beta'=\prob{\Delta x_{i,j}\neq0 \left| \Delta \hat x_{i,j}\neq0\right.}$ will typically be greater than $\beta$, since, in adaptive steganography, once a pixel is modified, the probability that a neighboring pixel be also modified is larger. 

Similarly, the joint probability that $\hat x_{i,j}$ be modified and $x_{i,j}$ be not modified is the following:
\begin{equation}
\label{eqn:prob2}
\prob{(\Delta x_{i,j}=0)\cap (\Delta \hat x_{i,j}\neq0)}=\beta(1-\beta').
\end{equation}

In the same way, the joint probability that $\hat x_{i,j}$ be not modified and $x_{i,j}$ be modified is:
\begin{equation}
\label{eqn:prob3}    
\prob{(\Delta x_{i,j}\neq 0)\cap (\Delta \hat x_{i,j}=0)} =(1-\beta)\beta'',
\end{equation}
for some $\beta''$. Now, we expect Expression (\ref{eqn:prob2}) and Expression (\ref{eqn:prob3}) to be equal, since they are totally symmetric cases, and hence:
\begin{equation}
\label{eqn:beta2}
   (1-\beta)\beta'' = \beta(1-\beta') \iff \beta''=\frac{\beta(1-\beta')}{1-\beta}. 
\end{equation}

Finally, the joint probability that neither $\hat x_{i,j}$ nor $x_{i,j}$ be changed is the following:
\[
\prob{(\Delta x_{i,j}=0)\cap (\Delta \hat x_{i,j}=0)}=(1-\beta)(1-\beta'').
\]
Now, substituting by the value of $\beta''$ obtained in Expression (\ref{eqn:beta2}), we have:
\[
\prob{(\Delta x_{i,j}=0)\cap (\Delta \hat x_{i,j}=0)}=1-2\beta+\beta\beta'.
\]

\begin{table}[ht]
\caption{Combined probabilities of changes in $x_{i,j}$ and $\hat x_{i,j}$ for adaptive steganography.}
\label{tab:prob2}
\centering 
\begin{tabular}{||c||c|c||} 
\hhline{~|t:==:t|} 
\multicolumn{1}{c||}{} & $\Delta\hat x_{i,j}=0$ & $\Delta\hat x_{i,j}\neq 0$ \\ \hhline{|t:=::==:|}
$\Delta x_{i,j}=0$ & $(1-\beta)(1-\beta'')=1-2\beta+\beta\beta'$ & $\beta(1-\beta')$ \\
\hhline{||-||-|-||} 
$\Delta x_{i,j}\neq0$ &$(1-\beta)\beta''=\beta(1-\beta')$ & $\beta\beta'$ \\
\hhline{|b:=:b:==:b|}
\end{tabular}
\end{table}

The results are summarized in Table \ref{tab:prob2}, and it can be easily checked that the sum of all four probabilities is 1.
From these four probabilities, we can proceed like in Lemma \ref{lemma:expected} to obtain:
\begin{multline*}
E[h'_k]=
\left(1-2\beta+\beta\beta'\right)h_k+\beta(1-\beta')\left(h_{k-1}+h_{k+1}\right)
\\+\frac{\beta\beta'}{2}h_k+\frac{\beta\beta'}{4}\left(h_{k-2}+h_{k+2}\right),
\end{multline*}
and, hence:
\begin{equation}
\begin{split}
E[h'_k]=\left(1-\alpha\right)h_k+\frac{\alpha}{2}\left(h_{k-1}+h_{k+1}\right)+\varepsilon',
\end{split}
\end{equation}
with
\begin{equation}
\label{eqn:betap}
\varepsilon'=\frac{3}{2}\beta\beta' h_k -\beta\beta' h_{k-1} - \beta\beta' h_{k+1} + \frac{\beta\beta'}{4}h_{k-2}+ \frac{\beta\beta'}{4}h_{k-2}. 
\end{equation}
Again, since both $\beta$ and $\beta'$ are close to zero, we approximate $\varepsilon'\approx 0$ and the proof is thus completed analogously as done for Lemma \ref{lemma:expected}.
\end{proof}

\color{black}

\section{Validation of the working hypothesis}
\label{sec:valhypo}
After introducing the method to detect classification inconsistencies in Section \ref{sec:prediction}, we can now proceed with an experiment to validate the working hypothesis presented in Section \ref{sec:directional}. The conducted experiment is the following, we have taken 500 cover and 500 stego images from the BOSS database, where the stego images have been created using HILL steganography with an embedding bit rate of $0.4$ bpp. For these 1,000 images, we have carried out four classification experiments, as detailed below:
\begin{enumerate}
    \item \label{case:dir} \textbf{Non-directional features only (NDFO):} For each image, all features that are directional for that image are replaced by 0 (erased).
    \item \textbf{Directional features only (DFO):} For each image, all features that are non-directional for that image are replaced by 0.
    \item \textbf{Random removal of features (RRF):} For each image, some features are randomly replaced by 0. The number of erased features is the same as in the first case (NDFO).
    \item \textbf{50\% of directional features removed (50\% DFR):} For each image, half the features that are directional for that image are replaced by 0.
\end{enumerate}

\begin{table*}
\centering
\textcolor{black}{\caption{Validation of the working hypothesis.}
\label{tab:validation}}
\begin{tabular}{||l||r|r|r|r|r||r|r|r|r|r|r||r||}
\hhline{|t:=:t:=====:t:======:t:=:t|}
\multirow{2}{*}{\textbf{Experiment}} & 
\multicolumn{5}{c||}{\textbf{Standard EC+RM}} & \multicolumn{6}{c||}{\textbf{Detection of inconsistencies}} & \multicolumn{1}{c||}{\textbf{Ratio of}} \\  \hhline{||~||-----||------||~||}
& \multicolumn{1}{c}{\textbf{Err}}& 
\multicolumn{1}{|c}{\textbf{TP}} & \multicolumn{1}{|c}{\textbf{TN}} & \multicolumn{1}{|c}{\textbf{FP}} & \multicolumn{1}{|c||}{\textbf{FN}} &
\multicolumn{1}{c}{$\bm{\widehat{\mathrm{Err}}_{0.5}}$} & 
\multicolumn{1}{|c}{\textbf{TP}} & \multicolumn{1}{|c}{\textbf{TN}} & \multicolumn{1}{|c}{\textbf{FP}} & \multicolumn{1}{|c}{\textbf{FN}} &
\multicolumn{1}{|c||}{$\bm{N_{\mathsf{NC}}}$} & $\pred^{\mathsf A}_{{\stego}_p,\mathcal F}(B'_k)=0$
\\ \hhline{|:=::=====::======::=:|} 
\textbf{NDFO} & $0.433$ & 246 & 321 & 179 & 254 & $0.466$ & 22 & 32 & 9 & 5 & 932 & $0.498$
\\ \hhline{||-||-----||------||-||}
\textbf{DFO} & $0.307$ & 334 & 359 & 141 & 166 & $0.293$ & 175 & 175 & 30 & 35 & 585 & $0.286$
\\ \hhline{||-||-----||------||-||}
\textbf{RRM} & $0.374$ & 315 & 311 & 189 & 185 & $0.357$ & 97 & 113 & 29 & 48 & 713 & $0.318$
\\ \hhline{||-||-----||------||-||}
\textbf{50\% DFR} & $0.368$ & 311 & 321 & 179 & 189 & $0.365$ & 88 & 111 & 39 & 33 & 729 & $0.326$ \\
\hhline{|b:=:b:=====:b:======:b:=:b|}         
\end{tabular}
\end{table*}

The results of these four experiments are shown in Table \ref{tab:validation}. The following discussion can be carried out regarding the obtained results:
\begin{itemize}
    \item It can be observed that the worst results are obtained when only non-directional features are used (NDFO). First, in the last column, we can see that the primary classifier ($\pred^{\mathsf A}_{{\stego}_p,\mathcal F}$) fails to classify stego and ``double stego'' images ($B'_k$). All of them should be classified as stego (1), but almost half of them (ratio equal to $0.498$) are incorrectly identified as cover (i.e. the images are classified at random). Consequently, the number of inconsistencies is huge. Almost all images (932 out of 1,000) produce an inconsistency. Besides, the classification error with standard Ensemble Classifiers and Rich Models (EC+RM) is large ($0.433$).
    \item On the other hand, the best results are obtained when only directional features are used (DFO). In that case, the ratio of images $B'_k\in \mathsf{B^T}$ that are incorrectly classified as cover is reduced to $0.286$, the number of inconsistencies is the lowest (585 out of 1,000), and the classification error with a standard EC+RM is $0.307$, meaning that almost 70\% of the images are classified correctly.
    \item The other two experiments (RRM and 50\% DFR) are carried out for control and represent intermediate situations between NDFO and DFO. In RRM, several features (either directional or non-directional) are removed randomly, whereas in 50\% DFR, half of the directional features are removed. When 50\% of the directional features are removed, the primary classifier can still deal with the images in $\mathsf{B^T}$ quite successfully, and the classification error ($0.326$) is much closer to the best result ($0.286$), obtained with directional features only, than the worse one ($0.498$), obtained with non-directional features only. 
\end{itemize}

These results are consistent with the working hypothesis that directional features are sufficient to prevent mistaking ``double stego'' with cover images, which is a condition required by the method of detection of inconsistencies to work properly. In addition, we can observe that it is not required that all features are directional. The results of the experiments RRM and 50\% DFR confirm that it is enough to have a relevant number of directional features.

Furthermore, these results also provide more insight. It can be observed that the higher the ratio of directional features, the better classification results are obtained, not only with the proposed method of detection of inconsistencies, but also with standard steganalyzers such as EC+RM. This seems to point out that directional features are \textbf{good features} for steganalys and opens new opportunities for future research. If this fact is confirmed, working to achieve better features that preserve directionality for most images and steganographic schemes is a very promising approach to move steganalysis closer to real-world conditions.

\color{black}
\section{Generalization for true feature models}
This appendix discusses the extension of the theoretical framework presented in Sections \ref{sec:simple} and \ref{sec:theoretical} to actual feature models and, more precisely to SPAM  \cite{SPAM} and ``minmax''  \cite{Fridrich:2011:minmax}  features, which are included in Rich Models \cite{Fridrich:2012:RM}.
\color{black}

Although the proofs given in \textcolor{black}{Section \ref{sec:theoretical}} are limited to the histogram of the residual image, it is not difficult to realize that a similar behavior is expected for other feature models, such as those of the SPAM \cite{SPAM} steganalysis. In fact, SPAM features can be viewed as an extension of the histogram residual image bins for multiple directions and different orders. SPAM features consider the co-occurrences of two (first-order) or three (second-order) adjacent pixel differences, limiting them within a given threshold $[-T,T]$ in order to reduce the dimension of the feature vectors. If the bins of the residual image exhibit directionality in subsequent embeddings, a similar behavior is expected for at least some of the SPAM features. \textcolor{black}{The extension for multiple dimensions is discussed in Section \ref{sec:multiple}, whereas truncation is analyzed in Section \ref{sec:truncation}.}

\textcolor{black}{Regarding the RM feature model, it must be taken into account that ``min'' and ``max'' operation are used to obtain several of its features. ``Min'' and ``max'' are non-linear operations and it is not obvious that the results presented for the simple model also hold in that case. This particular problem is considered in Section \ref{sec:minmax}.
In addition, some experiments are presented to confirm the generalization of the conditions for directionality of ``minmax'' features and truncation in Section \ref{sec:experimental}}.

\textcolor{black}{Finally, the extension to RM (which include SPAM) features is illustrated  experimentally in Section \ref{sec:true} (main document).}

\color{black}

\subsection{Multiple dimensions}
\label{sec:multiple}
In the simplest case, SPAM features consider three pixels, e.g. $x_{i-2,j}$, $x_{i-1,j}$ and $x_{i,j}$ for the horizontal direction, and the differences between them (two residuals), i.e. $r_{i,j}=x_{i,j}-x_{i-1,j}$ and $r_{i-1,j}=x_{i-1,j}-x_{i-2,j}$. Upon embedding, the pixels can be modified by adding 0 (no change), $+1$ or $-1$. For three pixels, this leads to $3\times3\times3=27$ cases. If the original residuals are $(r_{i-1,j},r_{i,j})=(a,b)$, the new pair of residuals after embedding become $(a',b')$ with $a'\in\{a-2,a-1,\dots,a+2\}$ and $b'\in\{b-2,b-1,\dots,b+2\}$. Among the possible 25 cases of the Cartesian product $\{a-2,a-1,\dots,a+2\}\times\{b-2,b-1,\dots,b+2\}$, only 15 actually occur, and their probabilities can be computed as done in Sections \ref{sec:non-adaptive} and \ref{sec:adaptive} for non-adaptive and adaptive steganography respectively. This procedure leads to a two-dimensional counterpart of the one-dimensional analysis presented in Sections \ref{sec:simple} and \ref{sec:theoretical}. Such analysis would be long and somewhat cumbersome by the larger number of cases, but feasible. In the end, the variation of a two-dimensional histogram would be examined, and the sign of those variations after one and two embeddings might be analyzed, yielding conditions for (soft) directionality. A similar procedure is possible for third-order SPAM features, moving to a three-dimensional histogram analysis. Possibly, a theoretical analysis of the multidimensional case would entail more ``geometrical'' complexity, since, apart from concavity and convexity areas, there might be some regions that are convex in some dimensions but concave in different ones. This could lead to some uncertainty in a few features as its directionality is concerned.

\subsection{``Minmax'' features}
\label{sec:minmax}

A second relevant difference between the simple model and RM features \cite{Fridrich:2012:RM} is the use of ``minmax'' operations in several directions to select some features. For example, if we take first-order residuals in four directions, we can define the following directional residuals:
\[
\begin{split}
   r^\mathrm{h}_{i,j}&=x_{i,j}-x_{i-1,j},\\ 
   r^\mathrm{v}_{i,j}&=x_{i,j}-x_{i,j-1},\\
   r^\mathrm{d}_{i,j}&=x_{i,j}-x_{i-1,j-1},\\
   r^\mathrm{m}_{i,j}&=x_{i,j}-x_{i-1,j+1},
\end{split}
\]
where the superscripts ``h'', ``v'', ``d'' and ``m'' stand for horizontal, vertical, diagonal and minor diagonal directions, respectively. 

\begin{figure}[ht]
\begin{center}
\begin{tabular}{|c|c|}
\hhline{|-|-|}
$x_{i-1,j-1}$ & $x_{i,j-1}$ \\
\hhline{|-|-|}
$x_{i-1,j}$ & $x_{i,j}$ \\
\hhline{|-|-|}
$x_{i-1,j+1}$ & \multicolumn{1}{c}{} \\
\hhline{-|~|}
\end{tabular}
\end{center}
\caption{\textcolor{black}{Relevant pixels at the horizontal, vertical, diagonal and minor diagonal directions.}}
\label{fig:4dirs}
\end{figure}

The pixels involved in the definitions of $r^\mathrm{h}_{i,j}$, $r^\mathrm{v}_{i,j}$, $r^\mathrm{d}_{i,j}$ and $r^\mathrm{m}_{i,j}$ are shown in Fig. \ref{fig:4dirs}. Now, ``minmax'' features can be defined as follows:
\[
\begin{split}
r^{\min}_{i,j}&=\min\left(\left\{r^\mathrm{h}_{i,j},r^\mathrm{v}_{i,j},r^\mathrm{d}_{i,j},r^\mathrm{m}_{i,j}\right\}\right),\\
r^{\max}_{i,j}&=\max\left(\left\{r^\mathrm{h}_{i,j},r^\mathrm{v}_{i,j},r^\mathrm{d}_{i,j},r^\mathrm{m}_{i,j}\right\}\right).
\end{split}
\]
Using these ``min'' and ``max'' residuals, we can build the corresponding ``min'' and ``max'' histograms $H^{\min}=\left(h^{\min}_k\right)$ and $H^{\max}=\left(h^{\max}_k\right)$, similarly as done in Section \ref{sec:histogram}. 
\begin{remark}
Because of these ``min'' and ``max'' operations, the features $h^{\min}_k$ and $h^{\max}_k$ are non-linear, which may significantly affect the theoretical analysis presented for the simple model in Sections \ref{sec:simple} and \ref{sec:theoretical}. It is not obvious whether the results of Lemmas \ref{lemma:expected} and \ref{lemma:adapt} hold for ``minmax'' features. 
\end{remark}

Next, we show that Lemma \ref{lemma:expected} extends for ``min'' features in the non-adaptive steganography case. The case of ``max'' features is completely analogous to the ``min'' counterpart and its proof is not presented for brevity.

In order to obtain a counterpart of Lemma \ref{lemma:expected} for the features $h^{\min}_k$,
taking a look at Fig. \ref{fig:4dirs}, it may appear that we would need to analyze $3^5=243$ cases, since for each of the five pixels represented in the figure, after data embedding, there are three different possibilities, namely a $-1$, a 0 or a $+1$ variation. However, there is a shortcut to make this analysis simpler. In order to compute the change in the values of the histogram bins $h^{\min}_k$, only two possibilities must be taken into account:
\begin{enumerate}
\item The first possibility is that the direction (``h'', ``v'', ``d'' or ``m'') of the pixel that yields the minimum residual $r^{\min}_{i,j}$ does not change after data embedding. For example, if the minimum residual is obtained for the horizontal direction before embedding, or $r^{\min}_{i,j}=r^{\mathrm{h}}_{i,j}$, then the direction for the new minimum is preserved after data embedding, or  
$\left(r^{\min}_{i,j}\right)'=\left(r^{\mathrm{h}}_{i,j}\right)'$, where ``$\left(\cdot\right)'$'' represents the new value obtained after data embedding.  This occurs with some average probability for all the samples of the residual image. This probability is denoted as $(1-\xi)$, with $0\leq(1-\xi)\leq 1$.
\begin{figure}[ht]
\centering
\resizebox{\columnwidth}{!}{
\subfloat[Original]{
\label{fig:original}
\begin{tabular}{|c|c|}
\hhline{|-|-|}
58 & 59 \\
\hhline{|-|-|}
60 & 63 \\
\hhline{|-|-|}
59 & \multicolumn{1}{c}{} \\
\hhline{-|~|}
\end{tabular}} \qquad
\subfloat[Sub-case 1]{
\label{fig:case1}
\begin{tabular}{|c|c|}
\hhline{|-|-|}
58 & \textcolor{black}{60$^\star$} \\
\hhline{|-|-|}
\textcolor{black}{59$^\star$} & 63 \\
\hhline{|-|-|}
59 & \multicolumn{1}{c}{} \\
\hhline{-|~|}
\end{tabular}}  \qquad 
\subfloat[Sub-case 2]{
\label{fig:case2}
\begin{tabular}{|c|c|}
\hhline{|-|-|}
58 & \textcolor{black}{60$^{\star}$} \\
\hhline{|-|-|}
\textcolor{black}{59$^{\star}$} & \textcolor{black}{62$^{\star}$} \\
\hhline{|-|-|}
59 & \multicolumn{1}{c}{} \\
\hhline{-|~|}
\end{tabular}} \qquad
\subfloat[Sub-case 3]{
\label{fig:case3}
\begin{tabular}{|c|c|}
\hhline{|-|-|}
58 & \textcolor{black}{60$^{\star}$} \\
\hhline{|-|-|}
\textcolor{black}{59$^{\star}$} & \textcolor{black}{64$^{\star}$} \\
\hhline{|-|-|}
59 & \multicolumn{1}{c}{} \\
\hhline{-|~|}
\end{tabular}}}
\caption{\textcolor{black}{Example of original data and three cases of variations with a change in the direction for the minimum residual.}}
\label{fig:3cases}
\end{figure}
\item Conversely, there is some probability ($0\leq\xi\leq 1$) that the direction of the pixel that yields the minimum residual changes after data embedding. For this to occur, a combination of two conditions must take place simultaneously:
\begin{enumerate}
\item First, if the pixel that yields the minimum residual for $x_{i,j}$ takes the value $x^{\min}$, there must be a pixel with value $x^{\min}-1$ in one of the other three relevant directions. This is illustrated in Fig. \ref{fig:original}, where the pixel in the horizontal direction (60) is the one that provides the minimum residual $r^{\min}_{i,j}=r^{\mathrm{h}}_{i,j}=63-60=3$, and the pixel in the vertical direction takes the value $59=60-1$. 
\item Second, the pixel that yields the minimum residual must have a $-1$ variation after embedding, whereas the pixel with value $x^{\min}-1$ must have a $+1$ variation. The reference pixel $x_{i,j}$ for the residual $r^{\min}_{i,j}$ can remain unchanged ($0$ variation), or have a $-1$ or a $-1$ variation. This yields another direction for the minimum residual after data embedding, with three different sub-cases illustrated in Figs. 
\ref{fig:case1}-\ref{fig:case3}, where a star sign ``$(\cdot)^{\star}$'' means that the corresponding pixel value has been modified after embedding data. In all three cases, the new direction for the minimum residual is the vertical one. The only difference between the three cases is the value of the minimum residual: $63-60=3$ in Fig. \ref{fig:case1}, $62-60=2$ in Fig. \ref{fig:case2}, and $64-60=4$ in Fig. \ref{fig:case3}.
\end{enumerate}
Since the maximum variation in the difference of two pixels with $\pm1$ changes after data embedding is $\pm2$ (which occurs if one pixel changes $-1$ and the other one changes $-1$), there is no other possibility that the direction of the pixel that yields the minimum (or maximum) residual changes after data embedding.
\end{enumerate}

\begin{remark}
\label{rem:several}
If several (more than one) direction pixels with the same value provide the (same) minimum residual for the reference pixel, all those direction pixels must have a $-1$ variation after embedding for a change in the direction of the minimum residual.
\end{remark} 

\begin{figure}[ht]
\centering
\subfloat[Original]{
\label{fig:original2}
\begin{tabular}{|c|c|}
\hhline{|-|-|}
58 & 60 \\
\hhline{|-|-|}
60 & 63 \\
\hhline{|-|-|}
59 & \multicolumn{1}{c}{} \\
\hhline{-|~|}
\end{tabular}} \qquad
\subfloat[Embedded]{
\label{fig:embedded}
\begin{tabular}{|c|c|}
\hhline{|-|-|}
58 & \textcolor{black}{59$^\star$} \\
\hhline{|-|-|}
\textcolor{black}{59$^\star$} & \textcolor{black}{62$^{\star}$} $|$ 63 $|$ \textcolor{black}{64$^{\star}$}  \\
\hhline{|-|-|}
\textcolor{black}{60$^{\star}$} & \multicolumn{1}{c}{} \\
\hhline{-|~|}
\end{tabular}}
\caption{\textcolor{black}{Example with several directions yielding the minimum residual.}}
\label{fig:several}
\end{figure}

The situation mentioned in Remark \ref{rem:several} is illustrated with an example in Fig. \ref{fig:several}, in which, for the original situation (Fig. \ref{fig:original2}), both the vertical and the horizontal pixels take the (same) value (60) that provides the minimum residual for the reference pixel (63). After data embedding (Fig. \ref{fig:embedded}), both the vertical and the horizontal pixels must change to 59 ($-1$ variation), and there should be another pixel with the value $60-1=59$ in the relevant directions that is changed with a $+1$ variation. In this case, the pixel in the minor diagonal changes from 59 to 60. Thus, when the minimum occurs at several directions at the same time, there must be at least three pixel changes (``h'', ``v'' and ``m'' in the example) for the direction of the minimum residual to be modified to a new one. The reference pixel can either remain the same (63), or have a $-1$ (62) or a $+1$ (64) variation.

Similarly, if instead of two direction pixels leading to the minimum residual there were three of them, at least four pixel changes would be required for obtaining a different direction for the minimum residual after data embedding.

After these observations, we can extend Lemma \ref{lemma:expected} to ``min'' and ``max'' features as follows:

\begin{lemma}
\label{lemma:minmax}
If we assume independent (Assumption \ref{assm:independent}) random $\pm1$ changes of the pixel values with probability $\beta=\alpha/2$   for $\alpha\in(0,1]$, under Assumption \ref{assm:bounded}, then the expected value of the histogram bins $\left(h^{\min}_k\right)'$ and $\left(h^{\max}_k\right)'$can be approximated as follows:
\[
\begin{split}
    E\left[\left(h^{\min}_k\right)'\right]&\approx (1-\alpha)h^{\min}_k+\frac{\alpha}{2}\left(h^{\min}_{k-1}+h^{\min}_{k+1}\right),\\
    E\left[\left(h^{\max}_k\right)'\right]&\approx (1-\alpha)h^{\max}_k+\frac{\alpha}{2}\left(h^{\max}_{k-1}+h^{\max}_{k+1}\right).
\end{split}
\]
\end{lemma}

\begin{remark}
This is exactly the same expression obtained in Lemma  \ref{lemma:expected} (and Lemma \ref{lemma:adapt}) for the  simple feature model, but replacing $h_{(\cdot)}$ by $h^{\min}_{(\cdot)}$ or $h^{\max}_{(\cdot)}$.
\end{remark}

\begin{proof}
The proof is given for ``min'' features only, since the case of ``max'' features is completely symmetric. To prove this lemma, we first obtain an upper bound of the probability $\xi$ that the direction for the the residual $r^{\min}_{i,j}$ changes after data embedding.  
In the simplest case (when there is only one direction pixel providing the minimum residual), three conditions must be met: 1) another pixel with value $x^{\min}-1$ is in the relevant directions (which occurs with some probability $\nu\in[0,1]$), 2) the pixel with value $x^{\min}$ is changed to $x^{\min}-1$ (which occurs with probability $\beta/2$), and 3) the pixel with value $x^{\min}-1$ is changed to $x^{\min}$ (which occurs with probability $\beta/2$). Now we can estimate the joint probability $\xi$ as follows:
\begin{equation}
\label{eqn:bound}
\xi=\nu\cdot\frac{\beta}{2}\cdot\frac{\beta}{2}=\frac{\nu\beta^2}{4}\leq{\frac{\beta^2}{4}},    
\end{equation}
where the inequality comes from the fact that $\nu\leq1$ (since $\nu$ is a probability). When there is more than one minimum (as in Fig. \ref{fig:several}), the probability would be even lower, since at least three pixel changes are required yielding an upper bound $\beta^3/8$ that is lower than $\beta^2/4$. Hence, the upper bound $\beta^2/4$ for $\xi$ is always valid.

Now, we can proceed similarly as done in the proof of Lemma \ref{lemma:expected} considering two cases:
\begin{enumerate}
\item When the direction of the pixel that yields the minimum residual does not change after data embedding (which occurs with probability $(1-\xi)$ as discussed above), we are in the same case that Lemma \ref{lemma:expected}, since we need to combine the variations of only two pixels (the reference pixel $x_{i,j}$ and the direction pixel that yields the minimum residual). Hence, we only need to include a factor $(1-\xi)$ in Expression (\ref{eqn:expect2}) and replace $h_{(\cdot)}$ by $h^{\min}_{(\cdot)}$:
$$(1-\xi)\left((1-\alpha)h^{\min}_k+\frac{\alpha}{2}\left(h^{\min}_{k-1}+h^{\min}_{k+1}\right)+\widetilde\varepsilon\right).$$
\item When the direction of the pixel that yields the minimum residual changes after data embedding (which occurs with some probability $\xi$ as discussed above), we can have three different cases:
\begin{enumerate}
\item If $\Delta x_{i,j}=0$, which occurs with probability $1-\beta$, then the residual $\left(r^{min}_{i,j}\right)'$ stays at the bin $\left(h^{\min}_k\right)'$. The joint probability of this case is $\xi(1-\beta)$.
\item If $\Delta x_{i,j}=1$, which occurs with probability $\beta/2$, then the residual $\left(r^{min}_{i,j}\right)'$ moves to bin $\left(h^{\min}_{k+1}\right)'$. The joint probability of this case is $\xi\beta/2$.
\item Similarly, if $\Delta x_{i,j}=-1$, which also occurs with probability $\beta/2$, then the residual $\left(r^{min}_{i,j}\right)'$ moves to bin $\left(h^{\min}_{k-1}\right)'$. The joint probability of this case is also $\xi\beta/2$.
\end{enumerate}
\end{enumerate}

Now, we can combine all these cases to write the expected value of the bin $\left(h^{\min}_k\right)$ as follows:
\begin{multline}
\label{eqn:expectedmin}
E\left[\left(h^{\min}_k\right)'\right]=\\(1-\xi)\left((1-\alpha)h^{\min}_k+\frac{\alpha}{2}\left(h^{\min}_{k-1}+h^{\min}_{k+1}\right)+\widetilde\varepsilon\right)\\+
\xi(1-\beta)h^{\min}_k+\xi\frac{\beta}{2}h^{\min}_{k-1}+\xi\frac{\beta}{2}h^{\min}_{k+1},
\end{multline}
where $\widetilde\varepsilon$ is defined in the same way as $\varepsilon$ in  Expression (\ref{eqn:epsilon}), but replacing $h_{(\cdot)}$ by $h^{\min}_{(\cdot)}$. Expanding some operations and making use of the definition $\beta=\alpha/2$, Expression (\ref{eqn:expectedmin}) can be rewritten as follows:
\begin{multline}
\label{eqn:expectedmin2}
E\left[\left(h^{\min}_k\right)'\right]=(1-\alpha)h^{\min}_k+\frac{\alpha}{2}\left(h^{\min}_{k-1}+h^{\min}_{k+1}\right)+(1-\xi)\widetilde\varepsilon\\ +\xi\beta h^{\min}_k-\xi\frac{\beta}{2}\left(h^{\min}_{k-1}+h^{\min}_{k+1}\right).
\end{multline}
Here, we can prove that $\widetilde\varepsilon\approx 0$ following the same procedure used to show that $\varepsilon\approx 0$ in the proof of Lemma \ref{lemma:expected}. Consequently, $(1-\xi)\widetilde\varepsilon\approx 0$, since $1-\xi\leq 1$. To complete the proof of the lemma, we only need to show that the final terms of Expression (\ref{eqn:expectedmin2}) can also be neglected. 

This latter step is also similar to the proof that $\varepsilon\approx 0$ in Lemma \ref{lemma:expected}, as detailed below. If we make an assumption analogous to Assumption \ref{assm:bounded} for $h^{\min}_k$, we have $\abs{h^{\min}_{k+1}-h^{\min}_k}\leq \widetilde\delta,\forall k\in[v_{\min},v_{\max}]$, for some relatively small $\widetilde\delta$, and then:
\begin{multline}
\label{eqn:vanish}
\abs{\xi\beta h^{\min}_k-\xi\frac{\beta}{2}\left(h^{\min}_{k-1}+h^{\min}_{k+1}\right)}=\\
\abs{\xi\frac{\beta}{2}\left(2h^{\min}_k-h^{\min}_{k-1}-h^{\min}_{k+1}\right)}=\\
\xi\frac{\beta}{2}\abs{h^{\min}_k-h^{\min}_{k-1}+h^{\min}_k-h^{\min}_{k+1}}\leq\\
\xi\frac{\beta}{2}\left(\abs{h^{\min}_k-h^{\min}_{k-1}}+\abs{h^{\min}_{k+1}-h^{\min}_k}\right)\leq\\
\xi\frac{\beta}{2}(2\widetilde\delta)=\xi\beta\widetilde\delta \leq \frac{\beta^2}{4}\beta\widetilde\delta = \frac{\widetilde\delta\beta^3}{4},
\end{multline}
where the last inequality comes from the bound for $\xi$ given in Expression (\ref{eqn:bound}). Finally, since $\beta^3$ is very small and $\widetilde\delta$ is a relatively small (bounded) quantity, we have $\widetilde\delta\beta^3/4\approx 0$ and thus:

$$\abs{\xi\beta h^{\min}_k-\xi\frac{\beta}{2}\left(h^{\min}_{k-1}+h^{\min}_{k+1}\right)}\approx 0,$$
which completes the proof.
\end{proof}

For the adaptive case, we can state an equivalent lemma as follows:
\begin{lemma}
\label{lemma:minmaxadapt}
If we assume dependent random $\pm1$ changes of the pixel values, the expected value of the histogram bins $\left(h^{\min}_k\right)'$ and $\left(h^{\max}_k\right)'$ can be approximated as follows:
\[
\begin{split}
    E\left[\left(h^{\min}_k\right)'\right]&\approx (1-\alpha)h^{\min}_k+\frac{\alpha}{2}\left(h^{\min}_{k-1}+h^{\min}_{k+1}\right),\\
    E\left[\left(h^{\max}_k\right)'\right]&\approx (1-\alpha)h^{\max}_k+\frac{\alpha}{2}\left(h^{\max}_{k-1}+h^{\max}_{k+1}\right).
\end{split}
\]
\end{lemma}
\begin{proof}
In this case, the computations of the probabilities of joint variations of several pixels are somewhat more cumbersome compared to the non-adaptive scenario, but not too complex. We can follow the same steps carried out in the proof of Lemma \ref{lemma:adapt} to define an auxiliary probability $\widetilde\beta$ (and its complementary counterpart $1-\widetilde\beta$) to consider the variation of the reference pixel when two neighboring direction pixels also change. More precisely, $\widetilde\beta$ is the probability that the variation of the reference pixel is different from zero after data embedding ($\Delta x_{i,j}\neq0$) when the pixels at the two relevant directions required for a change in the  direction that yields the minimum residual also change. With this definition, for the adaptive case, we obtain the following expression:
\begin{multline*}
\label{eqn:expectedmin2}
E\left[\left(h^{\min}_k\right)'\right]=(1-\alpha)h^{\min}_k+\frac{\alpha}{2}\left(h^{\min}_{k-1}+h^{\min}_{k+1}\right)+(1-\xi')\widetilde{\varepsilon'}\\ +\xi'\widetilde\beta h^{\min}_k-\xi'\frac{\widetilde\beta}{2}\left(h^{\min}_{k-1}+h^{\min}_{k+1}\right),
\end{multline*}
where $\xi'$ is the probability that the direction pixel that yields the minimum residual changes after data embedding \textbf{in the adaptive case}, and $\widetilde{\varepsilon'}$ is a small number that can be neglected ($\widetilde{\varepsilon'}\approx 0$) compared to the dominant terms of the expression.

\begin{remark}
Note that $\widetilde{\varepsilon'}$ is defined as $\varepsilon'$ in Expression (\ref{eqn:betap}), but replacing $h_{(\cdot)}$ by $h^{\min}_{(\cdot)}$ (or $h^{\max}_{(\cdot)}$). The term  $\widetilde{\varepsilon'}$ depends on $\beta'$, which is, in this case, the conditional probability that the reference pixel $x_{i,j}$ be modified when the neighboring pixel $\hat x_{i,j}$ that yields the minimum residual $r^{\min}_{i,j}$ is also changed.
\end{remark}

Following the same ideas used to prove Lemma \ref{lemma:minmax}, we can first show that:
\[
\xi'=\nu\frac{\beta}{2}\frac{\beta'}{2}\leq\frac{\beta\beta'}{4},
\]
and, finally, conclude that:
$$\abs{\xi'\widetilde\beta h^{\min}_k-\xi'\frac{\widetilde\beta}{2}\left(h^{\min}_{k-1}+h^{\min}_{k+1}\right)}\approx 0,$$
using steps similar to those of Expression (\ref{eqn:vanish}), which completes the proof.
\end{proof}

The main difference between the proofs of the adaptive and the non-adaptive cases is that conditional probabilities $\beta'$ and $\widetilde\beta$ are introduced in the adaptive case. In fact, the non-adaptive case is a particular instance of the adaptive counterpart with $\beta=\beta'=\widetilde\beta$ (i.e., independent pixel changes).

Now, the following final remark can be made with respect to ``minmax'' features and the theoretical analysis presented in Section \ref{sec:theoretical}:
\begin{remark}
The theoretical analysis presented in Section \ref{sec:theoretical} only requires the results obtained in Lemmas \ref{lemma:expected} and \ref{lemma:adapt}, which are generalized in this section with identical expressions for ``minmax'' features through Lemmas  \ref{lemma:minmax} and \ref{lemma:minmaxadapt}. Hence, all the theorems, expressions and results presented in Section \ref{sec:theoretical} are valid for ``minmax'' features simply replacing $h_{(\cdot)}$ by $h^{\min}_{(\cdot)}$ or $h^{\max}_{(\cdot)}$.
\end{remark}

%as shown in the proof of Lemma \ref{lemma:adapt}, but the result is basically the same. Hence, the extension to adaptive steganography would be carried out in a similar way as done in Lemma \ref{lemma:adapt}. 

%However, the analysis performed in Sections \ref{sec:simple} and \ref{sec:theoretical} can still hold if the ``min'' and ``max'' functions select \textbf{the same direction for an image and for the same image after embedding a new random message into it}. More formally, the analysis carried out in Section \ref{sec:theoretical} is still valid if $r^{\min}_{i,j}=r^\mathrm{D}_{i,j}$ and $r'^{\min}_{i,j}={r'}^\mathrm{D}_{i,j}$, where $D$ denotes the same direction (``h'', ``v'', ``d'' or ``m''), $r^{(\cdot)}_{i,j}$ stands the residuals of an image $Z$ and $r'^{(\cdot)}_{i,j}$ are the residuals of the embedded image $Z'=\stego^{\#}_p{(Z)}$. If, after embedding data, the selected ``min'' or ``max'' direction for a residual is preserved, the analysis carried out in Section \ref{sec:theoretical} still applies, because the ``min'' or ``max'' feature is identical to one of the four directional features for both $Z$ and $Z'$.

\subsection{Truncation}
\label{sec:truncation}
Apart from dimensionality (considered in Section \ref{sec:multiple} and ``minmax''  operations (discussed in Section \ref{sec:minmax}), there is another difference between the simple model and RM features, namely the use of the threshold $T$ that truncates the differences between pixels to $T$ or $-T$ if the absolute value of the difference is larger than $T$. A truncation like this can be viewed as using saturated residuals $\overline{r_{i,j}}$ according to the following definition: 
$$\overline{r_{i,j}}=\mathrm{trunc}_T(r_{i,j})=\min\left(\max(r_{i,j},-T),T\right),$$ 
which also introduces  uncertainty in the theoretical analysis presented in Sections \ref{sec:simple} and \ref{sec:theoretical}. This kind of operation can be easily understood in the one-dimensional case, since it is equivalent to grouping all histogram bins lower than or equal to $-T$ in a single bin, and all bins greater than or equal to $T$ in another one. This will produce two larger bins at positions $\pm T$, and the theoretical analysis would not change too much if $T$ is outside the concavity interval of the histogram. This will typically be the case, since the concavity interval is usually very narrow and centered in the zero bin, as shown in Section \ref{sec:validation}. 

\begin{figure*}[ht]
  \centering
  \subfloat[``Min'' features.] {\label{fig:minmax1} \includegraphics[width=.8\columnwidth]{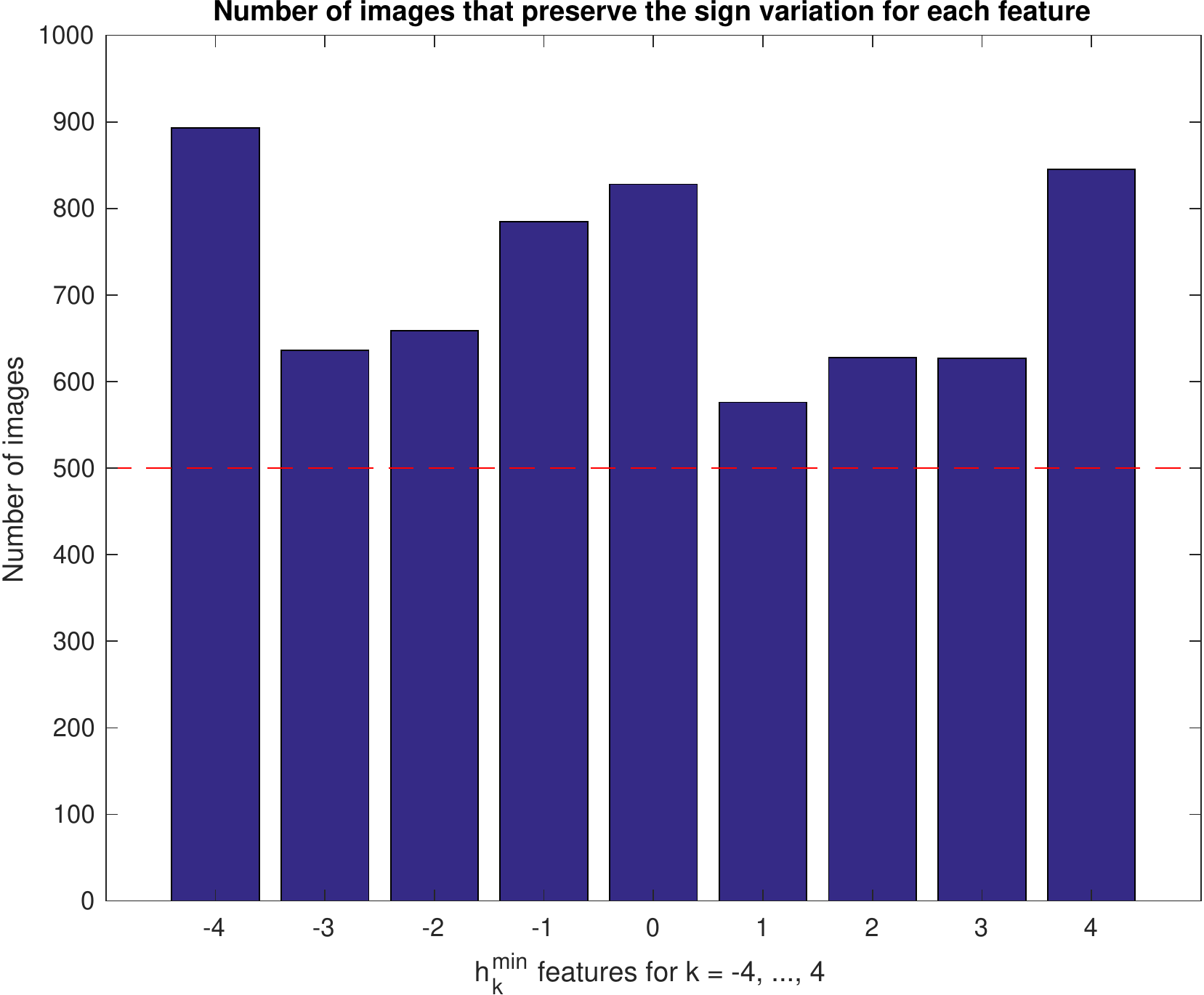}} \qquad
  \subfloat[``Max'' features.]{\label{fig:minmax2} \includegraphics[width=.8\columnwidth]{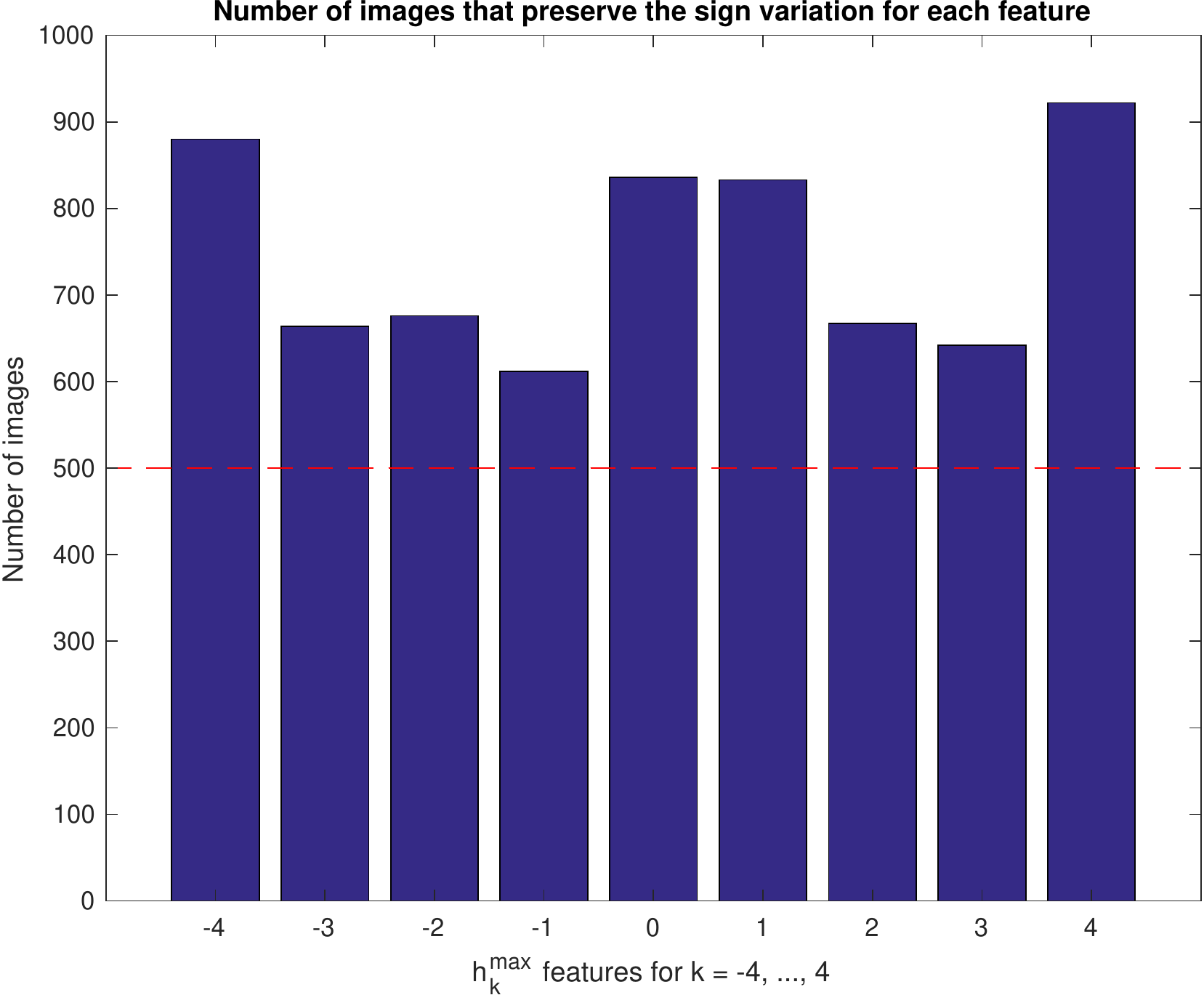}}
  \caption{Number of images that preserve the sign variation for each ``minmax'' feature in two subsequent embeddings.}
  \label{fig:minmax} 
\end{figure*}

Figs. \ref{fig:cauchy}, \ref{fig:concave} and \ref{fig:allvariations} show that the number of bins in the concave area of the histogram distribution is only four. With $T=3$ (the lowest value suggested for SPAM features), there would be five inner bins and two outer bins. These five inner bins would be enough to cover the concave area of the histogram distribution if it only includes four bins. With $T=4$, there would be nine bins, four of which would be in the concavity interval. Hence, a few of the inner bins will satisfy the condition for negative variation after one and two embeddings (concavity), whereas the outer bins, including $\pm T$, will satisfy the condition for a positive variation. In fact, adding several bins in a single one at positions $-T$ and $T$ will provide more robustness for the positive variation of those bins after embedding data, since the summation of several bins will be statistically more stable than a bin related to a single value. If, for some particular image, the concavity region was larger than $[-T,T]$ (very unlikely), there would be uncertainty for the extreme bins at $\pm T$, but the inner bins at $[-T+1,T-1]$ would satisfy the (concavity) condition for a negative variation after one and two embeddings and, thus, would keep the soft directionality condition proved for the simple model without truncation. 

\subsection{Experimental validation}
\label{sec:experimental}
Now, we present the results of an experiment to validate the discussion presented in Sections \ref{sec:minmax} and \ref{sec:truncation}. We consider $h^{\min}_k$ and $h^{\max}_k$ features for truncation with $T=4$, i.e., we have nine features (bins) for ``min'' and ``max'' features: $H^{\min}=\{h^{\min}_{-4},h^{\min}_{-3},\dots,h^{\min}_{4}\}$ and $H^{\max}=\{h^{\max}_{-4},h^{\max}_{-3},\dots,h^{\max}_{4}\}$. We have taken 1,000 images from the BOSS database and have made two subsequent random embedding to those images, with HILL steganography and an embedding rate of $0.4$ bpp. The objective of this experiment is to check if the majority of these 18 features (nine in $H^{\min}$ and nine in $H^{\max}$) are directional. 

The results of the experiment are shown in Fig. \ref{fig:minmax}. It can be observed that all the (truncated) ``minmax'' features are directional for more than half of the tested images. The average percent of directionality of these 18 features is 73.4\%, which is consistent with the analysis presented in Sections \ref{sec:minmax}, \ref{sec:truncation} and \ref{sec:theoretical}. It can also be observed that the ``min'' and ``max'' cases are quite symmetric. The most directional features in the ``min'' case are the extreme ones ($h^{\min}_{-4}$ and $h^{\min}_4$), which correspond to the convexity interval of the distribution of the residuals for most images, and the inner ones ($h^{\min}_{-1}$ and $h^{\min}_0$), which are typically in the concavity interval. For ``max'' features, the most directional features are also the extreme ones ($h^{\max}_{-4}$ and $h^{\max}_4$) and the inner ones ($h^{\min}_0$ and $h^{\min}_1$). 

It is worth pointing out the positive effect of truncation on directionality. The four most directional features are $h^{\min}_{-4}$, $h^{\min}_4$, $h^{\max}_{-4}$ and $h^{\max}_4$, which are directional for 
893, 845, 880 and 922 images, respectively. These four features are, in fact, summation bins of features of the simple model produced by the truncation described in Section \ref{sec:truncation}.

\subsection{Other feature models}
\color{black}

Finally, if \textcolor{black}{directionality} occurs for SPAM and \textcolor{black}{``minmax''} features \textcolor{black}{and, in general for RM features,} there is no reason why other feature models, either explicit or implicit (i.e. those extracted by a CNN), \textcolor{black}{should not} exhibit the same behavior, justifying the use of subsequent embedding in steganalytic systems. The directionality of features in subsequent embedding is \textcolor{black}{already} exploited in \cite{Lerch-Hostalot:2019} for Rich Models and CNN, showing that it is indeed a usual characteristic of true feature models and steganalysis methods. \textcolor{black}{An empirical analysis of directionality of the (implicit) features of CNN steganalysis would be a convenient area for future research.}

\color{black}

\section{Towards universal steganalysis}
\label{sec:towards}
The applications presented in Section \ref{sec:practical} are focused on targeted steganalys. As shown in Section \ref{sec:prediction}, in order to build the secondary training ($\mathsf B^{\mathsf L}$) and testing ($\mathsf B^{\mathsf T}$) sets, the targeted steganographic scheme $\stego_p$ is required to embed a new (random) message in all training and testing images. There is no direct application of the suggested method when the targeted steganographic function is unknown (that is, moving towards universal steganylsis), but there are still some possibilities to apply the proposed method.

The first possibility is already presented in Section \ref{sec:ssm}. If two steganographic schemes are different, like HILL and UNIWARD, but are based on similar principles, it is still possible to apply the proposed practical application. Indeed, Table \ref{tab:SSM} shows that the classification accuracy, after removing inconsistencies, is 84.5\% or better (since the classification error  $\overline{\mathrm{Err}}\leq 0.155$ in all the experiments) when trying to detect UNIWARD at $0.4$ bpp using HILL at $0.4$ bpp as the hypothesized (targeted) steganographic scheme. This occurs because different steganographic schemes share the same objective (undetectability) and tend to affect the same areas of the images. In particular, in adaptive steganography, most pixel changes are brought together in textured areas and edges. Hence, the proposed methods have some intrinsic robustness against a mismatch between the targeted and the actual steganographic functions if they are ``similar''. 

Going beyond the above mentioned ``robustness'', the next four sections outline other possibilities to apply the proposed applications when the steganographic function is unknown or even if the stego images within a testing set have been created with a diversity of steganographic functions and embedding bit rates.

\subsection{Multiple embedding functions}
\label{sec:mulembed}
As detailed in Section \ref{sec:prediction}, to apply the method based on consistency filters, it is essential to have the steganographic function $\stego_p$ so that the secondary training and testing sets can be obtained using $B_k=\stego^{\#}_p(A_k)$ (for the secondary training set) and  $B'_k=\stego^{\#}_p(A'_k)$ (for the testing set). If the steganalyst does not know which steganographic function and parameters have been used by the adversary, he/she may use a collection of hypothesized steganographic functions and parameters, as follows:
\begin{multline*}
    \mathbb{S}=\left\{
    \stego^{[1]}_{p_{1,1}},\stego^{[1]}_{p_{1,2}},\dots,\stego^{[1]}_{p_{1,m_1}},\stego^{[2]}_{p_{2,1}},\stego^{[2]}_{p_{2,2}},\dots,\right.\\ \left.\stego^{[2]}_{p_{2,m_2}},\dots,\stego^{[n]}_{p_{n,1}},\stego^{[n]}_{p_{n,2}},\dots,\stego^{[n]}_{p_{n,m_n}}\right\}.
\end{multline*}
where the superscript ``$^{[k]}$'' denotes a different steganographic function and the subscript $p_{k,j}$ denotes a particular value for the parameters of $\stego^{[k]}$.

To construct the training set, the steganalyst first selects a set of cover images $\mathcal{X}=\{X_i\}\subset\mathcal{C}$ and then needs to build a set of stego images $\mathcal{X}'$ to be used for training. Now, there are two possibilities:
\begin{enumerate}
    \item For each cover image $X_i$, all possible stego images are computed: $$X'_{i,k,j}=\left(\stego^{[k]}_{p_{k,j}}\right)^{\#}(X_i),$$ 
    using all the steganographic functions and parameters of the collection $\mathbb{S}$. This produces a set of stego images $\mathcal{X}'=\left(\mathbb{S}^*\right)^{\#}(\mathcal{X})\subset\mathcal{S}$, where the double superscript ``$*\#$'' indicates that, for each cover image, all steganographic functions and parameters are used, but only a random message and key is chosen for each function. This obviously leads to a set of stego images larger than the set of cover images, which is not typically recommended for a good performance of the classifier.
    \item Alternatively, we can select only one steganographic function and parameters for each cover image, in such a way that each cover image leads to only one stego image. This selection must be carried out ensuring that approximately the same number of stego images are created for each of the functions in $\mathbb{S}$:
        $$X'_i=\left(\stego^{[k]}_{p_{k,j}}\right)^{\#}(X_i), \text{for some } k,j.$$ 
    This possibility makes the training step easier, compared to the former alternative, since the number of cover and stego images in the training set is balanced. This yields a set of stego images $\mathcal{X}'= \mathbb{S}^{\#}(\mathcal{X})$, where ``$^{\#}$'' denotes the selection of a single embedding function, a single value of the parameters, a random message, and a random key for each image in $\mathcal{X}$.
\end{enumerate}
In general, we recommend the second alternative, since the classifiers can be challenging to tune if the training set is unbalanced. In the rest of this section, the second alternative is used. Hence, we have a training set $\mathsf{{A}^L}=\{(\mathcal{X};\text{`0'}),(\mathcal{X}';\text{`1'})\}$, where the labels `0' and `1' are used for cover and stego images, respectively.

Now, to apply the filters to detect inconsistencies defined in Section \ref{sec:prediction}, we need the secondary training set $\mathsf{B^L}$. The images $B_i$ in the secondary training set can be created by choosing a random embedding function, parameter, stego key and message and carrying out the embedding step on the corresponding image $A_i$ of the primary training set, i.e.:
  $$B_i=\left(\stego^{[k]}_{p_{k,j}}\right)^{\#}(A_i), \text{for some } k,j,$$ 
which can be written in abbreviated form as $\mathsf{B^L}= \mathbb{S}^{\#}(\mathsf{A^L})$. The labels for the secondary training images $B_i$ are the same as those of the corresponding primary training images, $A_i$, where a `0' and `1' denote stego and ``double stego'' images, respectively. Regarding the primary and secondary testing sets, we can proceed analogously and obtain $\mathsf{B^T}= \mathbb{S}^{\#}(\mathsf{A^T})$. 

Once we have a counterpart of the secondary training and testing sets, we can apply the practical methods described in Section \ref{sec:practical}, as shown in the next section.

\subsection{Experimental results with multiple embedding functions}
\label{sec:expmult}
The method described in the previous section is tested in this section using two different embedding functions, namely HILL and S-UNIWARD, with five possible embedding bit rates: $\{0.1, 0.2, 0.3, 0.4, 0.5\}$ bpp. Hence, we actually have ten different possibilities to obtain a stego image (two possible embedding algorithms multiplied by five possible embedding bit rates). For ``double stego'' images, $10\times10=100$ different combinations are possible, which is a relatively large number. If the primary and secondary training sets must have enough samples of each case, a large training set is required. Note that, in the secondary training set, half of the samples will correspond to stego images and the other half to ``double stego'' images. Stego images must represent ten different possibilities, whereas ``double stego'' images must represent 100 different combinations. To ensure an appropriate number of images of each type (combination), the primary training set must contain thousands of images. 

For this experiment, we have used the images from the Alaska2 data set \cite{Yousfi:2020:alaska2}, consisting of 75,000 JPEG images that we have converted to the spatial domain. 73,000 images have been used for training, 1,000 images for validation and, finally, 1,000 images for testing. With such an extensive training data set, it is not convenient to use RM features with Ensemble Classifiers, since they cannot handle such an extensive data set consistently. Instead, we have used the EfficientNet B0 CNN classifier \cite{Tan:2019}. The validation set has been used to choose the best model with an early stopping of 10 iterations. Each iteration uses 1,000 batches and each batch consists of 16 images. The Adamax optimizer has been used with a learning rate of $0.001$. With these settings, the best models for the $\pred^{\mathsf A}_{{\stego}_p,\mathcal F}$ and $\pred^{\mathsf B}_{{\stego}_p,\mathcal F}$ have been obtained after 23 and 8 iterations, respectively.

\begin{table}
\centering
\textcolor{black}{\caption{Experimental results with multiple embedding functions and bit rates, with the standard (primary) classifier ($\pred^{\mathsf A}_{{\stego}_p,\mathcal F}$) and the detection of classifier inconsistencies (DCI).}
\label{tab:universal}}
\resizebox{\columnwidth}{!}{
\begin{tabular}{||l||l|r|r|r|r|r|r||}
\hhline{|t:=:t:=======:t|}
\textbf{Classifier} & \textbf{Err}& \textbf{TP} & \textbf{TN} & \textbf{FP} &\textbf{FN} & {$\bm{N_{\mathsf{NC}}}$}  &
$\bm{\widehat{\mathrm{Err}}_{0.5}}$ \\
\hhline{|:=::=======:|} 
$\pred^{\mathsf A}_{{\stego}_p,\mathcal F}$ & $0.1$ & 438 & 462 & 38 & 62 & \multicolumn{1}{c|}{---} & \multicolumn{1}{c||}{---} \\
\hhline{||-||-------||}
DCI & $0.048$ & 288 & 384 & 1 & 33 & 294 & $0.147$ \\
\hhline{|b:=:b:=======:b|}         
\end{tabular}}
\end{table}

The classification accuracy results obtained with $\pred^{\mathsf A}_{{\stego}_p,\mathcal F}$ are 93.53\% for the training set, 91.80\% for the validation set and 90.00\% for the testing set. Similarly, with the secondary classifier $\pred^{\mathsf B}_{{\stego}_p,\mathcal F}$, the accuracy is 87.41\% for training, 84.10\% for validation and 80.70\% for testing. The results for the standard (primary) classifier and for the prediction method based on detecting classifier inconsistencies (DCI), proposed in Section \ref{sec:prediction}, are summarized in Table \ref{tab:universal}. It can be observed that 1) the primary classifier works with a high accuracy (the classification error is only $0.1$); 2) the classification error is further reduced to $0.048$ when the DCI method is applied (at the cost of labeling 294 out of 1,000 images as inconsistent); and 3) the predicted error of the primary classifier ($\widehat{\mathrm{Err}}_{0.5}$) is 0.147, which is quite a good approximation of the true classification error ($0.1$). 

This experiment illustrates that there are ways to use the proposed approach even if the steganalyst does not exactly known the embedding function or bit rate. This brings the proposed method closer to real-world universal steganalysis. However, we still need a collection of hypothesized embedding functions to apply the method. Another possibility, whose details are left for future research, is outlined in the following section.

\begin{figure*}[ht]
  \centering
  \subfloat[Original testing set ($\mathsf{A^T}$).] {\label{fig:ATS1} \includegraphics[width=.3\textwidth]{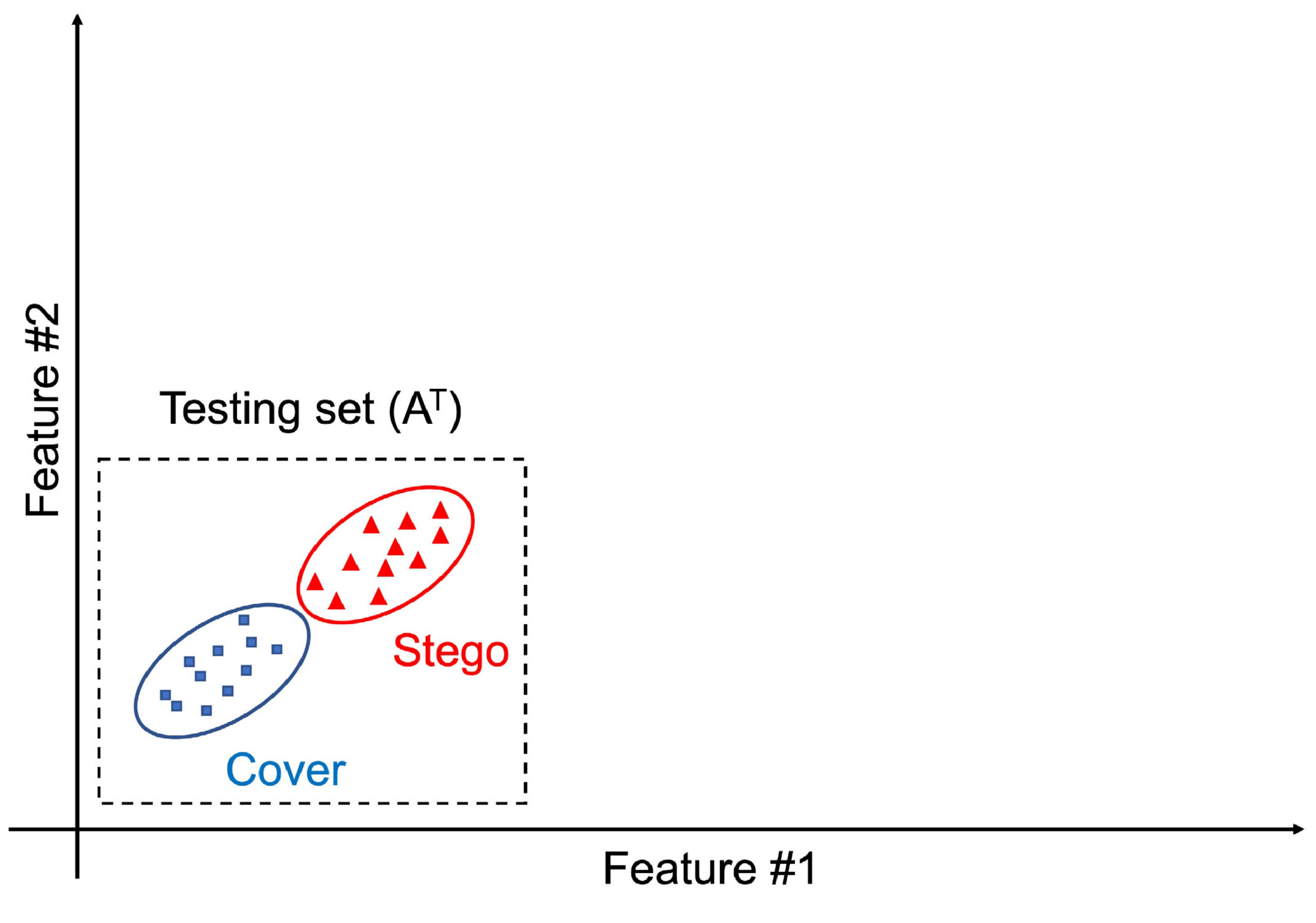}} \qquad
  \subfloat[Transformed testing set ($\mathsf{B^T}$).]{\label{fig:ATS2} \includegraphics[width=.3\textwidth]{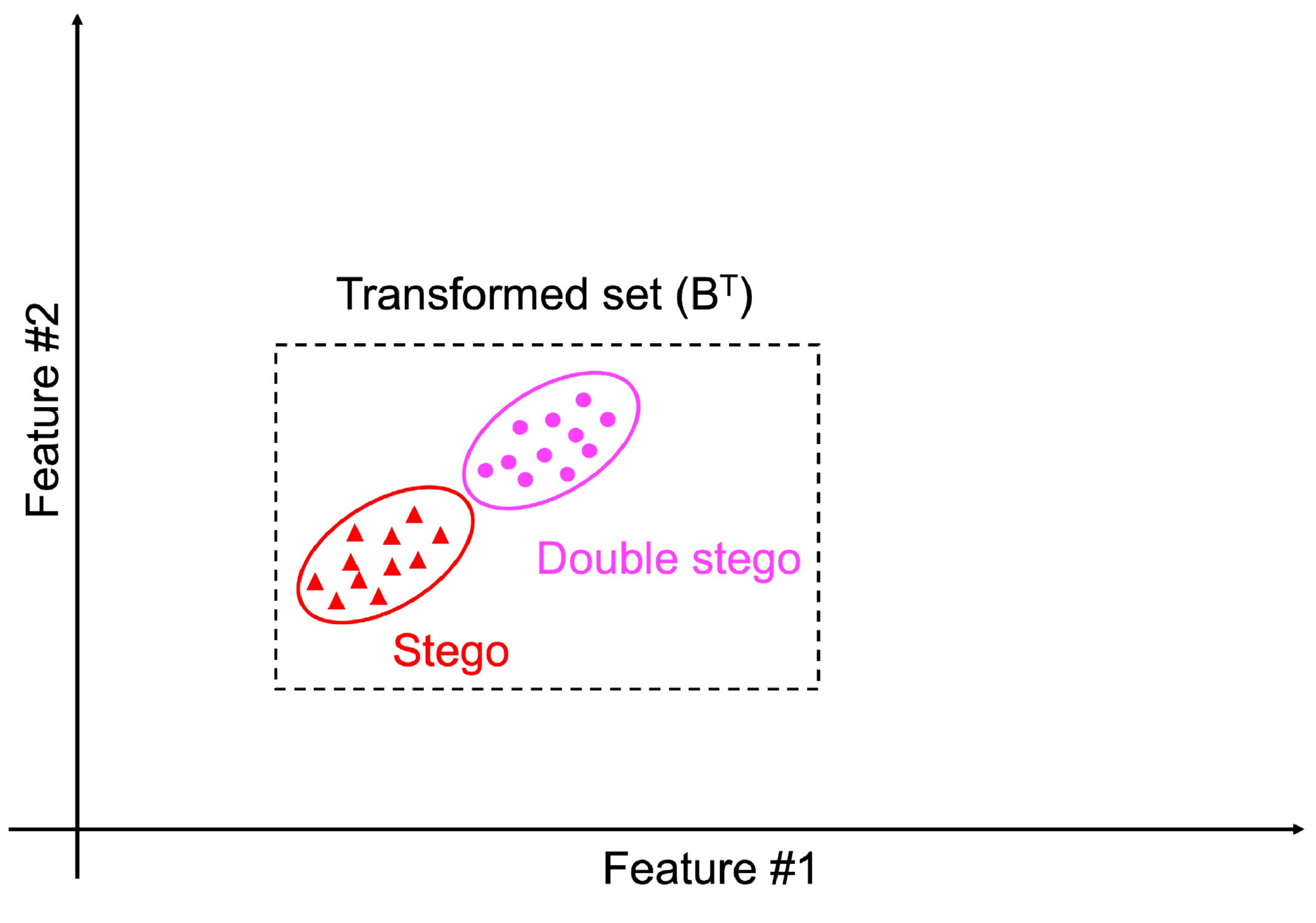}} \qquad
  \subfloat[``Double transformed'' testing set ($\mathsf{C^T}$).]{\label{fig:ATS3} \includegraphics[width=.3\textwidth]{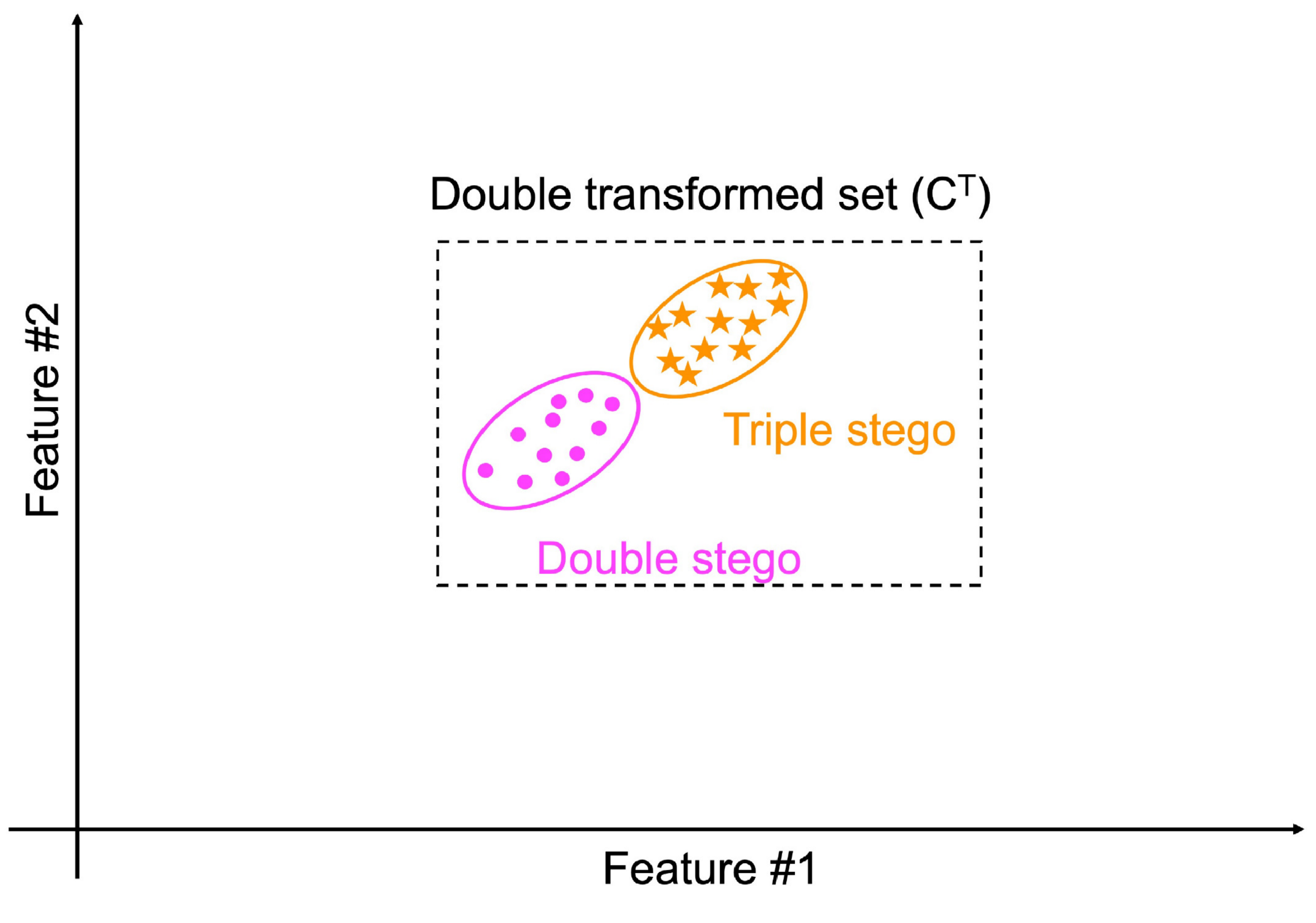}} \qquad \\
  \subfloat[Classifier trained with the sets $\mathsf{A^T}-\mathsf{C^T}$.]{\label{fig:ATS4} \includegraphics[width=.45\textwidth]{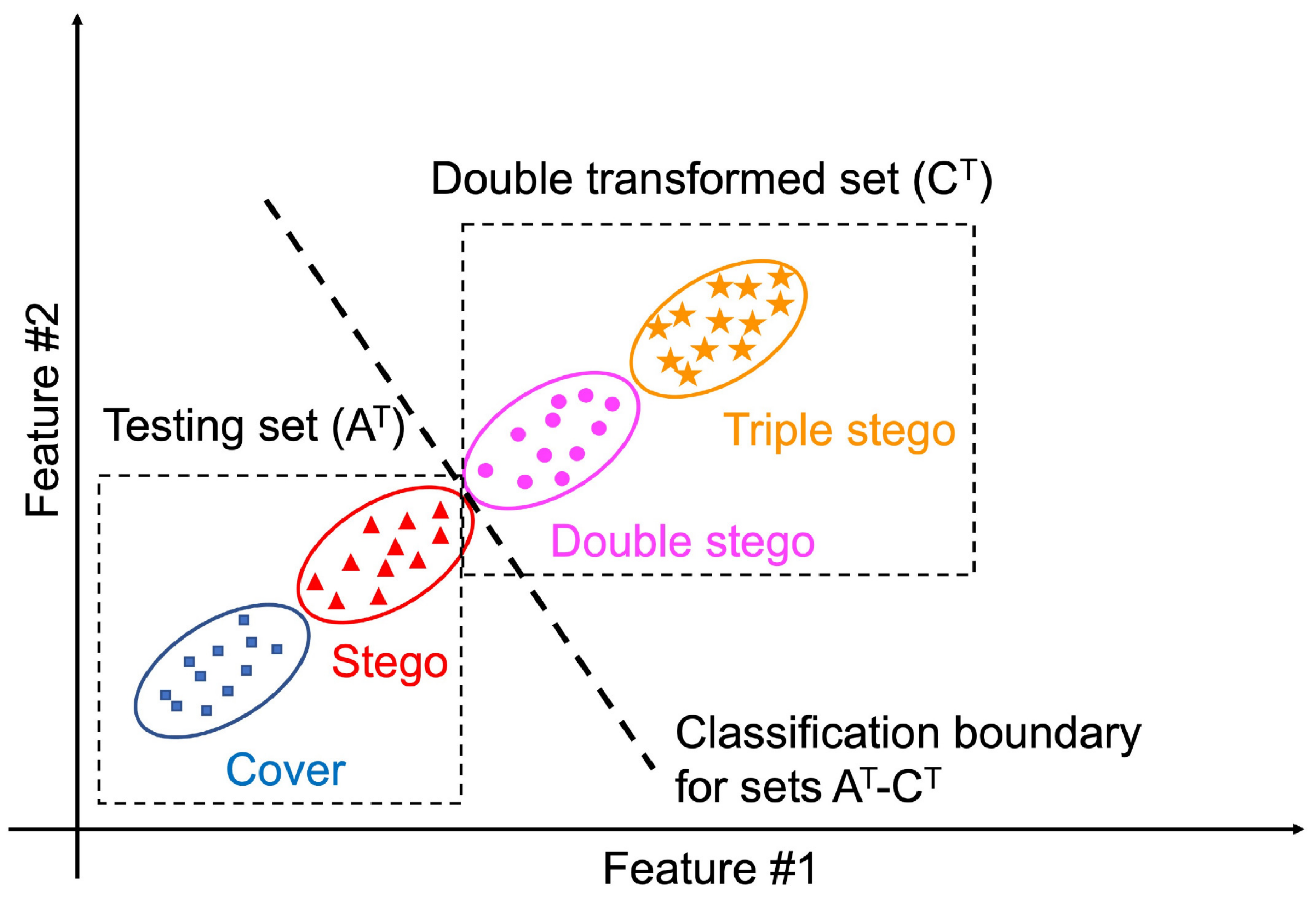}} \qquad
  \subfloat[Classification of $\mathsf{B^T}$ using the trained classifier.]{\label{fig:ATS5} \includegraphics[width=.45\textwidth]{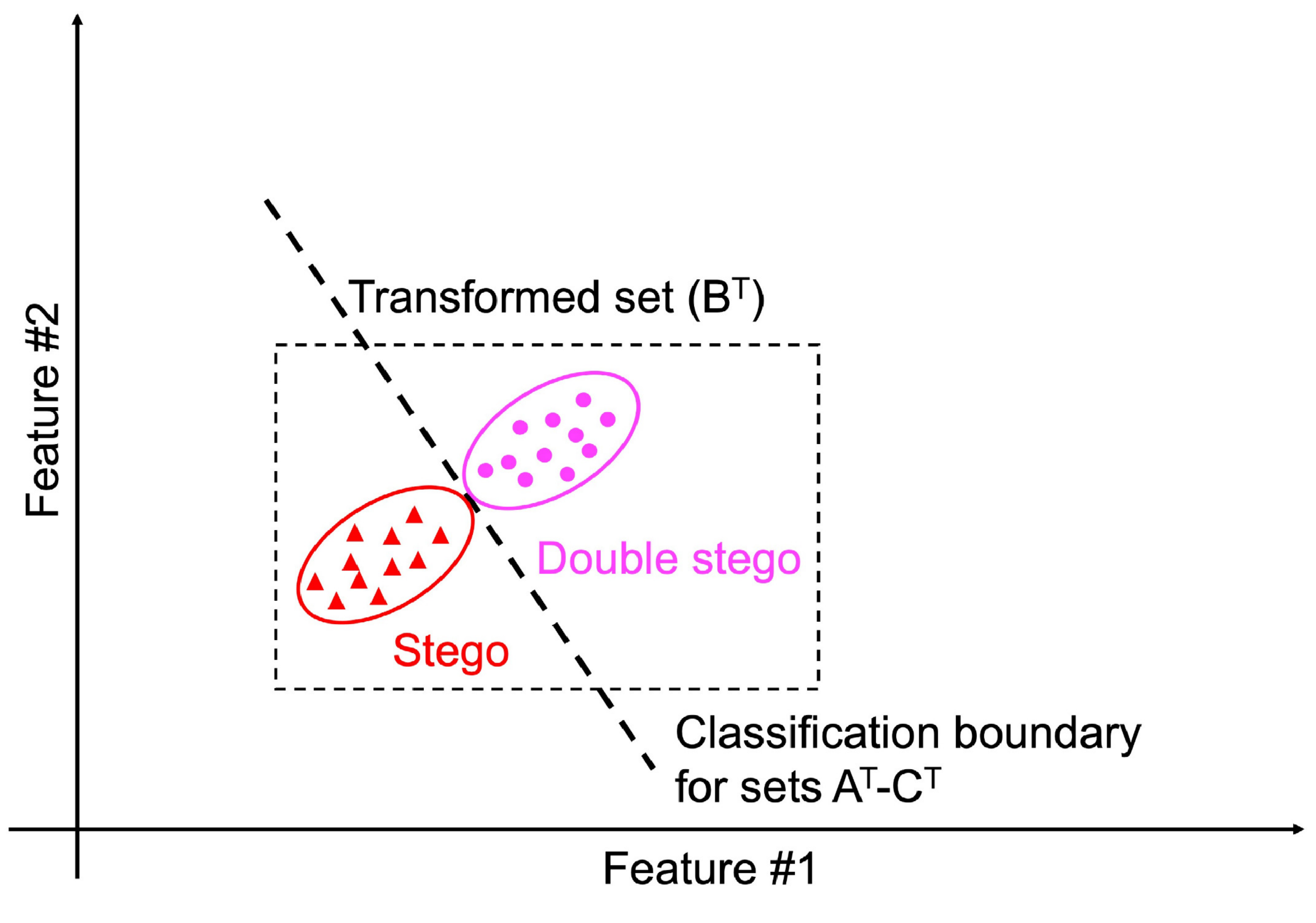}} \qquad
  \caption{{Unsupervised steganalysis based on directional features: Artificial Training Sets (ATS).}}
  \label{fig:ATS}
\end{figure*}

\subsection{Generalization of the embedding function}
\label{sec:generalization}
The previous section shows how the proposed DCI method can be used even if it is assumed that the steganographer can select among different embedding functions and parameters. The approach presented above is based on building complete (primary and secondary) training sets to include all those possibilities. This is a practical alternative that can work reasonably well at the cost of a much larger training time (and using classifiers that can handle thousands of images). However, there is another possibility that may be used to apply the DCI method without needing to cover the whole spectrum of embedding functions and parameters. The idea we outline here is, instead of trying to build an exhaustive set $\mathbb{S}$ will all the possible embedding functions and parameters, which leads to some explosion in the number of cases and requires an extensive training set, we may try to design a theoretical generalization of multiple embedding functions. This generalized embedding function, called $\boldsymbol G$, could be used to represent at least several different embedding methods (e.g. HILL and S-UNIWARD) with a single mathematical model. 

When applying steganography, as discussed in this paper repeatedly, there are some probabilities $\alpha$ of selecting a pixel and $\beta=\alpha/2$ of applying a $\pm1$ change on that pixel. Such probabilities differ for different embedding methods (and parameters, messages and stego keys), but still a generalized model of this probability may be obtained by analyzing the particular details of each embedding algorithm. That generalization would represent the probability that a given pixel of the image is selected (or modified) by a collection of steganographic methods and parameters, and not a single one. Once we have such a generalized probability, the generalized embedding function $\boldsymbol G$ may be implemented, making it possible to simulate a generalized embedding process. Thus, the individual embedding $\stego_p^{\#}$ used in this paper would be replaced by a generalization $\boldsymbol G^{\#}$ acting as a representative of multiple embedding functions and parameters.

All adaptive steganographic schemes try to find the areas of the images where the pixel changes are unnoticed by steganalysis. This typically corresponds to textured areas and edges. Hence, a detector of textures and edges could conveniently replace the true embedding methods, constituting the basis for the generalized function $\boldsymbol G$. We plan to explore this idea in our future research.

The most appealing side effect of using a generalized embedding function is the fact that numerical explosion, illustrated by the 100 combinations obtained for ``double stego'' images in the experiment presented in Section \ref{sec:expmult}, would be prevented. This would considerably reduce the number of samples required for training and, consequently, the computational burden would be similar to that of the standard (targeted) case, as long as the generalized embedding function can be simulated efficiently. 

Of course, some embedding methods may be too different from being generalized in this way. To cope with this situation, possibly different generalization functions would be required but, even in that case, we expect that the final method would entail less computational complexity compared to the case presented in Section \ref{sec:expmult}. 

The idea of designing a generalized embedding function is too intricate to explore it here and, as already remarked, it deserves the specific attention of a separate work in our future research.

\subsection{Unsupervised steganalysis}
The applications of subsequent embedding presented in the previous sections and appendices consider supervised steganalysis, which requires an explicit training set ($\mathsf{A^L}$). However, the directionaly of features can also be exploited for unsupervised steganalysis. This idea, called, steganalysis with Artificial Training Sets (ATS), was already proposed in our previous work \cite{Lerch-Hostalot:2015}, but is outlined here for completeness. 
A graphical representation of the ATS unsupervised steganalytic scheme based on directional features is shown in Fig. \ref{fig:ATS}. The algorithm is summarized below:
\begin{itemize}
    \item Let $\mathsf{A^T}$ be the testing data set; $\mathsf{B^T}=\stego^{\#}_p({\mathsf{A^T}})$ be the transformed testing set, obtained after embedding data in all the images of $\mathsf{A^T}$; and $\mathsf{C^T}=\stego^{\#}_p({\mathsf{B^T}})=\stego^{\#}_p({\stego_p^{\#}(\mathsf{A^T}}))$ be the ``double transformed'' testing set, obtained after embedding data in all the images of $\mathsf{B^T}$. As a result, $\mathsf{A^T}$ contains cover and stego images, $\mathsf{B^T}$ contains stego and ``double stego'' images, and $\mathsf{C^T}$ contains ``double stego'' and ``triple stego'' images. This is illustrated in Figs. \ref{fig:ATS1}, \ref{fig:ATS2}, and \ref{fig:ATS3}, respectively, for a subspace of two directional features.  
    \item Now, we can create an artificial training set formed by the sets $\mathsf{A^T}$ (`0' label) and $\mathsf{C^T}$ (`1' label) to train a classifier to learn the boundary between $\mathsf{A^T}$ and $\mathsf{C^T}$, as illustrated in Fig. \ref{fig:ATS4}. A sufficient condition to allow such a classification scenario is the existence of a subspace of directional features that keep the cover, stego, ``double stego'' and ``triple stego'' separated enough to learn the corresponding classification boundary.
    \item By exploiting directionality, the same boundary can be used to classify the transformed set $\mathsf{B^T}$ into stego  and ``double stego'' images, as shown in Fig. \ref{fig:ATS5}. With the classifier trained as shown in Fig. \ref{fig:ATS4}, the stego images in $\mathsf{B^T}$ should be classified with the same label (`0') as the images of $\mathsf{A^T}$ (which contains cover and stego images), whereas the ``double stego'' images in $\mathsf{B^T}$ should be classified with the same label (`1') as the images of $\mathsf{C^T}$ (which contains ``double stego'' and ``triple stego'' images).
    \item Finally, since there is a bijection, $\mathsf{B^T}=\stego^{\#}_p({\mathsf{A^T}})$, between the elements of $\mathsf{B^T}$ and $\mathsf{A^T}$ , it is possible to match each stego image of $\mathsf{B^T}$ with a cover image in $\mathsf{A^T}$, and each ``double stego'' image in $\mathsf{B^T}$ with a stego image in $\mathsf{A^T}$. Hence, the classification of the images in  $\mathsf{B^T}$ as stego or ``double stego'' is equivalent to the classificaion of the images in  $\mathsf{A^T}$ as cover or stego. The bijection  between the elements of $\mathsf{A^T}$ and $\mathsf{B^T}$ is recorded in order to complete the classification of the original testing set $\mathsf{A^T}$. 
\end{itemize}

In summary, the classification of $B'_i=\stego^{\#}_p(A'_i)$ as stego (`0') or ``double stego'' (`1')  is equivalent to the classification of $A'_i$ as cover (`0') or stego (`1'), respectively. Consequently, we can classify $\mathsf{A^T}$ without labeled samples and, by definition, this is \textbf{unsupervised classification}.

Theoretical proofs and experimental results for the ATS steganalytic scheme in different scenarios, including CSM and unknown message lengths, can be found in \cite{Lerch-Hostalot:2015}.

\color{black}

\end{document}